\documentclass[pdflatex,sn-mathphys-num]{sn-jnl}

\usepackage{graphicx}%
\usepackage{multirow}%
\usepackage{amsmath,amssymb,amsfonts}%
\usepackage{amsthm}%
\usepackage{mathrsfs}%
\usepackage[title]{appendix}%
\usepackage{xcolor}%
\usepackage{textcomp}%
\usepackage{manyfoot}%
\usepackage{booktabs}%
\usepackage{wasysym}
\usepackage{algorithm}%
\usepackage{algorithmicx}%
\usepackage{algpseudocode}%
\usepackage{listings}%
\usepackage{subcaption}%
\usepackage{rotating} 

\theoremstyle{thmstyleone}%
\theoremstyle{thmstyletwo}%

\theoremstyle{thmstylethree}%

\usepackage{xspace}
\newcommand{\degree}{\ensuremath{^\circ}\xspace}

\newcommand{\htp}{\ensuremath{\mathrm{H}_2^+}\xspace}
\newcommand{\nuebar}{\ensuremath{\bar{\nu}_e}\xspace}

\newcommand{\smco}{Sm$_2$Co$_{17}$\xspace}
\newcommand{\ndfe}{Nd$_2$Fe$_{14}$B\xspace}

\begin{document}

\title[IsoDAR@Yemilab: PDR - Vol. I]{IsoDAR@Yemilab: Preliminary Design Report - Volume I: Cyclotron Driver}



\author[]{\sur{The IsoDAR Collaboration}}
\author*[1]{\fnm{Daniel} \sur{Winklehner}}\email{winklehn@mit.edu}
\author[2]{\fnm{Michel} \sur{Abs}}
\author[1]{\fnm{Jose R.} \sur{Alonso}}
\author[1]{\fnm{Janet M.} \sur{Conrad}}
\author[1]{\fnm{Samuel J.} \sur{Engebretson}}
\author[2]{\fnm{Eric} \sur{Forton}}
\author[2]{\fnm{Alexander T.} \sur{Herrod}}
\author[2]{\fnm{Denis} \sur{Joassin}}
\author[1]{\fnm{Jarrett} \sur{Moon}}
\author[2]{\fnm{S\'ebastien} \spfx{de} \sur{Neuter}}
\author[2]{\fnm{Erik} \spfx{Van der} \sur{Kraaij}}
\author[2]{\fnm{Gil} \sur{W\'ery}}
\author[1]{\fnm{Eleanor} \sur{Winkler}}
\author[3]{\fnm{Andreas} \sur{Adelmann}}
\author[4]{\fnm{Spencer N.} \sur{Axani}}
\author[1]{\fnm{William A.} \sur{Barletta}}
\author[5]{\fnm{Roger} \sur{Barlow}}
\author[6]{\fnm{Larry} \sur{Bartoszek}}
\author[1]{\fnm{Adriana} \sur{Bungau}}
\author[7]{\fnm{Luciano} \sur{Calabretta}}
\author[8]{\fnm{Pedro} \sur{Calvo}}
\author[9]{\fnm{Georgia} \sur{Karagiorgi}}
\author[8]{\fnm{Concepti\'on} \sur{Oliver}}
\author[9]{\fnm{Michael H.} \sur{Shaevitz}}
\author[10]{\fnm{Jon} \sur{Ameel}}
\author[10]{\fnm{Andrew} \sur{Chan}}
\author[10]{\fnm{Emilie} \sur{Lavoie-Ingram}}
\author*[10]{\fnm{Joshua} \sur{Spitz}}\email{spitzj@umich.edu}

\affil[1]{\orgdiv{Laboratory for Nuclear Science}, \orgname{Massachusetts Institute of Technology}, \orgaddress{\street{77 Massachusetts Ave}, \city{Cambridge}, \postcode{02139}, \state{MA}, \country{USA}}}

\affil[2]{\orgname{Ion Beam Applications}, \orgaddress{\street{3 chemin du cyclotron}, \city{Louvain-la-Neuve}, \postcode{1348}, \country{Belgium}}}

\affil[3]{\orgname{Paul Scherrer Institut}, \orgaddress{\city{Villigen PSI}, \postcode{5232}, \country{Switzerland}}}

\affil[4]{\orgdiv{Department of Physics \& Astronomy}, \orgname{University of Delaware}, \orgaddress{\street{210 South College Ave.}, \city{Newark}, \postcode{19716}, \state{DE}, \country{USA}}}

\affil[5]{\orgname{University of Huddersfield}, \orgaddress{\city{Huddersfield}, \postcode{HD1 3DH}, \country{UK}}}

\affil[6]{\orgname{Bartoszek Engineering, P.E.}, \orgaddress{\street{818 W. Downer Place}, \city{Aurora}, \postcode{60506}, \state{IL}, \country{USA}}}

\affil[6]{\orgname{INFN Laboratori Nazionali di Legnaro}, \orgaddress{\city{Legnaro}, \country{Italy}}}

\affil[7]{\orgname{Bartoszek Engineering, P.E.}, \orgaddress{\street{818 W. Downer Place}, \city{Aurora}, \postcode{60506}, \state{IL}, \country{USA}}}

\affil[8]{\orgname{Centro de Investigaciones Energ\'eticas, Medioambientales y Tecnol\'ogicas}, \orgaddress{\city{Madrid}, \country{Spain}}}

\affil[9]{\orgdiv{Department of Physics}, \orgname{Columbia University}, \orgaddress{\city{New York}, \state{NY} \postcode{10027}, \country{USA}}}

\affil[10]{\orgdiv{Physics Department}, \orgname{University of Michigan}, \orgaddress{\street{450 Church Str}, \city{Ann Arbor}, \postcode{48109}, \state{MI}, \country{USA}}}

\abstract{This Preliminary Design Report (PDR) describes the IsoDAR electron-antineutrino
source in two volumes which are mostly site-independent and describe the cyclotron driver providing a 60 MeV, 10~mA proton beam (this Volume); and the medium energy beam transport line (MEBT) and target (Volume II). The IsoDAR driver and target will produce about $1.15\cdot10^{23}$ electron-antineutrinos over five years. Paired with a kton-scale 
liquid scintillator detector, it will enable a broad particle physics
program including searches for new symmetries, new interactions and new particles. 
Here in Volume I, we describe the driver, which includes the ion source, low energy beam transport, and cyclotron. The latter features Radio-Frequency Quadrupole (RFQ) direct axial injection and represents the first accelerator purpose-built to make use of so-called vortex motion.}




\maketitle

\tableofcontents

\clearpage
\section{Introduction}\label{sec1}

\begin{figure}[t!]
\begin{center}
    \includegraphics[width=1.00\textwidth]{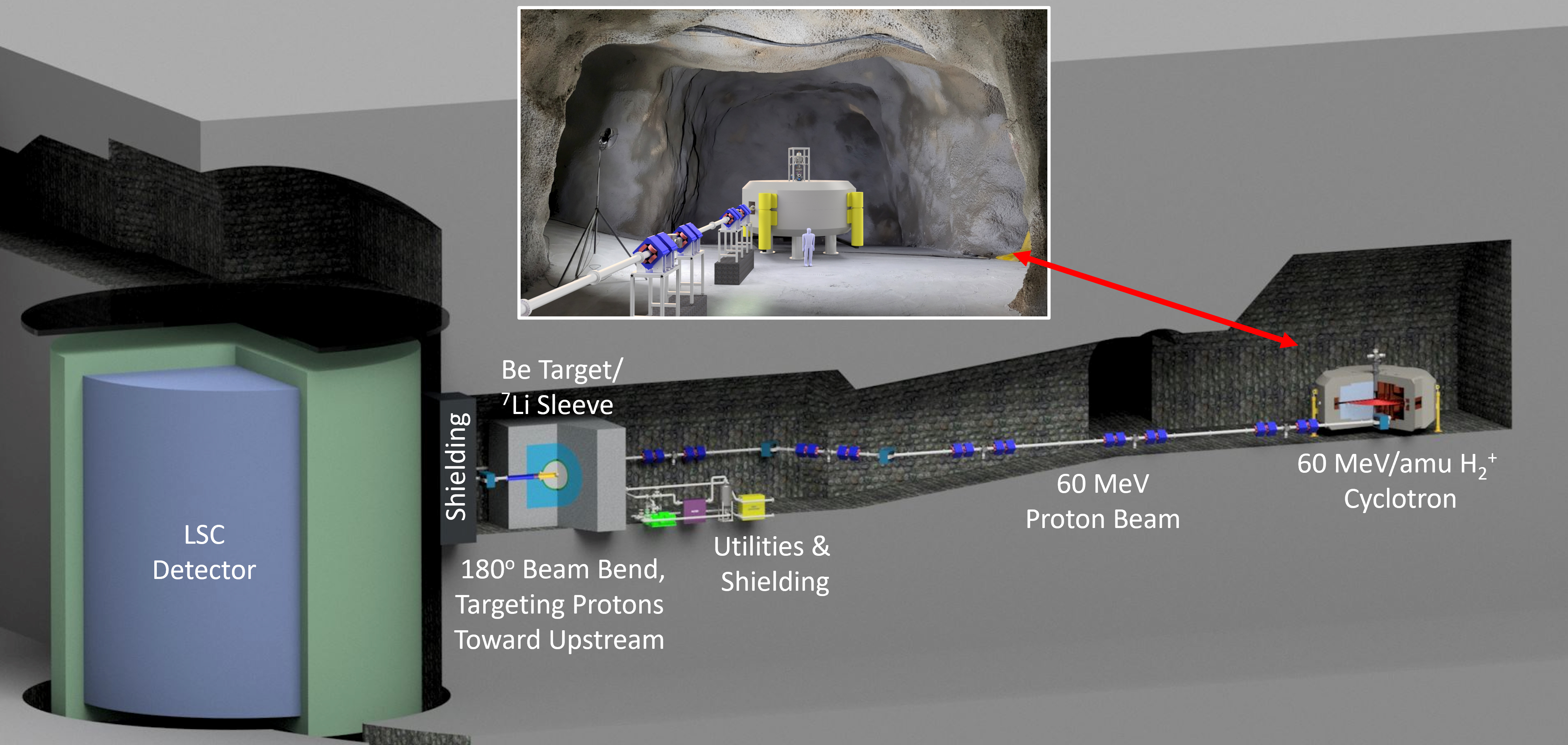}
    \caption{Main: The IsoDAR accelerator and target system adjacent to the LSC detector at Yemilab. Top: A to-scale rendering of the cyclotron overlaid on a photo of the cavern. The project comprises four major systems, three of which are shown: 1) the cyclotron; 2) the 60~MeV beam transport; and 3) the target and sleeve.  The fourth is a data acquisition and monitoring system for all components.}
    \label{fig:isodar_yemilab}
\end{center}
\end{figure}

If successful, the Isotope Decay-At-Rest experiment (IsoDAR) will be the first high-intensity, proton-driven source to be located alongside an underground multi-kiloton-scale scintillator detector.
The first version of IsoDAR is planned to be constructed at the Yemilab facility~\cite{alonso_isodaryemilab_2022} within the Handuk mine, located in South Korea where the ``Liquid Scintillator Counter'' (LSC)\footnote{The LSC was renamed $\nu$EYE in March 2025, but for consistency throughout this document, we will refer to the detector as the LSC.} is planned to be installed~\cite{seo_physics_2023}. 
The combination of proton driver (a compact cyclotron with direct radio-frequency quadrupole, or ``RFQ'', injection), beam transport line, and target/sleeve arrangement provides the particles (neutrinos, neutrons, and photons) for the physics experiments described in Section~\ref{sec:physics}. Hence, we call the full system (shown in Fig.~\ref{fig:isodar_yemilab})
the ``IsoDAR source'', or simply ``source''. We estimate that in five calendar years (four live years), the source produces 1.67 million $\bar{\nu}_e + p \rightarrow e^+ + n$ (inverse beta decay or ``IBD'') events and 7000 $\bar \nu_e + e^- \rightarrow \bar \nu_e + e^-$ (elastic scattering or ES) events in the LSC that can be used for searches for new physics. Here, \nuebar are electron antineutrinos, $p$ are protons, $n$ are neutrons, and $e^-$ and $e^+$ are electrons and positrons, respectively. In addition, neutrons and mono-energetic photons are also produced in the target (and largely contained there) while the accelerator is running. These may act as progenitors for new particles that enter the LSC and leave signatures from interactions and decays. IsoDAR can address a broad range of new physics manifested as new symmetries, new interactions, and new particles that are inaccessible to traditional experiments (cf. Section~\ref{sec:physics} and references therein).

This document is Volume I in a two-volume Preliminary Design Report (PDR) for IsoDAR that provides the engineering details necessary to establish credible solutions to the key issues and cost-drivers identified during the past decade of conceptual design and R\&D. The contents of the PDR will cover the Technical Facility:  the cyclotron, the beam transport and the target/sleeve. This document builds on the structure and text of the Conceptual Design Report (CDR) of the IsoDAR Technical Facility~\cite{abs_isodarkamland_2015}.  
This step of producing a PDR substantially reduces the risk related to IsoDAR approval. 
The remaining detailed engineering that will follow will build on these results and will be straightforward to estimate in terms of scope and cost.
  
This document presupposes a basic understanding of cyclotrons. For an introduction to cyclotrons and a cost/benefit analysis of this choice of machine for IsoDAR, see the discussion in the CDR.

The 60~MeV/amu IsoDAR Proton Driver is exceptional in that it is designed and engineered to produce an order of magnitude higher current proton beams at extraction than commercial machines of similar energy. Three novel concepts set this accelerator apart from commercial ones: First, the cyclotron accelerates \htp, mitigating ``space charge effects" (electromagnetic interactions between beam particles). Second, the beam is axially injected through an RFQ, acting as a high-efficiency buncher. Third, the design harnesses ``vortex motion'', a complex effect previously observed at the Paul Scherrer Institute at the PSI Injector II cyclotron \cite{stetson:vortex, baumgarten:vortex1}. Vortex motion induces spiraling within the individual beam bunches, providing compression and stability, which leads to substantially lower losses during extraction. For more information concerning the accelerator physics that shapes the IsoDAR design, see Refs.~\cite{winklehner_order--magnitude_2022, winklehner_high_2018}.

These three concepts drive many of the engineering choices presented in this PDR. The other design driver is the need to assemble the cyclotron underground. The cavern for the IsoDAR source has already been constructed at Yemilab, and so the dimensions and constraints are understood, leading to realistic engineering plans.

This PDR is structured as follows:
\begin{itemize}
\item {\bf Chapter 2:} Proceeds with an overview of the physics
searches that will be possible with IsoDAR.
\item {\bf Chapter 3:} Presents a brief overview of the requirements for the proton driver.
\item {\bf Chapter 4:} Describes engineering related to the IsoDAR front end.
\item {\bf Chapter 5:} Describes engineering related to the general IsoDAR cyclotron design.
\item {\bf Chapter 6:} Describes specifics of the design related to installation underground.
\end{itemize}

\section{IsoDAR Physics That Informs Design Goals \label{sec:physics}}

The physics case for IsoDAR has driven the design decisions that are reported in this Preliminary Design Report.     In this section, we provide a relatively high-level overview of the physics goals and connect them to the design needs, providing references for the reader to pursue more details.

Today, particle physicists build massive underground detectors to capture low-energy interactions with unparalleled precision, while accelerator physicists develop powerful machines for new physics above ground; but these instruments have never been joined in close proximity.  IsoDAR aims to change that paradigm.  It will be the first high-intensity accelerator located underground, next to a multi-kiloton-scale detector, opening entirely new windows for particle physics.

The designs presented in this document are largely site-independent. Still, we will use details of the planned initial run 
at Yemilab in Korea to provide a concrete description of the physics impact. Figure~\ref{fig:isodar_yemilab} shows a 3D rendering of IsoDAR's planned installation at Yemilab. The volumes of this PDR focus on the IsoDAR source, which consists of the elements to the right of the LSC (cylinder at left). More information on the LSC is available in Refs. \cite{seo_physics_2023, alonsoIsoDARYemilabReportTechnology2022, kimYemilabNewUnderground2024}. Important quantities within this layout that impact the physics are listed in Table~\ref{tab:assumptions}. 
Although this first volume concentrates only on the accelerator, the physics case relies on all systems, and this chapter will also be relevant to Volume 2 of the PDR.

\begin{table}[t] 
\centering
\caption{Specifications for the IsoDAR@Yemilab design that directly impact the physics opportunities. See text for explanations of the live-time, fiducial volume cut and energy minimum cut.}
\label{tab:assumptions}
\begin{tabular}{|c|c|} 
\hline
\multicolumn{2}{|c|}{\textbf{Accelerator Run Parameters}} \\
\hline
Runtime  &  5 years  \\ 
\hline
Duty factor  &  80\%  \\ 
\hline
Livetime  &  4 years  \\ \hline
Protons on IsoDAR target/year  &  $1.97\cdot 10^{24}$  \\ \hline
\multicolumn{2}{|c|}{\textbf{Expected total antineutrino flux in 4 yr livetime}} \\
\hline
$\bar{\nu}_e$  &  $1.15\cdot 10^{23}$  \\ \hline
Source production radius (1$\sigma$) & 0.41 m \\ \hline
\multicolumn{2}{|c|}{\textbf{LSC design parameters}} \\
\hline
LSC fiducial mass  & 2.26 kton \\ \hline
Energy resolution at 3 MeV & 3.7\% \\ \hline
Baseline distance from source  &   17~m  center-to-center (9.5-25.6 m) \\ \hline
Depth of LSC and source & 2700 m.w.e. \\ \hline
\multicolumn{2}{|c|}{\textbf{SM rates with livetime, energy and fiducial volume assumptions}} \\
\hline
IBD ($\bar{\nu}_e+p \rightarrow e^+ + n$) &  1.67 million \\ \hline
ES ($\bar \nu_e + e^- \rightarrow \bar \nu_e +e^-$) & 7000 \\ \hline
\end{tabular}
\end{table}

\subsection{Introduction to Design for BSM Searches in IsoDAR}

 One typically searches for Beyond Standard Model (BSM) physics through its effect on Standard Model (SM) particles in three ways.  First, through the production of a new particle via coupling to an SM particle that subsequently interacts or decays.   Second, through the observation of new symmetries that introduce unexpected behavior of SM particles. Third, modifications to SM interactions due to new forces that couple to SM particles.     From this, one sees that a crucial key to the discovery of new physics lies in the SM flux available to an experiment.     The higher the quality and quantity of the SM flux, the deeper the reach in the search for new physics.

\begin{figure}[tb!]       
\begin{center}
\includegraphics[width=\textwidth]{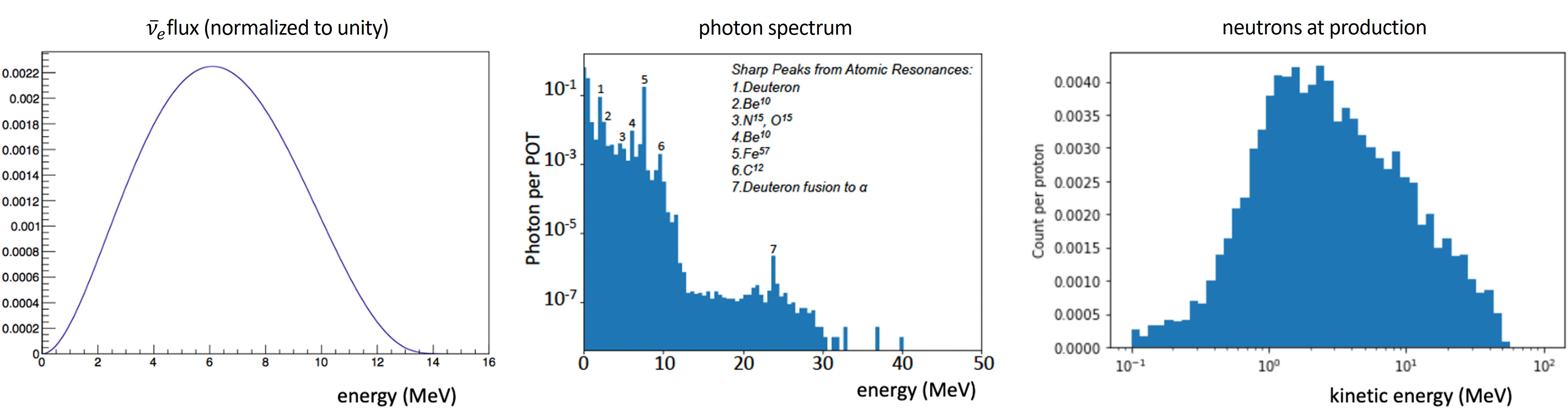}
\end{center}  
\caption{\label{threeflux} Energy dependence of fluxes.  Left: $\bar \nu_e$, normalized to unity;  Middle: photons per Proton On Target (POT) from excited nuclei. Right: kinetic energy of neutrons at production in the target and sleeve, normalized per proton.}
\end{figure}

The IsoDAR system focusses 60 MeV protons onto a beryllium target that is surrounded by a 99.99\% pure $^7$Li sleeve that, in turn, is embedded in substantial shielding.   Given this design, IsoDAR 
produces three unique SM fluxes for BSM searches:  electron antineutrinos, monoenergetic photons and neutrons.   We discuss the qualities and quantities of these fluxes below, and they are shown in Fig.~\ref{threeflux}.   These three SM fluxes fall into two categories of BSM physics that can be explored.   
The first category consists of SM particles that can exit the IsoDAR source and cross the shielding, ultimately producing signatures in the detector — electron antineutrinos.    
The second category is the set of copiously produced SM particles — the photons and neutrons that interact electromagnetically and strongly, respectively.  A new BSM particle can couple to these contained fluxes and, if it has a very low interaction cross-section, will traverse the shielding and enter the detector.  To be observed, the particle must subsequently mix with, interact with, or decay to SM particles that are then observable in the LSC.    The IsoDAR design has three crucial advantages for BSM searches using these particles:
\begin{enumerate}
\item The rate of SM particle production is very high.
\item The solid angle to the LSC detector, 17~m from the source, is very large.
\item The cosmogenic and environmental backgrounds are very low.
\end{enumerate}
We consider the importance of each of these points below.

\subsection{Designing IsoDAR's Electron Antineutrino Flux}

\begin{figure}[t!]
\begin{centering}
\centering
\includegraphics[width=0.6\columnwidth]{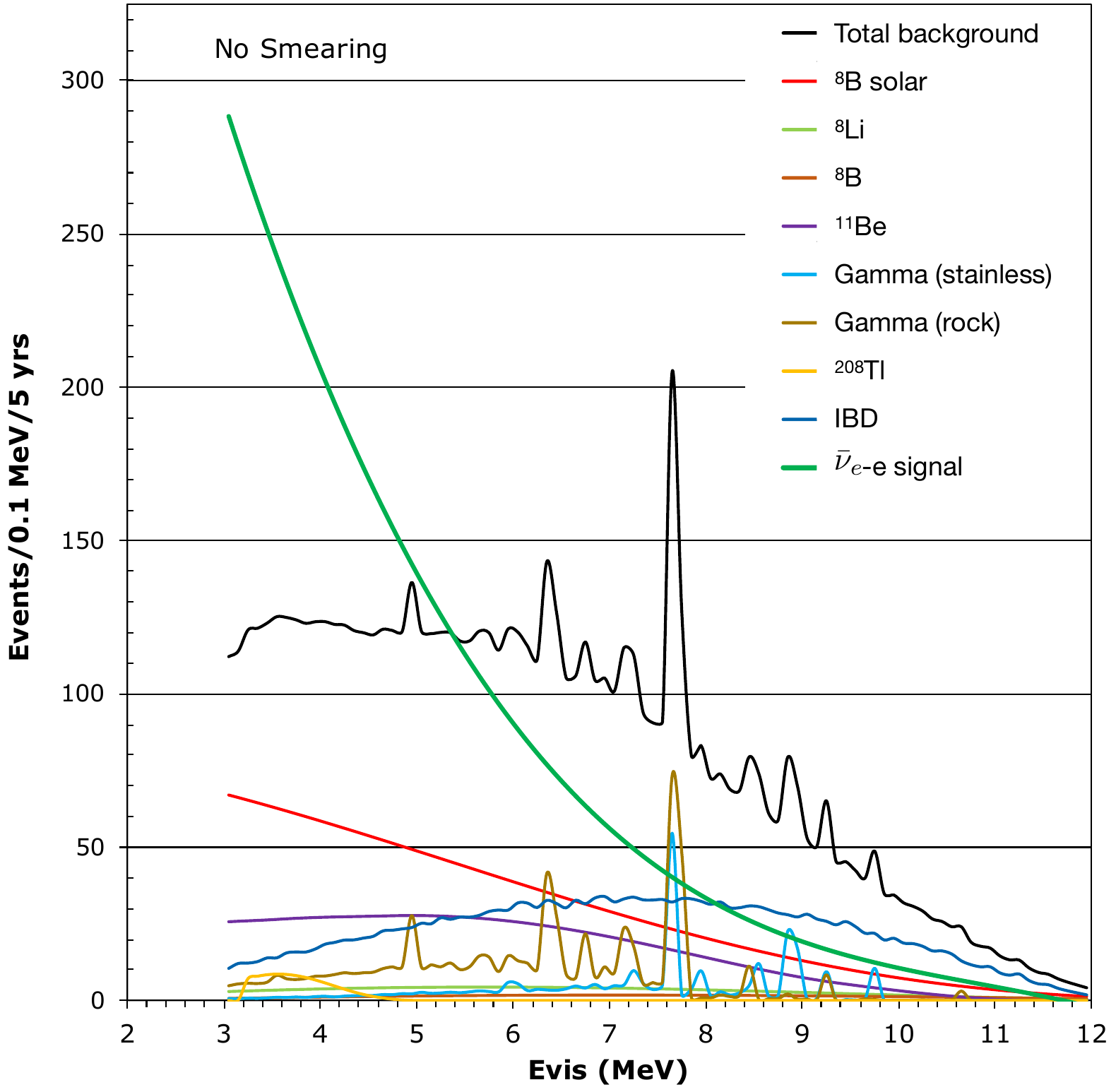}
\caption{The ES signal and background event rates expected in 5~years of running IsoDAR@Yemilab. No energy smearing has been applied to the distributions. From Ref.~\cite{alonso_neutrino_2022}.}
\label{fig:ESbackgrounds}
\end{centering}
\end{figure}

IsoDAR was originally designed for BSM searches using the electron-antineutrino flux.  Although the physics program has since expanded to include state-of-the-art searches using both photon and neutron fluxes, the collaboration has prioritized the quality of the electron-antineutrino flux as a primary design goal.    The quality issues that have been maximized by the design are:  understanding the spatial dependence of the flux, understanding the energy dependence of the flux, and purity of antineutrino content (i.e., minimizing backgrounds).  The quantity of the electron-antineutrino flux is also a design priority, where rates ahve been balanced against cost and practicality of installing and running at an underground site.    The outcome of the design is a world-leading rate of electron-antineutrino interactions, which we quote for a live time of 4 years.   IsoDAR will collect 1.67 million $\bar{\nu}_e+p \rightarrow e^+ + n$ (inverse beta decay or ``IBD’’) events and 7000 $\bar \nu_e + e^- \rightarrow \bar \nu_e +e^-$ (elastic scattering or ``ES’’) events in five calendar years—unprecedented sample sizes to probe new interactions \cite{alonso_neutrino_2022}.      These are the only two interactions open within the SM for $\bar \nu_e$ scattering at $\sim 6$ MeV.   The IBD signature requires a coincidence between the initial positron light-flash and the subsequent neutron capture, which greatly reduces backgrounds compared to the single-flash ES signature.  To further minimize background, both the positron (for IBD) and electron (for ES), signals are required to be reconstructed to $>3$ MeV and to be within the 2.26~kt (1.15~kt) fiducial volume for IBD (ES) events.   The resulting backgrounds to the IBD signal are found to be negligible, while the assumed backgrounds for physics from the ES signal, with all cuts, are shown in Fig.~\ref{fig:ESbackgrounds}.

Some of the neutrino physics goals require precise knowledge of the spatial distribution of the beam.    To this end, an isotope decay-at-rest (hence the name IsoDAR) flux was selected.   This flux has the advantage of being produced in $4\pi$ steradians,  and the relative flux within a given solid angle is straightforward to determine.    

An isotope that $\beta$-decays with high endpoint (or Q-value) is desirable since 
the interaction rate in the detector increases with energy.   After studying many isotopes, the collaboration has chosen $^8$Li, which has the well-predicted $\beta$ decay spectrum~\cite{bungau_proposal_2012}, shown in Fig.~\ref{threeflux}, that peaks at approximately 6 MeV and has an endpoint of 16 MeV, minimizing systematic uncertainties on the energy dependence of the flux.
By choosing a flux with high endpoint,  we can apply a visible energy cut above 3 MeV, removing most environmental background events while retaining 10\% of the total flux.

The tradeoff is that a high Q-value corresponds to a short beta decay lifetime—839 ms for $^8$Li—requiring continuous replenishment via an accelerator-based system that produces neutrons that capture on $^7$Li. This leads to several design choices for the target, the surrounding ``sleeve'' and the accelerator that we discuss below.  

The design requires a target that generates high quantities of neutrons at energies that must then be moderated efficiently to allow capture. After evaluating alternatives and considering radiation handling issues, we selected $^9$Be as the target, since its final neutron is bound by only 1.7~eV. A tungsten target would release more neutrons per proton on target, but would also generate unwanted neutrinos in the flux. An important design choice was to cool the target using a 17-liter closed-loop system of D$_2$O, as opposed to H$_2$O, which has a high neutron capture cross section.  

The neutrons exit the target and enter a surrounding region with $^7$Li, which we refer to as the ``sleeve.'' This sleeve serves three purposes: to multiply neutrons, to moderate fast neutrons, and to provide the $^7$Li target for capture.   Natural lithium is a mixture of $^6$Li and $^7$Li. Due to the very high $^6$Li neutron capture cross section, we require isotopic purification to 99.99\% $^7$Li. Accessing 480 kg of $^7$Li at this purity has required coordination with industry.   A cost-effective method to 
further enhance neutron production is to add beryllium to the sleeve. As the fast neutrons moderate, they inelastically scatter off the beryllium, releasing additional neutrons. We found the $^8$Li production rate is maximized with a  25\% $^7$Li / 75\% $^9$Be mixture. 
The goal is to maximize neutron absorption in $^7$Li while minimizing losses in all other elements of the target and sleeve and minimize background neutrino flux rates, hence the design focuses on low-Z materials with low capture cross-sections, including, for example, a carbon neutron reflector surrounding the sleeve.   With the design presented here, the background neutrino production rate above the 3 MeV analysis cut is negligible compared to the signal anti-neutrino rate.

Having selected the isotope, the next design choice is the best accelerator to drive the system.   The 839 ms lifetime of $^8$Li is very long compared to the RF structure in nearly all accelerators, opening a wide array of choices---including cyclotrons that operate in ``continuous wave'' mode.    Ref.~\cite{adelmann_cost-effective_2012} provides a detailed analysis of the cost-effectiveness leading to our choice of a 60 MeV/amu H$_2^+$ compact cyclotron design.
In comparison, Linacs are expensive to build and operate, with high power demands. They produce low-emittance beams, which is actually a drawback for us, as we need to spread the beam power to protect the target. FFAs remain in the R\&D stage, lack industry partners, and have a large footprint. No high-rate DT generator currently meets our neutron flux requirements, though designs are improving. However, enclosing a large DT generator box with the required $^7$Li sleeve would significantly increase costs.
An H$_2^+$ cyclotron outperforms a D$^+$ accelerator. Simulations show that fast neutrons from accelerated D$^+$ often escape the sleeve without being captured, thereby reducing efficiency. The 60~MeV/amu energy level provides the optimal balance between a compact underground-friendly footprint and sufficient neutron production to bathe the $^7$Li sleeve, generating the necessary antineutrino flux.
In summary, we have chosen a design that targets 10~mA of protons at 60~MeV, as described in volume 1 of this preliminary design report.

In the case of the ES single-flash signal, beam-off running allows measurement of environmental backgrounds above 3~MeV.   This data can be collected during the 20\% time period when the accelerator is down for maintenance. In the future, we may also investigate introducing beam chopping, which allows for nearly simultaneous beam-on and beam-off running.

\subsubsection{The Contained Monoenergetic Photon Flux}

As the neutrons moderate in the sleeve,  nuclei are excited through collisions.  The excited nuclei decay, producing monoenergetic photons.  These are absorbed within the target and sleeve material; however, prior to absorption, the photons can potentially mix with BSM particles that can subsequently enter the detector and decay or interact inside.   

The total energy of the resulting BSM particle is also monoenergetic.   This energy signature allows for discovery through searches for sharp peaks in the energy measured within the detector if the BSM particle decays to electrons.  Broad peaks are produced if the BSM particle interacts to produce scattered electrons.    These signatures have a single electromagnetic flash and, therefore, must be differentiated from those environmental backgrounds with energies higher than the 3 MeV analysis cut that are seen in Fig.~\ref{fig:ESbackgrounds}.     However, beam-off running allows measurement of the environmental backgrounds and, if peaks are relatively narrow, the background estimate can be obtained by extrapolation from above and below the peak energies.

The set of monoenergetic peaks produced by photons in the source region is seen in Fig.~\ref{threeflux}.  The highest flux is from isotopes produced by neutrons that are captured in the iron shielding that surrounds the source.  We discuss design considerations for the shielding under the neutron flux, below.      The second-highest peak, resulting from the breakup of deuterons in the cooling water, at 2.2~MeV, is below the 
3~MeV analysis cut.   Thus, at present, we do not consider possible BSM physics searches from this peak, but may do so in the future.   Below these, a range of peaks extending above 10 MeV can provide progenitor photons for BSM particles.  In summary, although the design has not been optimized for this physics (it is optimized for the electron antineutrino flux), this flux is sufficiently high to open doors to new physics.

\subsubsection{The Contained Neutron Flux}

The neutrons may also be progenitors of BSM particles that can penetrate the shielding around the source region.   Neutron--dark sector couplings are motivated by the neutron beam/bottle experiment lifetime discrepancy \cite{Nico:2004ie,Yue:2013qrc,Serebrov:2017bzo,UCNt:2021pcg} . This
allows them to disappear from the source area and reemerge in the LSC's central and far side through \mbox{$n \rightarrow n^\prime \rightarrow n$} transitions ($n\ldots$ neutron, $n^\prime\ldots$ dark sector neutron). IsoDAR has much higher sensitivity than surface experiments to this signature due to the large LSC detector and reduced backgrounds \cite{Hostert:2022ntu}.  

The signature for the neutron reappearance is a single flash from neutron capture in the center of the LSC detector. The 2.2 MeV capture on H is below the 3 MeV cut for analysis, but the 4.95 MeV capture on C, which occurs 1\% of the time, will be observable within the analysis cut.

The most obvious background that must be controlled for this analysis is  neutron punch-through from the source. IsoDAR has performed extensive GEANT-4 studies of shielding designs for Yemilab.  These are presented in Ref.~\cite{bungauNeutrinoYieldNeutron2024}, but not discussed in detail in this preliminary design report, which discusses the elements of the design that are not Yemilab specific. (Yemilab specific designs will be collated into a separate design report.)
The present design reduces neutron levels above 3 MeV reaching the outside of the Yemilab detector to just a few neutrons per year.   For neutrons of any energy to penetrate from the outer wall of the detector to the fiducial volume, they must pass through 1.5~m of oil in the veto region, and 1.0~m of oil in the buffer region without capturing. If further protection from penetrating neutrons is desired, cuts can be placed in the fiducial region. In addition, we note that beam-off running allows studies of the environmental backgrounds.

\subsection{Connecting the Fluxes to the IsoDAR Physics Program}

The purpose of the design is to allow 
IsoDAR to offer a rich BSM physics program if paired with a kton-scale liquid scintillator detector like the 
LSC at Yemilab~\cite{seo_physics_2023}. 
Here we briefly discuss some of the unique capabilities of the IsoDAR BSM physics program, dividing the discussion into three categories: new particles, new symmetries, and new interactions. Details of these example analyses are beyond the scope of this design report, but can be found in the references.

\subsubsection{New Particles}

\begin{figure}[t!]
\begin{center}
\vspace{-0.2in}
{\includegraphics[width=0.6\textwidth]{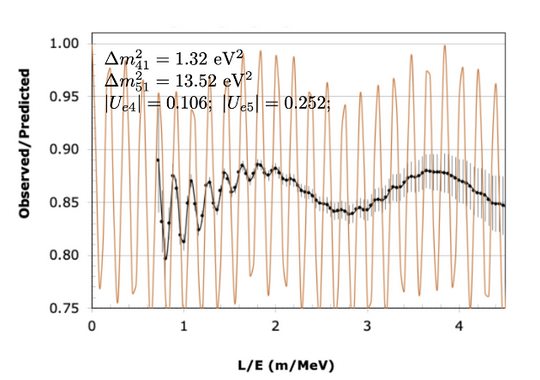}}
\vspace{-0.2in}
\end{center}
\caption{An example of a potential 3+2 signal in IsoDAR, orange--true signature, black--IsoDAR reconstruction with uncertainty shown. The true oscillations are dampened due to limited statistics and energy smearing. The oscillation parameters chosen for this example are allowed by today's experiments \cite{Diaz:2019fwt}.}
\label{3+2}
\vspace{-0.1in}
\end{figure}

\paragraph{\it Example 1:  Additional neutrinos---}
Many new physics models introduce additional neutrinos beyond the three Standard Model (SM) flavors that have feeble or even zero SM couplings, though they may interact through new BSM couplings~\cite{Arguelles:2022tki}.  For light new neutrinos (with masses less than 10 MeV) it is feasible to design sensitive neutrino oscillation experiments.  The probability that a neutrino interacts through the SM will vary with $L/E$, where
 $L$ is the distance traveled, and $E$ is the neutrino energy.  The signature in the case of IsoDAR is a disappearance pattern of the $\bar \nu_e$ flux as a function of $L/E$ that will be distinct depending on the model.  For example, Fig.~\ref{3+2} shows the complex flux disappearance pattern that arises for the case where two additional light, non-interacting neutrinos are added to the three SM neutrinos, hence called ``3+2,'' for an example set of oscillation parameters allowed by today's experiments \cite{Diaz:2019fwt}.  Table~\ref{tab:assumptions} gives the assumptions for the IsoDAR measurements in Fig.~\ref{3+2}.  The fine measurement capability in $L/E$ will allow the observed data to precisely constrain the oscillation pattern [i.e., separating (3+1) versus (3+2)], greatly narrowing the theoretical models that can produce an effect.

 This analysis uses the $\bar \nu_e$ flux.   Most of the design choices for the IsoDAR source and the pairing with the LSC were driven by optimizing the sensitivity to this physics.    In addition to optimizing the detector parameters associated with $E$ and $L$, very high rates are crucial, as it allows for the fine $L/E$ reconstruction that may reveal the signature of new physics.   The use of the IBD signature, which exhibits coincident flashes, reduces backgrounds to a negligible level, and the measurement is statistics-limited.  The figure of merit typically used to compare experiments is the capability to resolve the amplitude of disappearance in the simplest case of one additional neutrino (``3+1'').   At 5$\sigma$, IsoDAR can resolve signals that are well beyond today's best limits across large swaths of parameter space, which far exceeds competing proposals.    See Ref.~\cite{Arguelles:2022tki} for further context within the broader particle physics program.

\begin{figure}[t!]
\begin{center}
{\includegraphics[width=\textwidth]{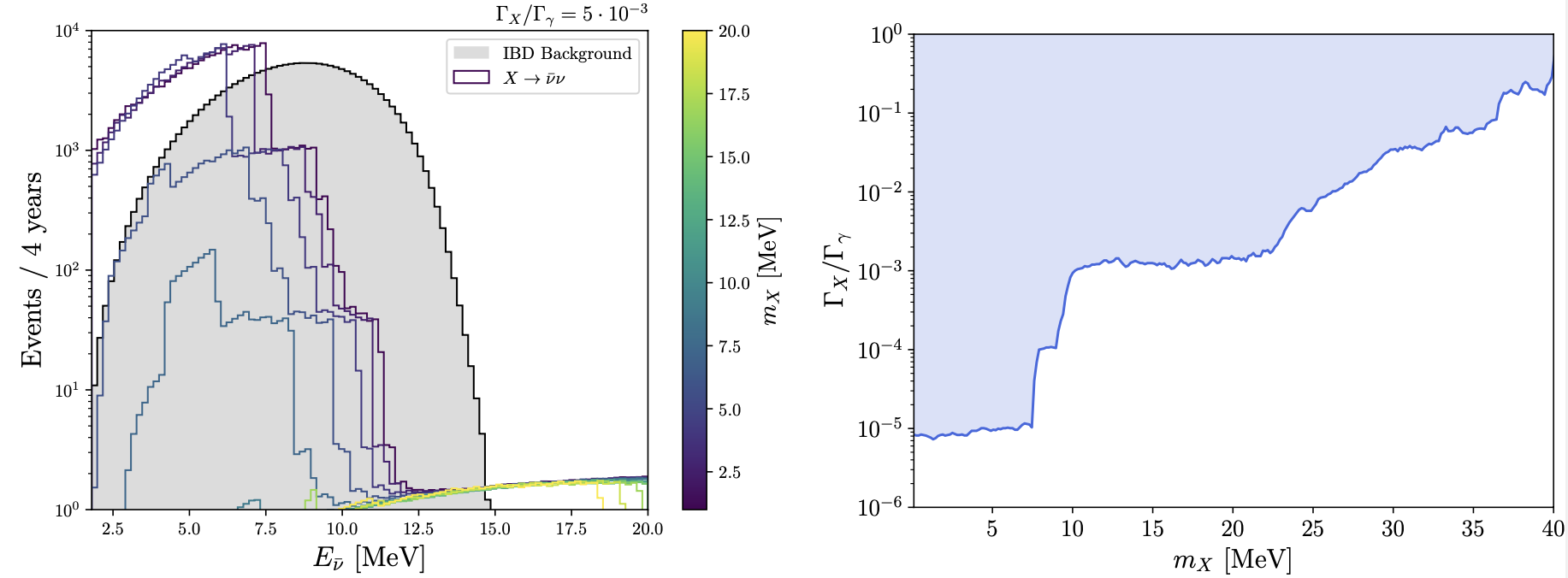}}
\end{center}
\caption{Left:  Example $X\rightarrow \nu \bar\nu$ signatures for 0.5\% photon mixing with $^8$Li-produced $\bar \nu_e$ background (gray).   Right: $X$ mass and $X$-to-photon production ratio 90\% CL sensitivity. From Ref.~\cite{alonso_neutrino_2022}.}
\label{IsoDARbump}
\vspace{-0.1in}
\end{figure}

\paragraph{\it Example 2:  Production of a new particle coupling to photons---}
An entirely new design enables searches for previously unknown particles.
For an IsoDAR example, consider a new 
low mass mediator, which we will call ``$X$'', that mixes photons produced in the 
target and decays to $\bar \nu \nu$, where the $\bar \nu$ is observable through the IBD interaction  \cite{alonso_neutrino_2022}.  

This analysis used the contained monoenergetic photon flux.   The monoenergetic nature of the flux, hence the total energy of the $X$, leads to distinct peaks in the energy deposited by the $\bar {\nu}_e$ interactions. Fig.~\ref{IsoDARbump} (left) displays resulting signatures for various masses (see color bar) assuming 0.5\% $X$-to-photon production. The IBD background from $\bar \nu_e$ stemming from $^8$Li decays (gray) constitutes the primary background and has a distinct shape. The sensitivity is depicted in Fig.~\ref{IsoDARbump} (right), where the blue region is excluded in the case of the SM. No existing limits are shown because this search is a first-of-its-kind. 

\subsubsection{New Symmetries}
\begin{figure}[b!]
\begin{center}
{\includegraphics[width=0.6\textwidth]
                 {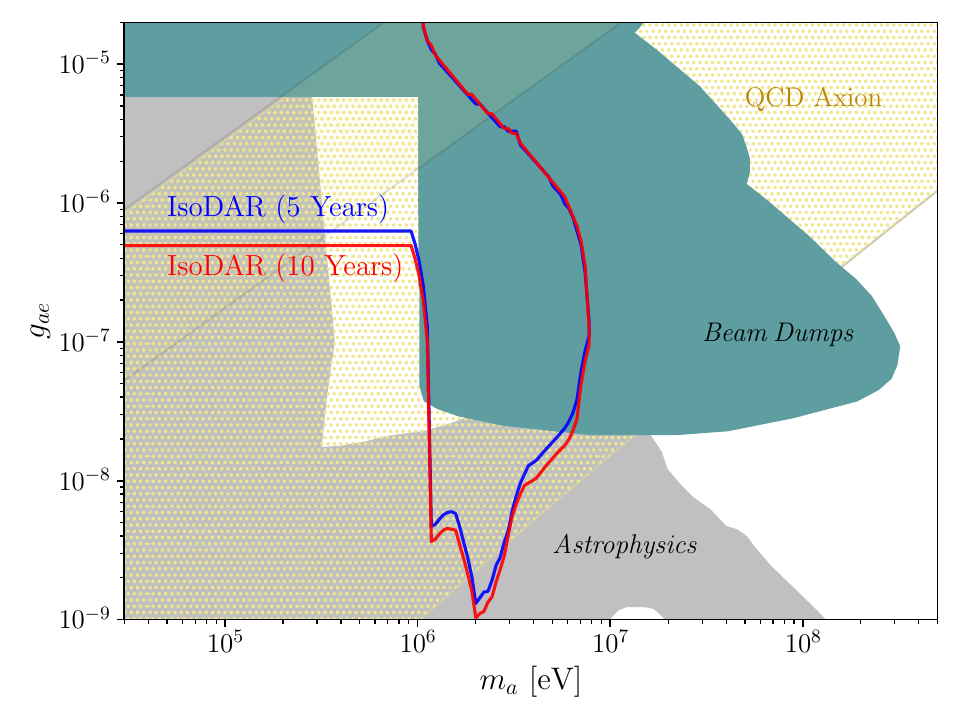}}
\end{center}
\caption{IsoDAR sensitivity for axions (a) with mass $m_a$ coupling ($g$) to electrons  (lines) \cite{waites_axionlike_2023}.
Gold dots--axion allowed; Teal--excluded regions; Gray--parameters with astrophysical implications.}
\label{ae}
\end{figure}

\paragraph{\it Example 1:  A Dark Sector---}

Dark sector models were initially motivated by the existence of dark matter, which cannot be accommodated within the SM. The model postulates a ``dark sector'' of particles and forces that interact with the SM through ``portals.''  Although motivated by the dark matter puzzle, dark sector models have expanded to address other theoretical and experimental puzzles.   IsoDAR can search for evidence of the dark sector through mixing of the SM neutron, $n$, to a dark-sector neutron, $n^\prime$. Along with general connections to dark matter, by way of a dark sector, this effect is motivated by the discrepancy in neutron lifetime measurement between neutron beam and neutron bottle experiments \cite{Nico:2004ie,Yue:2013qrc,Serebrov:2017bzo,UCNt:2021pcg}. 

This analysis utilizes the contained neutron flux. Through $n \rightarrow n^\prime \rightarrow n$ transitions,  neutrons can disappear from the target and reappear in 
the LSC. Along with the essential aspects of the source design discussed above, the underground environment that provides a low background and the LSC's size, allowing tight fiducial cuts  
add to the sensitivity~\cite{Hostert:2022ntu}.
Alternatively, if the $n^\prime$ are Majorana, then $n \rightarrow n^\prime \rightarrow \bar n$ may occur ($\bar{n}\ldots$anti-neutron), depositing $\sim$2 GeV of energy in the LSC upon annihilation. This process is a $B$ violating effect with implications for the early universe \cite{Berezhiani:2020vbe}.   This contribution to dark-sector studies is unique within a wide range of proposed dark-sector experiments, as discussed in Ref.~\cite{Batell:2022xau}.

\paragraph{\it Example 2:  The QCD Axion---}
The QCD axion arises in theories that introduce a new symmetry to explain why  CP violation is suppressed in  QCD \cite{Peccei:1977hh, Backhouse:2021qca}.  The introduction of the symmetry-breaking mechanism to make the new model parameter sufficiently small to agree with experimental data leads to the introduction of a new particle, the axion, $a$.    There has been a long history of searches for the QCD axion at accelerators~\cite{Backhouse:2021qca}, and only a small open window of parameter space remains.  IsoDAR can probe this territory~\cite{waites_axionlike_2023}. 

This search employs the contained monoenergetic photon flux, generating axions by their coupling to photons.    Axions can also couple to electrons, and the signature in the LSC is from interactions and decays involving both electrons and photons. In Fig.~\ref{ae} we show QCD axions coupling to electrons as an example of the capability.   

\subsubsection{New Interactions}

\paragraph{\it Example 1: The Weak Mixing Angle ---}

Despite many open questions within the SM, the theory is remarkably internally self-consistent within the electroweak sector~\cite{ParticleDataGroup:2020ssz}. As a result,   
 LEP and LHC data can be used to predict the rate and energy dependence of neutrino scattering precisely. Deviations might indicate new physics. The NuTeV experiment observed a deviation of the weak mixing angle ($\sin^2 \theta_W$) parameter from the SM prediction \cite{NuTeV:2001whx}. If this is new physics, it must be associated with neutrino properties, as polarized electron experiments have not observed the effect. However, as a neutrino-nucleon experiment, the NuTeV anomaly may stem from unaccounted-for nuclear physics effects; however, numerous possibilities have been investigated, but none fully explain the deviation and so the deviation remains unexplained~\cite{ParticleDataGroup:2020ssz}. 

\begin{figure}[tb!]       
\begin{center}
\includegraphics[width=\textwidth]{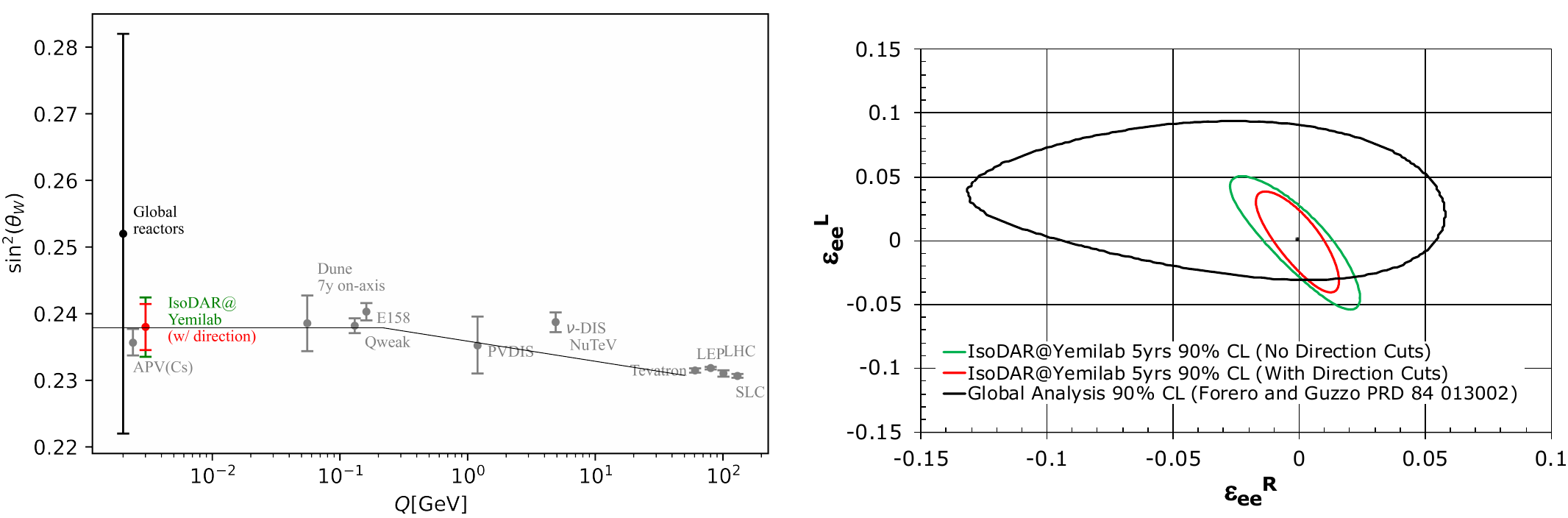}
\end{center}  
\vspace{-0.1in}
\caption{Left:  Weak Mixing Angle expectation for IsoDAR compared to present results and the future DUNE expectation. Right:  expectation for IsoDAR in NSI parameter space. From Ref.~\cite{alonso_neutrino_2022}.}
\label{IsoDARESphys}
\end{figure}

 IsoDAR offers a new approach to this question through measurement using neutrino-electron elastic scattering---a purely leptonic process that evades the concerns related to nuclear physics in NuTeV.  
IsoDAR will record $\sim$7000 ES events over its 4-year accumulated up-time. This constitutes the world's largest sample, allowing the most precise measurement of $\sin^2 \theta_W$ at low momentum transfer, $Q$.
Fig.\ref{IsoDARESphys}, right, presents two expectations for IsoDAR, where the red assumes background reduction from Cherenkov ring identification in LSC, while the green indicates the expectation without this additional information---both are presented since this design capability of LSC is yet to be determined. Total statistics limit the outcome, and further running can proceed with IsoDAR and DUNE as it comes online if a NuTeV-like deviation is observed.

This result makes use of the $\bar \nu_e$ flux.   An important design point is that the contemporaneous $\approx 1.7$M IBD events, which have a very well understood cross section, as discussed above, can be used to determine the flux normalization, allowing measurement of $\sin^2\theta_W$ with a precision of 1.9\% (see Fig.\ref{IsoDARESphys} left).

\paragraph{\it Example 2: Non-Standard Interactions of Neutrinos ---}

The accurately predicted energy dependence of the ES sample enables a precise phenomenological search for non-standard neutrino interactions (NSIs). This powerful search for lepton-lepton couplings complements the lepton-quark coupling searches by the active, global coherent neutrino scattering program. The SM right- and left-handed couplings, $g_R$ and $g_L$, are altered by NSI corrections ($\epsilon_{e e}^{e R}$ and $\epsilon_{e e}^{e L}$) for electron flavor antineutrino interactions with electrons (hence ``ee'') to $\tilde g_R= g_R+\epsilon_{e e}^{e R}$ and $\tilde g_L=g_L+\epsilon_{e e}^{e L}$. The SM expectation for IsoDAR@Yemilab in the parameter space of these deviations is seen in Fig.~\ref{IsoDARESphys} (right), where red assumes background reduction using Cherenkov light, and green assumes only scintillation light reconstruction. This can be compared to the world result \cite{guzzo} in black, which is dominated by $\nu_e$--electron ES scattering from beam dump experiments. 

This analysis uses the $\bar \nu_e$ flux.   Because the IsoDAR result is very pure antineutrino, the expectation seen in Fig.~\ref{IsoDARESphys} (right) is rotated with respect to the past results, where the fluxes had a mix of $\nu_e$ and $\bar \nu_e$ content.
The result complements searches for NSIs from coherent scattering experiments, which aim to discover new couplings between neutrinos and nucleons. For a review of NSI searches, see Ref.~\cite{Arguelles:2022tki}.

\section{Requirements for the Proton Driver}

As shown in Fig.~\ref{fig:isodar_yemilab}, the Proton Driver consists of a 60~MeV/amu cyclotron accelerating and extracting 5~mA of \htp, a transport line that strips the \htp ions close to the extraction point, analyzes the stripped beam to monitor the condition of the stripping foil, and transports the 10~mA of protons to the target.  The target consists of three nested hemispherical beryllium shells, the outermost being 20~cm in diameter, cooled with heavy water flowing around the shell structure.  The proton beam is shaped, with quadrupole magnets and wobbler magnets to produce a distribution that allows for optimal heat transfer to the cooling water.  Neutrons produced in the target system are moderated and flood into a roughly-spherical sleeve, about 1 meter in diameter, containing a mixture of beryllium chips and highly-enriched $^7$Li. The beta-decay of the $^8$Li resulting from the capture of neutrons produces the neutrinos of interest in the experiment.  

The principal requirements for the accelerator system are to reliably produce a 10~mA, 60~MeV beam of protons in the most efficient manner possible with the lowest possible loss of particles in the acceleration process, particularly above the Coulomb threshold. The experience of the cyclotron system at PSI~\cite{Stammbach} is that keeping the total power of beam loss during cyclotron extraction to less than 200~W still allows access for component maintenance. Considering that the total beam power is 600~kW, this is a very stringent requirement. To be decisive within five years, we require a CW beam with 80\% availability. Secondary requirements like compactness stem from the underground installation. Section~\ref{sec:installation} addresses the complex issue of transport and assembly of large and heavy cyclotron components in the underground environment at Yemilab.


\subsection{Overview of the Accelerator Layout}
\begin{figure}[t!]       
\begin{center}
\includegraphics[width=0.95\textwidth]{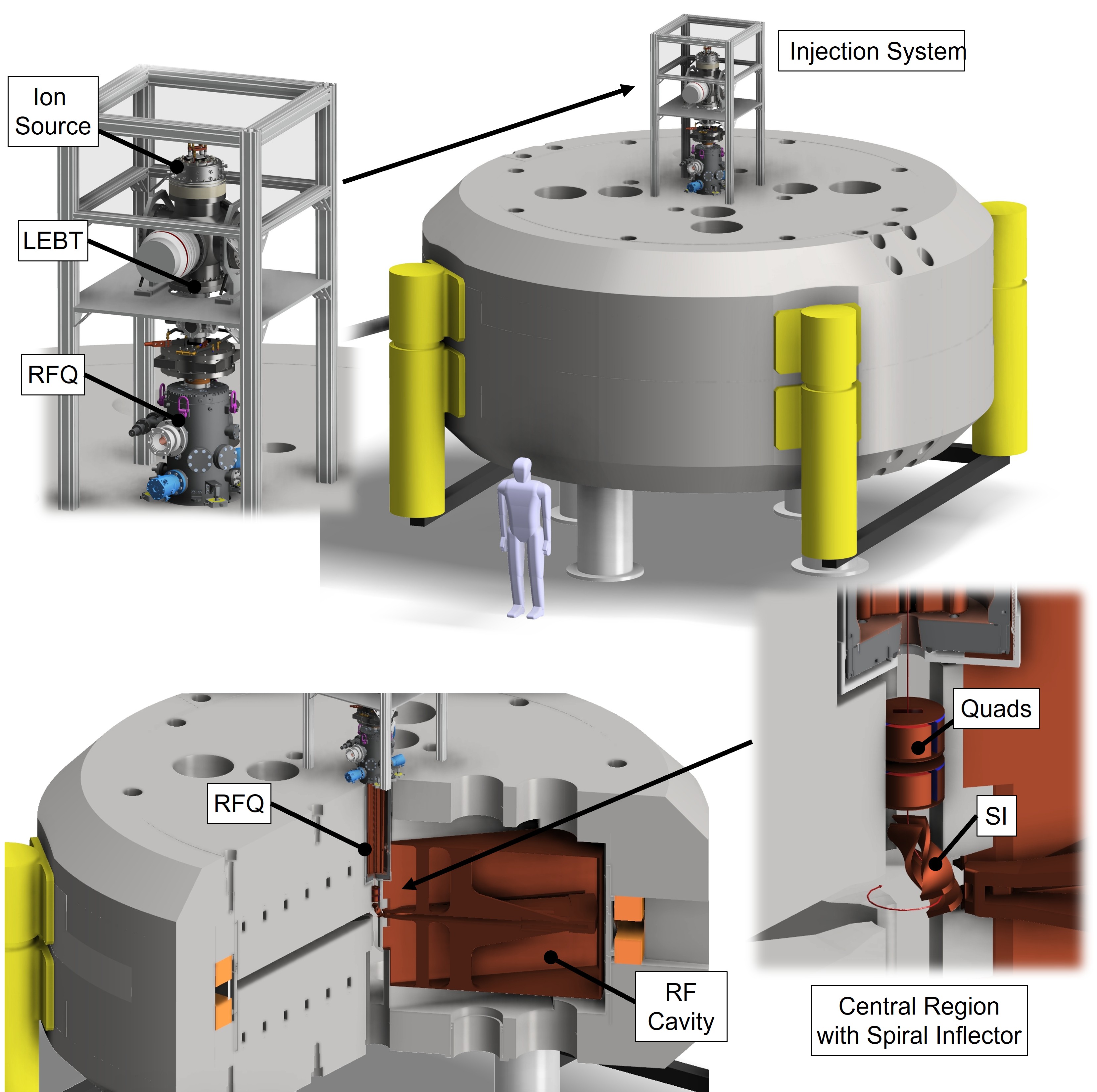}
\end{center}  
\caption{Overview of the components of the HCHC-60 cyclotron system. Only one RF cavity is shown to not clutter the image. ``Quads+SI'' refers to focusing quadrupoles leading into the spiral inflector (SI). The beam is produced in the ion source at the top, shaped and guided into the RFQ by the LEBT, which is partially embedded in the cyclotron, bent onto the horizontal plane by the spiral inflector, and then acceleratred in the cyclotron.}
\label{fig:MainCycloOverview}
\end{figure}

To guide the reader through this PDR and, indeed, the path of an ion from genesis 
in the ion source to the full 60~MeV/amu and extraction, we provide an overview in
Fig.~\ref{fig:MainCycloOverview}.
The main components of the system, which will be described in much detail in the following chapters, are:
\begin{itemize}
    \item The ion source, providing 10~mA of \htp at an emittance $< 0.1 \pi$-mm-mrad.
    \item The Low Energy Beam Transport (LEBT), shaping and guiding the beam to the RFQ and providing diagnostics, and chopping for commissioning and personnel and machine protection.
    \item The Radiofrequency Quadrupole (RFQ), embedded in the cyclotron yoke, to bunch and pre-accelerate the beam to 35~keV/amu. 
    \item The central region with electrostatic quadrupoles for transverse beam focusing, the spiral inflector to bring the beam from the vertical axis onto the median plane, and the specific RF electrode (aka ``dees'') shapes to center the beam during the first turns.
    \item The main magnet and RF Cavities for acceleration to 60~MeV/amu.
    \item (not shown) The extraction system consisting of electrostatic and magnetic channels guiding the beam out of the cyclotron with the goal of extracting 5~mA of \htp.
\end{itemize}

\subsection{Status of the HCHC-60 Development}
The HCHC-60 contains several subsystems that have never been used or combined 
in this way. To demonstrate the most critical aspects, we have begun an experimental
campaign, building and testing some of these subsystems. Some already exist and are being tested, others are at the preliminary design level. Hence the overall status of the project is the PDR stage. The following gives a brief overview of the status of each subsystem. 
\begin{itemize}
    \item Ion Source. We have designed and built two iterations of a multicusp ion source (MIST-1 and MIST-2). MIST-2 is currently being experimentally tested.
    \item RFQ. The RFQ technical design is finished and it is currently being fabricated by Bevatech GmbH in Germany. We expect delivery in spring 2026 and will commission the RFQ together with the ion source.
    \item HCHC-1.5 test cyclotron. A smaller version of the HCHC-60, this 1.5 MeV/amu test cyclotron is fully funded and currently in the Request For Proposal (RFP) stage. We expect vendor selection by the end of summer 2025.
    \item HCHC-60. The full 60~MeV/amu cyclotron has reached the PDR stage. We describe the design and simulations in detail in the later sections of this document. We have not yet begun procurement of any hardware for the full cyclotron. 
\end{itemize}
The ion source and RFQ could be reused directly in the final HCHC-60 cyclotron. The central region of the HCHC-1.5, while designed to mimic as much as possible the HCHC-60 central region, will have to be adapted to the magnetic field of the HCHC-60.

\section{Design of the Front End \label{sec:frontend}}

\subsection{The Ion Source}
The \htp beam to be injected into the cyclotron and accelerated to 60~MeV/amu will be produced by a filament-driven multicusp ion source. We have built a first prototype, the MIST-1 (Multicusp Ion Source at MIT - Version 1), the technical details and first measured results of which are reported in Refs.~\cite{axani_high_2015, winklehner_first_2018, winklehner_high-current_2021-1, winklehner_new_2022, weigel_epics_2023}. Here we summarize these results and present ongoing upgrade work (MIST-2). The upgrade to MIST-2 improves on MIST-1 by implementing refinements from lessons learned during the systematic measurements and runtime experiences described below. The design of the upgraded MIST-2 has been finalized and is in the final stages of fabrication and assembly. A publication on the MIST-2 design and performance is forthcoming.

\subsubsection{Operating Principle}
In a multicusp ion source \cite{ehlers:multicusp1}, a plasma is confined by a set of permanent magnets that are arranged around the plasma chamber with alternating polarity, thereby creating ``cusps''
in the magnetic field. On average, the electrons and ions produced in the center of the chamber see an increasing magnetic field as they travel outward, which eventually turns them back. One notable exception are the cusps, which are the natural loss points of
this configuration.

If the source is filament-driven\footnote{Other options exist, e.g., RF-driven.}, electrons are emitted from a hot filament. A plasma discharge is initiated and sustained by applying a potential difference between the filament
and (parts of) the source chamber. Thus, the filament acts as the cathode, and the source chamber acts as the anode of a discharge. Electron-impact ionization with neutral molecules and atoms in the source chamber produces secondary electrons and ions.

Neutral gas is fed into the source to facilitate the generation of a high-density plasma. The type of gas determines the ion species produced in the source. Here we use H$_2$ gas to form a plasma -- and beam -- comprising protons, \htp, and H$_3^+$.

\begin{table}[b!]
    \centering
        \caption{A summary of the parameters for the MIST-1 ion source.}
        \label{tab:IonSourceParameters}
    \begin{tabular}{||l | c||}
\hline\hline
Parameter & Nominal Value\\
\hline
Plasma chamber length & 6.5~cm\\
Plasma chamber diameter & 15~cm\\
Permanent magnet material & Sm$_2$Co$_{17}$\\
Permanent magnet strength & 1.05~T on surface\\
Front plate magnets & 12 bars (star shape)\\
Radial magnets & 12 bars \\
Back plate magnets & Four bars in three rows\\
Front plate cooling & Embedded steel tube\\
Back plate cooling & Embedded copper pipe\\
Chamber cooling & Water jacket\\
Water flow (total) & $\approx$2~L/min\\
Filament feedthrough cooling & Water cooled\\
Filament material & Water mixed with Cu and Ni\\
Filament diameter & $\approx$0.8~mm\\
Discharge voltage & Max. 150~V\\
Discharge current & Max. 24~A\\
Filament heating & voltage Max. 8~V \\
Filament heating & current Max. 100~A \\
\hline\hline
    \end{tabular}
\end{table}

A significant concern, which we will quantify in Section~\ref{sec:mist-performance}, is the production of ions other than H$_2^+$. Of particular concern are H$_3^+$, which can be produced via H$_2^+ +$H$_2 \rightarrow$H$_3^+ + $H, as well as the production of free protons arising from premature dissociation of the H$_2^+$. To mitigate this, the length of the plasma chamber is 6.5~cm, which is shorter than typical proton sources. This corresponds to the lower end of the range for the mean free path of H$_2^+$ ions within a neutral H$_2$ gas which we estimate to be between 5 - 20~cm depending on gas input flow rate. For more details on this computation, see Ref.~\cite{winklehner_high-current_2021-1}.

\subsubsection{Mechanical Design}
\begin{figure}[!t]
\centering
\includegraphics[width=0.9\linewidth]{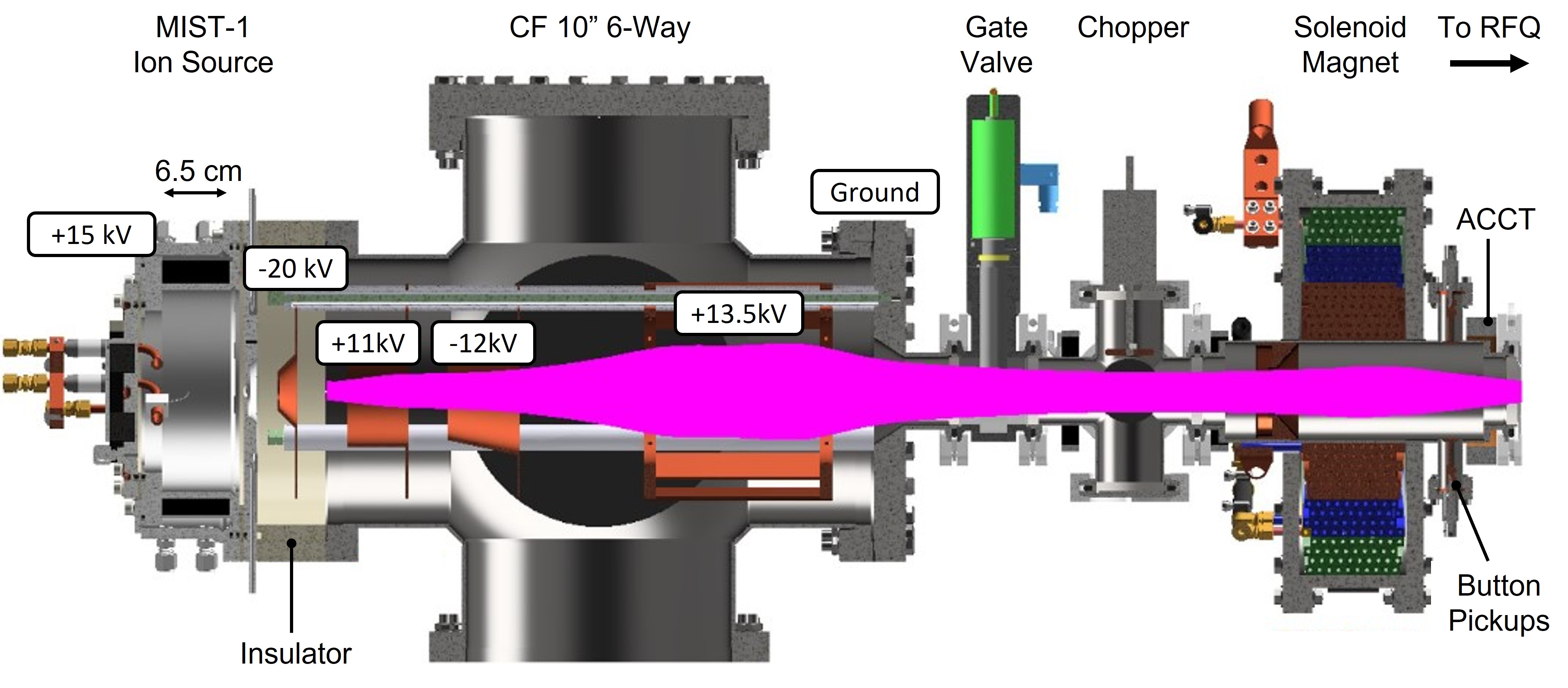}
\caption{Cross section of MIST-1 with extraction system, LEBT, and diagnostics. The diagnostics comprise an AC Current Transformer (ACCT) and a set of four capacitive button pickups right. Both are placed between the solenoid magnet and the RFQ at the end of the LEBT.
Also shown is an example set of ion trajectories simulated with IBSimu and WARP. Adapted from Ref.~\cite{waites_high_2022}.}
\label{fig:ExtractionSystemSimulation}
\end{figure}

\begin{figure}[!t]
\centering
\includegraphics[width=0.7\linewidth]{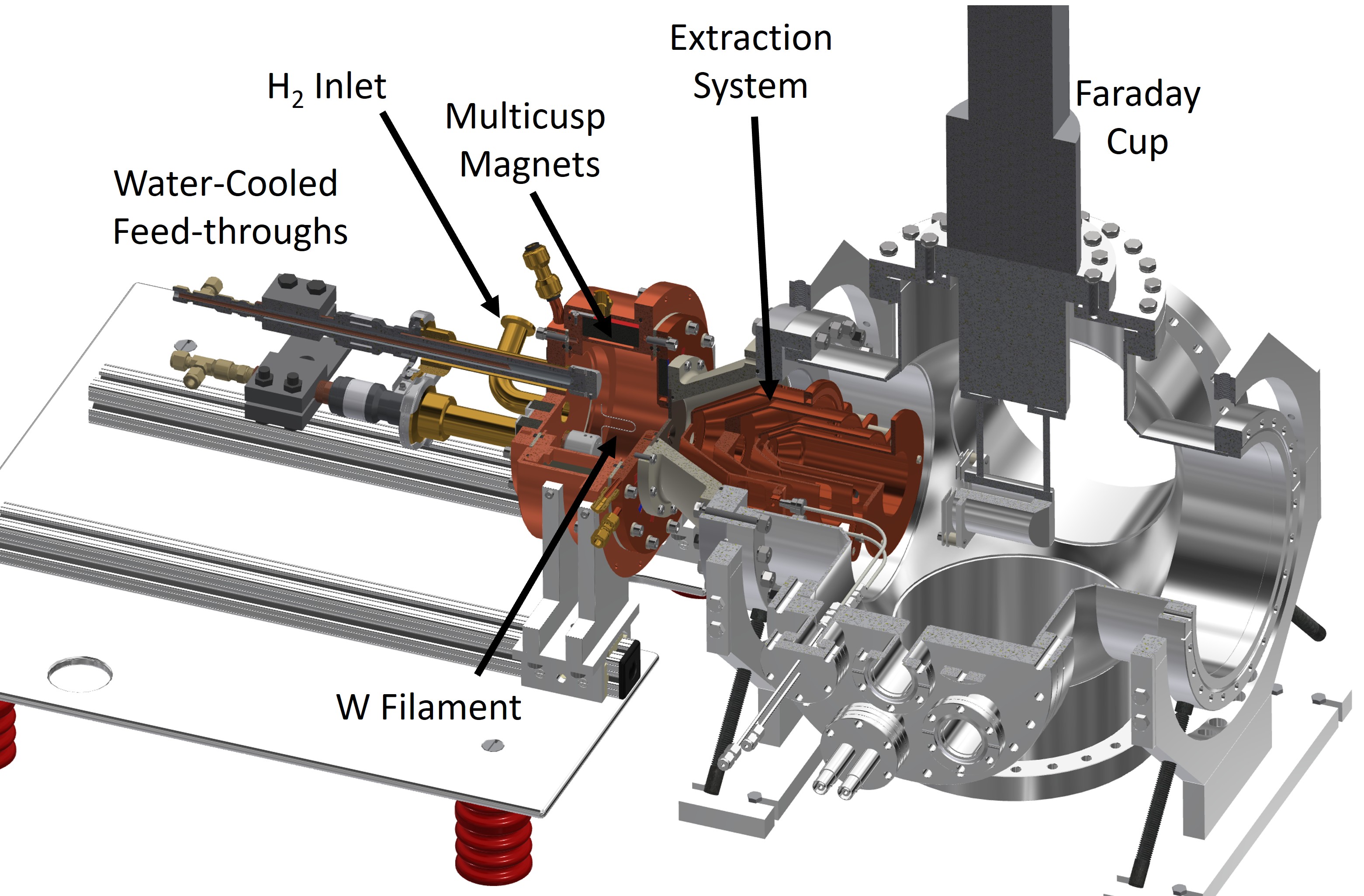}
\caption{MIST-2 CAD rendering calling out the most important features.}
\label{fig:MIST-2_Overview}
\end{figure}

For visual reference as we discuss the details of the ion source, 3D CAD renderings of MIST-1 (with subsequent LEBT) and MIST-2 are provided in Figs.~\ref{fig:ExtractionSystemSimulation} and \ref{fig:MIST-2_Overview}, respectively. The main ion source parameters are also summarized in Table~\ref{tab:IonSourceParameters}.

The MIST-type ion sources comprise a plasma chamber surrounded by a ``water-jacket,'' a back plate, and a front plate. The water jacket lets us pump de-ionized (DI) water around the plasma chamber for cooling. The radial confinement magnets are mounted inside the water-jacket and are cooled directly. At 6.5~cm, the plasma chamber is short compared to typical ion sources. This is a direct consequence of the short mean free path of \htp in the plasma/neutral gas of the ion source discussed above.

The nominal configuration for the confinement magnets is: Four rows of alternating magnets on the back plate, twelve alternating radial magnets, and a star formation of twelve smaller magnets on the front plate, also alternating in polarity. This leaves a small volume of low magnetic flux around the extraction aperture, allowing the ions to easily drift out of the source. All magnets are \smco. For MIST-2 we have begun studying different configurations (no magnets, \ndfe magnets). See below for more details. 

The back plate holds four strips of confinement magnets, as well as the water-cooled filament feedthroughs, and the gas inlet (a KF NW-16 elbow). Temperature sensors (thermocouples) can be attached to the back plate to monitor the temperature of feedthroughs and magnets. The back plate itself is also water-cooled with cooling channels embedded in the plate. The filament is mounted to the high-current feedthroughs using molybdenum filament holders. The filament itself is made from 
tungsten and can have different thickness as well as shapes (this is a parameter currently being optimized).

The front plate holds the front confinement magnets as well as the plasma aperture, a 
W-Cu (75\%~-~25\%) alloy (aka ``Elkonite'') plate that is embedded in the front plate. We have several plasma apertures with different-size openings ranging from 4~mm to 8~mm to test beam quality and total extractable current for various aperture diameters.

As seen in Figs.~\ref{fig:ExtractionSystemSimulation} and \ref{fig:MIST-2_Overview}, following the plasma aperture is the ``extraction system'' which shapes and guides the beam as it is initially formed. The MIST-1 extraction system comprises five electrodes which shape and accelerate the ion beam. These include a plasma electrode (+15~kV), a puller electrode (-3~kV to -20~kV), and three electrostatic lenses -- Lens 1 (+10~kV to +15~kV), lens 2 (-3~kV to -15~kV), and lens 3 (+10~kV to +15~kV).

\paragraph{MIST-2 versus MIST-1\label{sec:mist2}}
The changes we made in the MIST-2 design are mainly to improve cooling, assembly and maintenance, and diagnostics. They are:
\begin{itemize}
    \item Replace all stainless steel components with oxygen-free high-conductivity (OFHC) copper to improve heat transfer
    \item Increase the cross-sectional areas of all water cooling channels
    \item New high-power feedthroughs that are detachable from the back plate
    \item An additional port on the back plate for a Langmuir probe
    \item A redesigned extraction system. The new pentode extraction system comprises a water-cooled puller, an intermediate electrode, a negative electrode and a grounded final electrode.
    The more compact extraction system with negative potential followed by ground will help keep space charge compensation high in the LEBT and allows us to use a Faraday cup and emittance scanners for beam diagnostics directly after the source.
\end{itemize}

These changes will improve long-term running stability, reduce maintenance downtime, 
and allow for a better understanding of the plasma and beam conditions.
Our high-fidelity simulations using IBSimu~\cite{kalvasIBSIMUThreedimensionalSimulation2010a} and WARP~\cite{vayNovelMethodsParticleInCell2012} suggest that the beam dynamics downstream of the six-way cross will not change between the MIST-1 and MIST-2 configurations. All particle simulations and designs still hold for LEBT, RFQ, and cyclotron.

\subsubsection{Electrical Design}

Hydrogen gas is introduced via an inlet in the backplate of the source as indicated in Fig.~\ref{fig:MIST-2_Overview}. The amount of hydrogen entering the ion source is controlled using a mass flow controller (MKS Instruments Model GV50A) which has a 5 standard cubic centimeter per minute (SCCM) full range. This system interfaces with the EPICS PC control system which is documented in detail in Ref.~\cite{weigel_epics_2023}.

\begin{figure}[!t]
  \centering
  \includegraphics[width=0.5\linewidth]{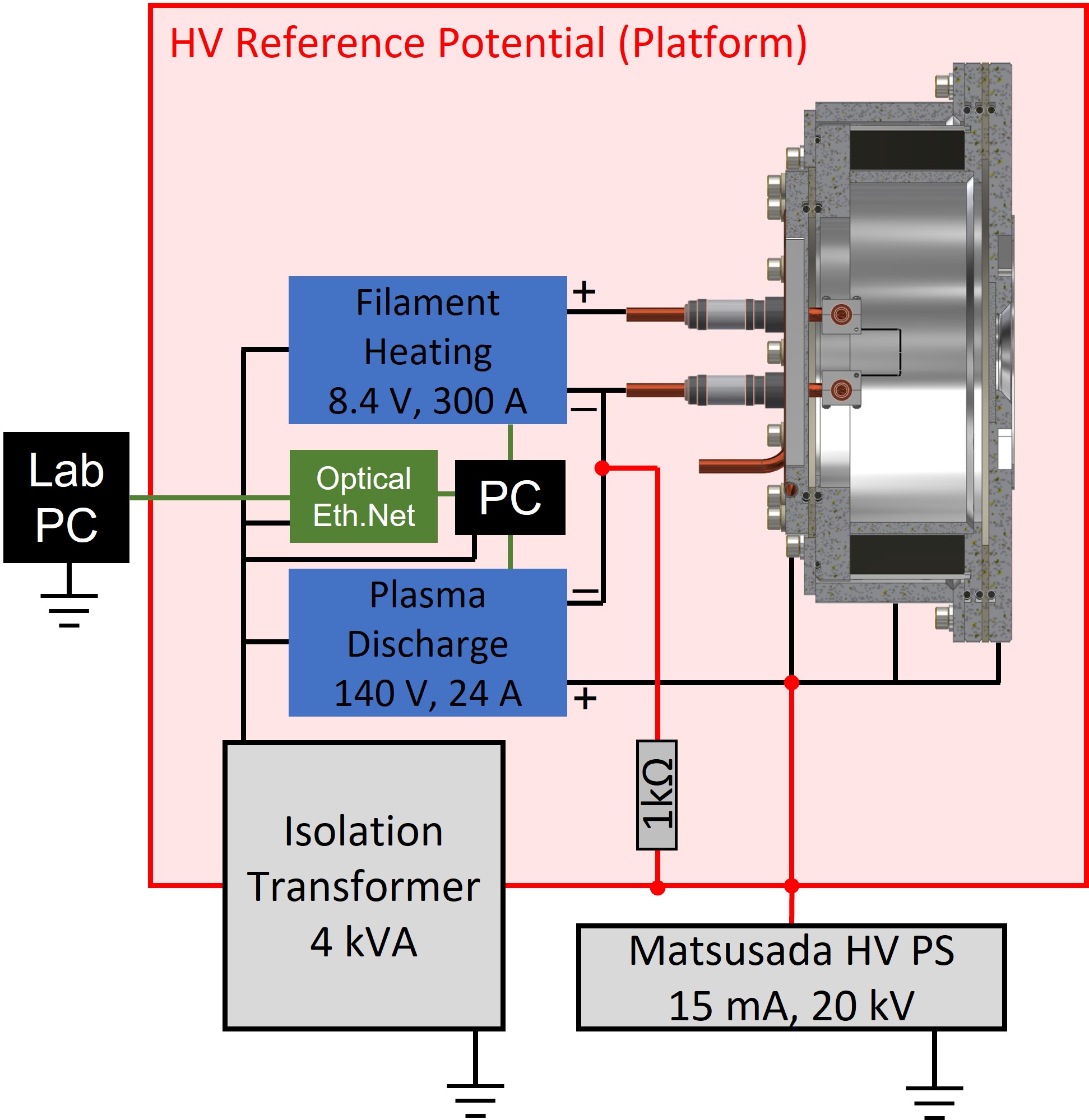}
  \caption{Wiring schematic for the ion source. The red square region corresponds to the HV reference potential. Green lines are data connections. Black lines are power cables.}
  \label{fig:FilamentWiringDiagram}
\end{figure}

The filament is heated by an TDK Lambda (8.4~V, 300~A max) power supply. In addition to this power supply which is used to heat the filament, a second power supply (Matsusada REK) is connected to the filament and to the source chamber to maintain a potential difference of 80-140~V between them in order to facilitate electron discharge of up to 24~A max (typical during our tests was 5-10~A). The source body, back plate, and front plate are all mutually electrically insulated from each other for added flexibility in the electric field shape. Power is transferred to the platform via two isolation transformers of 2~kVA each. A schematic of the wiring of the ion source is illustrated in Fig.~\ref{fig:FilamentWiringDiagram}.

The ion source (platform) is lifted to high voltage using a Matsusada~AU, 15~mA, 20~kV power supply, all lens elements in the extraction system are powered by Matsusada~AU, 7.5~mA, 20~kV power supplies of either negative or positive polarity.

The devices on the HV platform (control computer, filament discharge, filament heating power supply, etc.) are inside a Faraday cage on platform potential to reduce electromagnetic interference (EMF). Surrounding the platform and ion source is another Faraday cage at ground potential for personnel protection and EMF reduction.

\subsubsection{Beam Dynamics}
We simulated beam formation and shaping in the extraction system with the IBSimu 
package~\cite{kalvasIBSIMUThreedimensionalSimulation2010a},
which performs cylindrically symmetric 2D (aka ``RZ'') and 3D simulations including space charge through an iterative process. IBSimu also contains a 1D plasma sheath model which is a good approximation for the plasma-beam
interface at the source aperture. Our extraction simulations of the MIST-1 ion source followed by 
a WARP~\cite{vayNovelMethodsParticleInCell2012} simulation of the LEBT yielded excellent agreement with measurements~\cite{winklehner_first_2018, winklehner_new_2022}. Example trajectories of the new MIST-1 extraction system and RFQ matching LEBT are shown in Fig.~\ref{fig:ExtractionSystemSimulation}.

\begin{figure}[!t]
  \centering
  \includegraphics[width=1.0\linewidth]{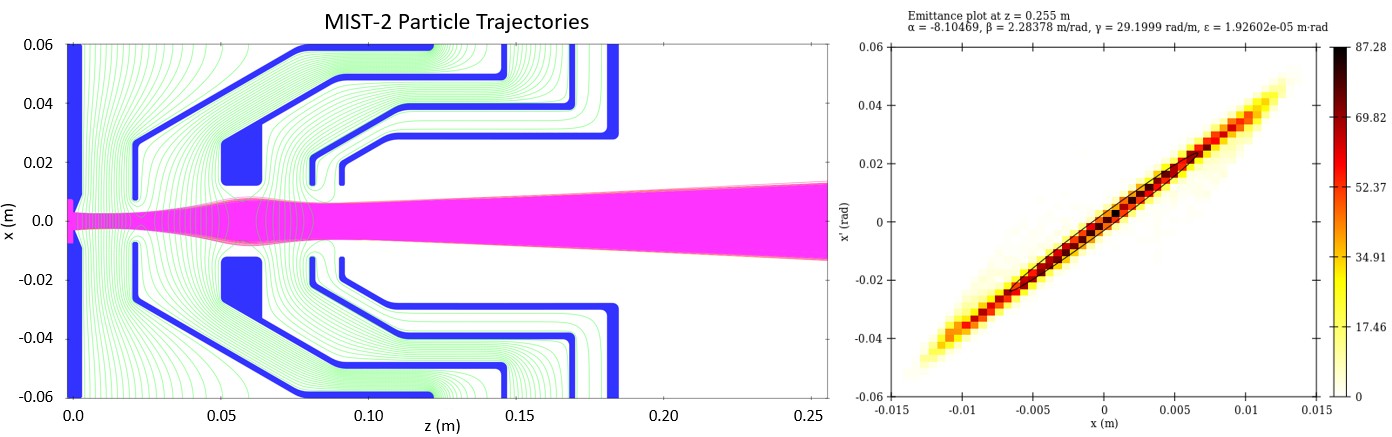}
  \caption{IBSimu Simulation of MIST-2 extraction. Left: particle trajectories in the z-x plane, Right: x-x' phase space 25.5~cm after the extraction aperture. Both MIST-1 and MIST-2 exhibit a very low emittance (see text).}
  \label{fig:MIST-2_Extraction}
\end{figure}

An extraction system simulation of the MIST-2 system is shown in Fig.~\ref{fig:MIST-2_Extraction}. The simulated normalized RMS emittance for the new MIST-2 extraction system using an 8~mm diameter plasma aperture is 0.077~$\pi$-mm-mrad. Simulated normalized RMS emittances for the MIST-1 extraction system were between 0.05 and 0.1~$\pi$-mm-mrad depending on source aperture and extracted current and agreed well with measurements~\cite{winklehner_first_2018}.

\subsubsection{Performance\label{sec:mist-performance}}

\paragraph{Goals \& Measurement Methods}
In this section we summarize the latest measured results as well as discussing the results of several optimizations that have been thus far performed as well as several that remain to be performed. The ion source target parameters that will be required for IsoDAR to achieve the planned decisive results for a sterile neutrino search include the following benchmarks:

\begin{itemize}
    \item High current: $\approx$10 mA of H$_2^+$
    \item High purity: H$_2^+$ fraction $>$\,80\%
    \item Low emittance: $<$\,0.1~$\pi$-mm-mrad (RMS, normalized)
    \item Ability to run continuously in DC mode
\end{itemize}

Characterization of the ion source is done using a short diagnostic beam line. This test beam line comprises two Faraday cups, a dipole magnet with separation slits, and a pair of Allison emittance scanners.
The total beam current is measured using the first Faraday cup located posterior to the extraction system. The dipole magnet separates the ions by mass and passes them to the second Faraday cup for mass analysis. Allison-type electrostatic emittance scanners measure each species' horizontal and vertical emittances.

\paragraph{Optimization of Ion Source Parameters}
The ion source is designed such that several parameters can be varied in order to optimize the magnitude, purity, and emittance tightness of the current from the ion source. 
\begin{itemize}
    \item H$_2$ input flow rate
    \item Filament discharge voltage
    \item Filament discharge current
    \item Filament size, shape, \& position \textit{(Optimization not yet fully studied)}
    \item Permanent magnet type
\end{itemize}

\textbf{H$_2$ input flow optimization}. Three series of optimizations on the input gas flow rate have been performed. The three series differ in how the filament discharge voltage is managed. In the first series, a proportional integral derivative (PID) loop which stabilizes discharge was disabled allowing the discharge current to rise to the source steady-state value and then stabilized at that value by reengaging the PID. In the second series, the discharge current was kept fixed at 4~A. In the third series, the discharge current was reduced to keep the total extracted current fixed at 1.25~mA. Under each of these three conditions, the gas input rate was varied from 0.25~-~1.25~SCCM. This study demonstrates a general trend of greater H$_2^+$ production at higher discharge currents and, as theoretically expected, a higher fraction of H$_3^+$ with higher flow rates.

\textbf{Discharge voltage optimization}. Two series of discharge voltage optimizations have been performed differing in how discharge current is managed. In the first, discharge current is permitted to freely find a steady state using the same method as used in the flow optimization described above. In the second, the discharge current was fixed at 4~A. 

\textbf{Discharge current optimization}. One series of current optimizations was explored in which the flow rate and discharge were kept constant while discharge current was varied from 2~A to 8~A. No strong dependence on current was observed.

\paragraph{Latest Optimized Results}
To date, only an incomplete optimization of the ion source has been performed. As such, parameters which simultaneously meet all the target goals have not yet been fully identified. The ion source's peak performance to date varies, therefore, depending on which of the performance metrics is optimized. Optimization has been performed for maximum total current density, maximum H$_2^+$ purity, and a hybrid option optimizing for the maximum H$_2^+$ current density with the constraint that H$_2^+$ be dominant.

\noindent\textbf{Highest total current density:} 41~mA/cm$^2$ with 31\% H$_2^+$ purity.\\
\textbf{Highest H$_2^+$ purity:} 76\% purity with 11.4~mA/cm$^2$ total current density.\\
\textbf{Highest H$_2^+$ current density with H$_2^+$ dominant}: 10.4~mA/cm$^2$ \htp at 62\% purity (16.8~mA/cm$^2$ including all species).

With current optimizations and a 4~mm aperture, the ion source is capable of delivering approximately 1.1~mA of H$_2^+$ with 76\% purity. With an 8~mm aperture this corresponds to 4.4~mA. This is a world-leading result and is only about a factor of 2 short of the target current.

The main bottleneck for an increased beam current historically has been the extraction system, not the actual initial plasma generation. The extraction system characterized in this work was designed for initial testing and commissioning. We have recently replaced it with the version presented in this chapter. Our high-fidelity simulations suggest
that the new extraction system will be able to transport and focus the nominal beam current the source can generate, removing this bottleneck.

\paragraph{Magnet Study}
The choice of permanent magnet material and strength is one of the optimization parameters of this type of ion source. Here, plasma density, confinement time, and magnet thermal properties are important parameters to consider. Originally, we selected \smco magnets for their higher Curie temperature. However, long-term running tests suggest that the magnet temperature is well controlled, which would allow the use of \ndfe magnets, if they yielded a better beam.

We aim to maximize the \htp fraction of the MIST-1 ion source. Within the plasma, atomic processes include ionization, dissociation, and recombination, leading to protons, \htp, and H$_3^+$ in the beam. Magnets with a higher field strength yield higher plasma density and longer confinement of the ions in the plasma, which can alter the rates of these processes. Thus, we are interested in how different magnet types alter the ion fraction and total current. We investigated the current and ion fraction of the beam with no magnets, \smco magnets (grade 28), and \ndfe (grade N50H). The configuration without magnets provides no confinement, and the \ndfe magnets provide the strongest confinement. We performed our measurements using the diagnostic beam line presented in Ref.~\cite{winklehner_first_2018}, containing a dipole magnet and a Faraday cup with 8~mm entrance slit for mass analysis. We investigated each magnet type at a range of H$_2$ mass flows (0.25~-~1.5~SCCM) and discharge voltages (60~-~140~V). 

In our preliminary analysis, we observed a trade-off between the total beam current and the \htp fraction: with lower confinement, the \htp fraction was high, but the current was low. This result was expected; lower confinement and lower plasma density leads to less time and fewer reactants for the ions to undergo processes that result in \htp dissociation or the pick-up of another proton (producing H$_3^+$). However, the lower plasma density also leads to a lower total beam current.

For MIST-2, the baseline design is now to use \ndfe magnets because of the higher current they provide. A paper on the detailed analysis of the magnet study is forthcoming.

\paragraph{Results with the MIST-2 back plate}
\begin{figure}[th]
\centering
\includegraphics[width=0.6\linewidth]{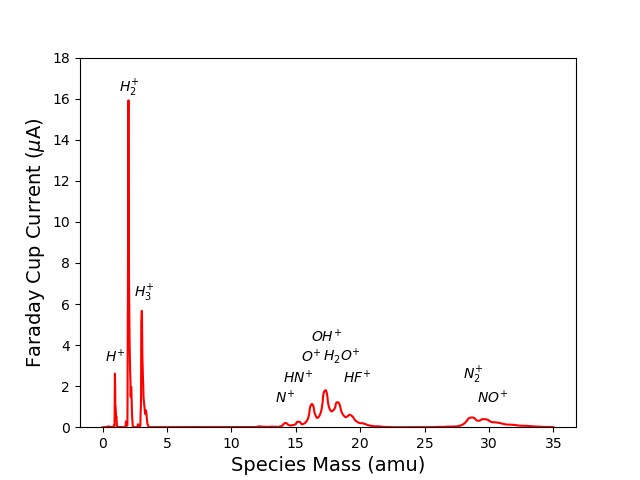}
\caption{A typical spectrum with the new MIST-2 back plate. The total beam current was 3.43~mA, the \htp fraction $\approx 60$\%. See main text for a discussion on the small contaminant peak.}
\label{fig:IonSpectraHiCurrent}
\end{figure}
We already performed the first test of the MIST-2 components described in Section~\ref{sec:mist2}
by installing the new back plate (which is backward compatible) on MIST-1. We were able to run the source up to a stable beam current of 3.43~mA. We did this with the source set to 7.5~kV (the nominal source voltage of 15~kV requires a full beam current of 10~mA to form the appropriate plasma meniscus), puller set to -5~kV, lens 1 at 1.5~kV, lens 2 at -3~kV, lens 3 at 0.5~kV and the filament heating set to 150~A.  

Due to the improved cooling and filament feedthroughs and holders, the source ran very stably; as we expected. A sample spectrum from this series of studies is shown in Fig.~\ref{fig:IonSpectraHiCurrent}. Here, a small cluster of beam contaminants can still be seen, which we attribute to insufficient baking and out-gassing of source components, including Viton O-rings. Past experience has shown that vacuum baking the O-rings before installation (done here), overall good vacuum conditions, and baking the source and beam line before operation (not done in this instance), decreases the contaminants to a negligible level after a few hours of operation~\cite{winklehner_first_2018}.

\subsubsection{Risks and Mitigation\label{sec:ion_source_risk}}
\noindent\textbf{Risk: Ion Source Long-Time Stability.} As IsoDAR is designed to run continuously
for 5 years, the long-time stability is a major concern for the ion source. Hands-on
maintenance of the ion source is going to be limited, because of the close proximity of
the cyclotron.

\noindent\emph{\textbf{Mitigation:} Before the final technical design, we will develop alternative heating methods (external antenna and 2.45 GHz microwaves). 
MIST-1 and MIST-2 were being designed with an exchangeable backplate for quick filament replacement and easy further development of alternative
heating methods. Another option is a dual-source operation using a switching magnet.}

\noindent\textbf{Risk: Ion source maximum \htp current is too low}. The current densities seen 
from the MIST-1 ion source are still a factor 2 below the requirements for IsoDAR.

\noindent\emph{\textbf{Mitigation:} Our current upgrade path should lead us comfortably to 
a total extracted current of 12~mA, particularly considering that MIST-1 has demonstrated
an exceptionally low emittance, which opens up the possibility of going to even larger extraction apertures (10~mm or 12~mm). However, we also have tested a 2.45~GHz flat-field ECR ion source (the VIS~\cite{alonso_isodar_2015}) in the past, from which we measured 15~mA of \htp. The low ratio of 50\% \htp (30~mA total current) deterred us, but as a mitigation, a VIS-type ion source paired with a dipole magnet is a viable alternative.}

\subsection{The Low Energy Beam Transport}
The low energy beam transport line (LEBT) serves the purpose of connecting the ion source to the RFQ. The LEBT steers and shapes the beam, guiding it into the RFQ. The LEBT also provides a location at which to take beam diagnostics. The requirements for the LEBT are:

\begin{enumerate}
    \item Transport beam with minimal losses,
    \item Provide space for diagnostics,
    \item Match the beam to the RFQ,
    \item Provide safety mechanisms for personnel and machine protection.
\end{enumerate}

Details of the LEBT design and performance were given in Refs.~\cite{waites_matching_2021, waites_low_2022, waites_high_2022} here we summarize and update these. It should be noted that the full LEBT design for the HCHC-XX as described here is at the level of a PDR. The only exception is the system of copper electrodes shown in Fig.~\ref{fig:ExtractionSystemSimulation}, which we already tested and reported on in the
previous section.

\subsubsection{Design Overview}
At this stage, ions emerge from the ion source. The beam now passes through a gap surrounded by a series of six electrodes arranged axially within a six-way cross 
(see Fig.~\ref{fig:ExtractionSystemSimulation}). These six electrodes can have their 
voltage optimized in order to best match the parameters for acceptance into the rest of the LEBT. 
A conical flange is connected to the end of the six-way cross. The conical shape minimizes beam
loss on exit. 

\begin{figure}[b!]
\centering
\includegraphics[width=1.0\linewidth]{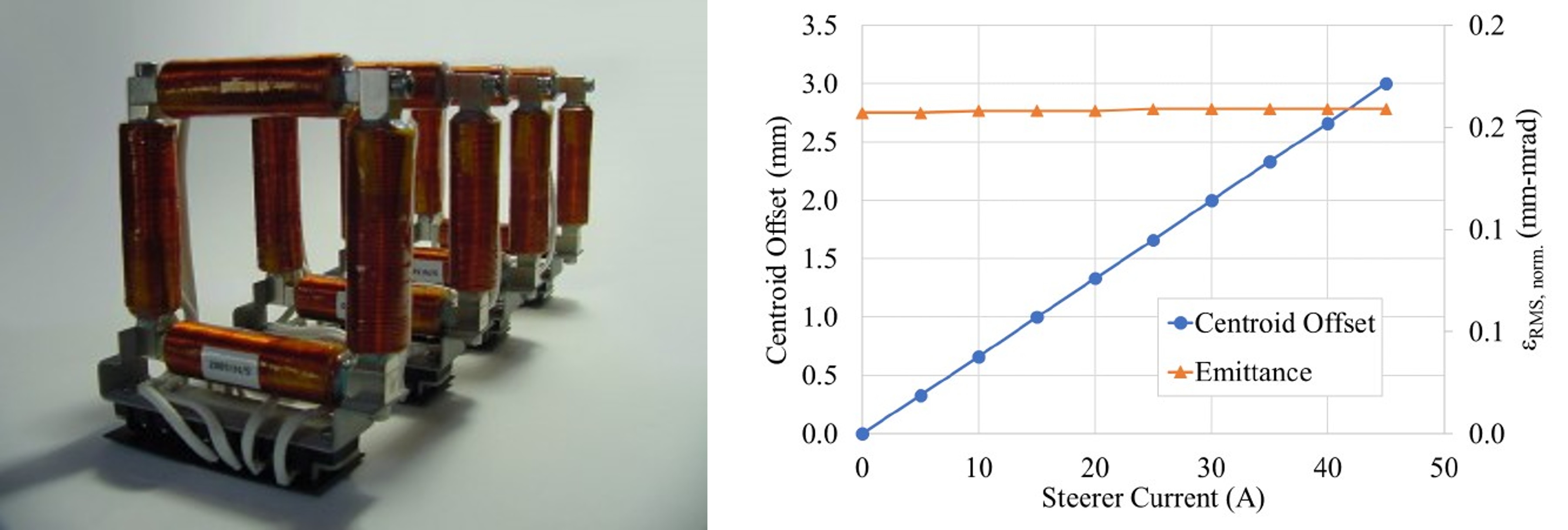}
\caption{Left: Photo of Alpha Magnetics, Inc. magnetic steerers~\cite{2562NotreDame}. Right: Result of LEBT steering simulations. It can be seen that several millimeters offset can be achieved without significant change in emittance. From Ref.~\cite{waites_high_2022}.}
\label{fig:LEBTSteering}
\end{figure}

Following the six-way cross lies a diagnostic segment. A gate valve separates the six-way cross from this segment. The diagnostic section comprises a second cross, which can be used for additional vacuum attachments, diagnostic readouts, and a chopping electrode. The diagnostic section also includes a solenoid and two magnetic steerers which can be used for additional beam tuning. A flange in place toward the end allows space for an AC Current Transformer (ACCT) \cite{bergozACCT} which characterizes the beam immediately prior to entering the RFQ. The diagnostic segment is illustrated in Fig.~\ref{fig:ExtractionSystemSimulation}.

\subsubsection{Beam Steering \& Chopping}
The diagnostic segment also serves to steer the beam into the RFQ or terminate the beam as needed. Even small, sub-millimeter misalignment can prevent efficient injection of the beam into the RFQ. In order to ensure that the beam enters the RFQ at both the correct position and angle, two magnetic steerers  and a solenoid magnet are placed around the beamline. These are controlled independently by an external power source to produce tunable dipole fields which optimize beam entry into the RFQ. The optimal fields were calculated using COMSOL and beam dynamics simulated in WARP. The effect of the steerers is demonstrated in Fig.~\ref{fig:LEBTSteering}.

\begin{figure}
\centering
\begin{minipage}{.5\textwidth}
  \centering
  \includegraphics[width=1\linewidth]{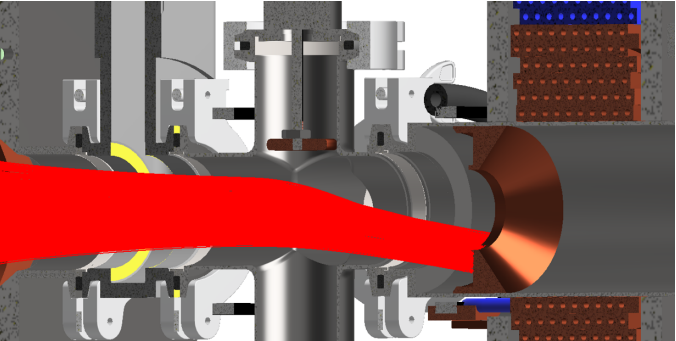}
  \caption{CAD model of the LEBT diagnostic segment with beam simulation (red) overlaid. If the copper electrode (top) is held at adequate potential the beam is pushed off its center course and terminates on the copper ring. From Ref.~\cite{waites_high_2022}.}
  \label{fig:BeamChop_LoydThesisFig5}
\end{minipage}%
\hfill
\begin{minipage}{.45\textwidth}
  \centering
  \includegraphics[width=.8\linewidth]{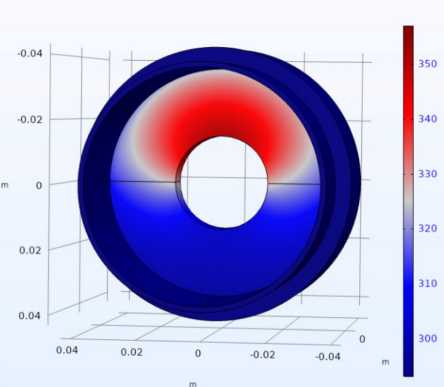}
  \caption{COMSOL simulation of the beam terminating on the copper aperture surrounded by the steel beampipe, with copper water cooling channel on the exterior. From Ref.~\cite{waites_high_2022}.}
  \label{fig:ThermalSafetyBeamDumpSim_LoydThesisFig6}
\end{minipage}
\end{figure}

In order to control the amount of power delivered from the ion source, a beam chopper is built into the cross of the diagnostic segment to allow for a duty factor in an otherwise DC beam. The beam chopper is built into the cross in the diagnostic section. The chopper comprises a deflection plate, followed by a copper ring for particle dumping press-fitted into a steel beam pipe which in turn is surrounded by a copper water cooling jacket. The deflection plate can be put at a high potential relative to the cross, which is grounded, diverting the beam such that it terminates on the copper ring. See Fig.~\ref{fig:BeamChop_LoydThesisFig5}. This method permits controlling the average beam power, without changing the charger per bunch resulting from the RFQ. This is required to maintain vortex motion in the cyclotron.

A 12~mA beam of 15~keV particles fully terminating on the beam stop corresponds 
to $\sim$180~W power delivered. We need to ensure that the dump is capable of 
dissipating this amount of power. The system was modeled by assuming a water cooling line at room temperature. To be conservative, the simulation was run using four times the actual anticipated beam power, i.e., 720~W. The beam dump never rose above 360~K in our simulation, significantly below the melting point of copper. The COMSOL simulation is shown in Fig.~\ref{fig:ThermalSafetyBeamDumpSim_LoydThesisFig6}

\subsubsection{Beam Dynamics}

Beam simulations are performed in order to characterize the performance of the LEBT with the ultimate goal of ensuring that beam's Twiss parameters matches the desired input parameters for the RFQ.

Beam simulations for the LEBT have been performed using IBSimu and Warp. IBSimu provides a more accurate plasma model, but is computationally intensive and so was used only for the region by the ion source where the plasma is extracted and the density is thus high, requiring this accuracy. IBSimu is used for the first several millimeters after the extraction aperture, where density remains high. After this, and for the remainder of the downstream region of the LEBT, the distributions from IBSimu are handed off to Warp which provides accurate results for the lower density regions of the beam with less computational burden. 

Space charge is included in all calculations. However, as is well-known, in the absence of external electrical fields, slow secondary charged particles from the interaction of the beam with the residual gas can be accumulated inside the beam envelope, effectively neutralizing the beam charge. This process is called ``space charge compensation''~\cite{winklehnerSpaceChargeCompensation2015}. In all LEBT simulations, we estimate space charge compensation according to the model presented in Ref.~\cite{winklehnerSpaceChargeCompensation2015}. It is usually above 80\%, alleviating the detrimental effect of space charge in some parts of the LEBT. However, we also test our designs with no compensation and find that, while beam losses are slightly higher, we can still obtain the required beam inside the cyclotron, albeit requiring about 20\% higher beam from the source.

\begin{figure}[th]
\centering
\includegraphics[width=0.5\linewidth]{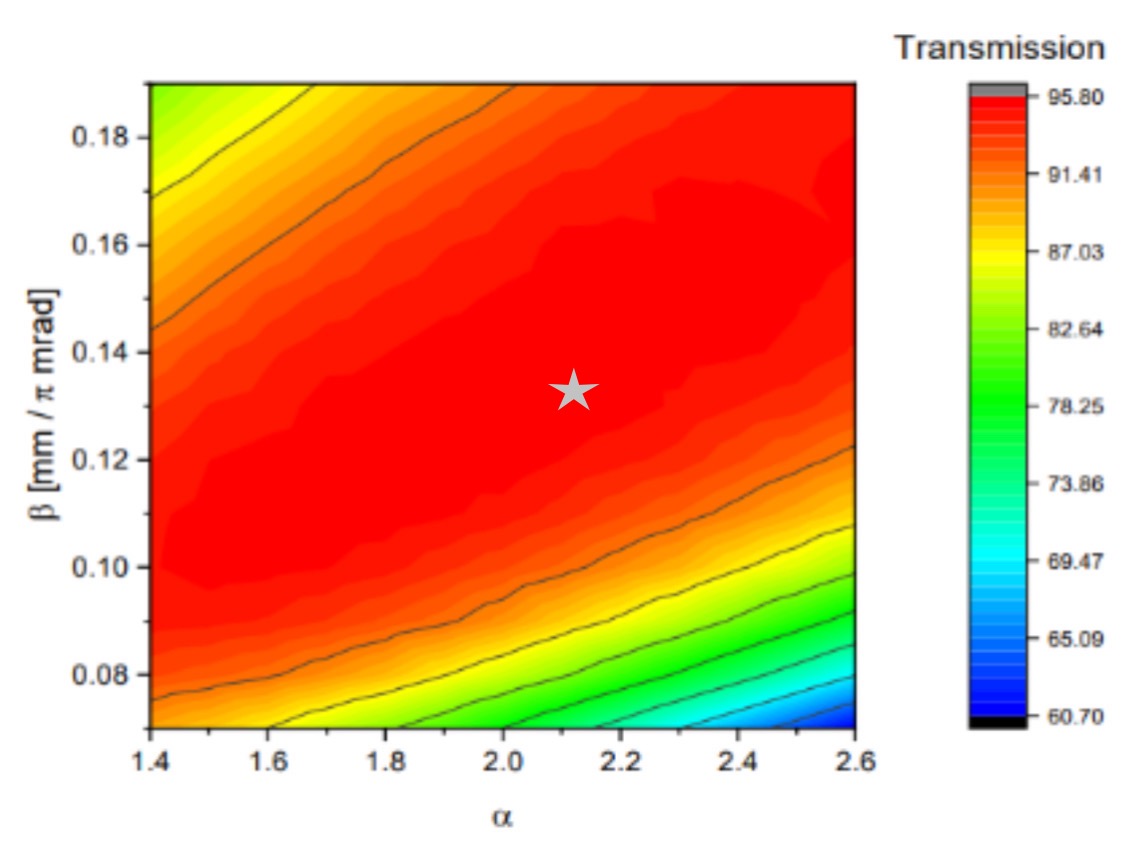}
\caption{A contour plot showing the Twiss parameters required for optimal transmission through the RFQ. The indicated point shows the parameters at the end of the LEBT.
From Ref.~\cite{waites_high_2022}}.
\label{fig:RFQTwissParametersTempPlot}
\end{figure}

\begin{figure}[th]
\centering
\includegraphics[width=0.8\linewidth]{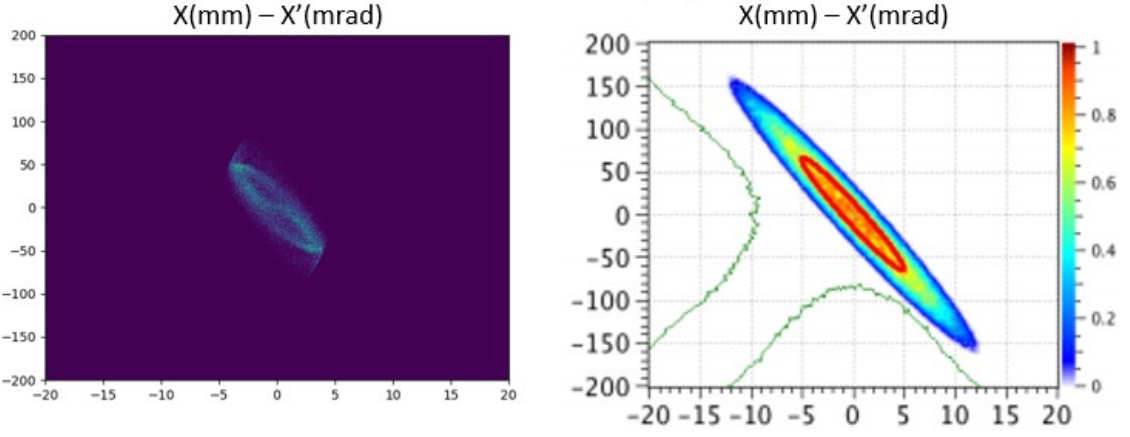}
\caption{Phase space from the LEBT simulation (left) versus the phase space for the ideal beam input to the RFQ (right). From Ref.~\cite{waites_high_2022}.}
\label{fig:EmittanceComparisonLEBTvsRFQ}
\end{figure}

For the IBSimu simulations a 12~mA beam composed of 80\% H$_2^+$, 10\% H$^+$, and 10\% H$_3^+$ was used. This provides ample current for IsoDAR assuming 50\% global transmission efficiency while using current from the ion source that has already been empirically demonstrated. As illustrated in Fig.~\ref{fig:RFQTwissParametersTempPlot} and in Table~\ref{tab:CompareLEBTTwiss} the LEBT can match the required input parameters for the RFQ. Moreover, as illustrated in Fig.~\ref{fig:EmittanceComparisonLEBTvsRFQ} this match is achieved with emittance that surpasses that used for baseline RFQ simulations. 

\begin{table}[th]
    \centering
        \caption{A comparison of LEBT Twiss parameters and baseline RFQ parameters.}
        \label{tab:CompareLEBTTwiss}
    \begin{tabular}{l | c | c} \\    
        Parameter & LEBT Output & Baseline \\
        \hline\hline
        Normalized RMS Emittance & 0.175~$\pi$-mm-mrad & 0.3~$\pi$-mm-mrad \\
        $\alpha$ & 1.3 & 1.0 \\
        $\beta$ & 0.13~mm/mrad & 0.17~mm/mrad \\
    \end{tabular}
\end{table}


\subsubsection{Risks and Mitigation}
\noindent\textbf{Risk: Unexpected Power Loss.} In the case of an unexpected loss of power to one of the power supplies, the beam dynamics could be very different from nominal and lead to damage
downstream (e.g., the RFQ).
     
\noindent\emph{\textbf{Mitigation:} Careful simulation of various failure modes showed that the LEBT
is robust under such changes. The time constant involved with heating up of materials are such that the beam can be switched off before damage occurs. However, for the MIST-2 design, we are envisioning the addition of a Faraday cup directly downstream of the extraction system. Operating spring-loaded, it can be inserted quickly to intercept the beam early on.}

\noindent\textbf{Risk: Lower than anticipated space charge compensation.} Optimal beam transport was 
calculated for a high compensation factor. If this is not the case, transport efficiency and beam dynamics may deteriorate.

\noindent\emph{\textbf{Mitigation:} The LEBT design was double-checked with no space charge compensation and care was taken that injection into the RFQ is nevertheless possible. Typically, higher losses, which are still manageable, are seen. The ion source must provide about 20\,\% more beam current in this case.}

\subsection{The Radiofrequency Quadrupole}
\begin{figure}[!t]
\centering
\includegraphics[width=0.6\linewidth]{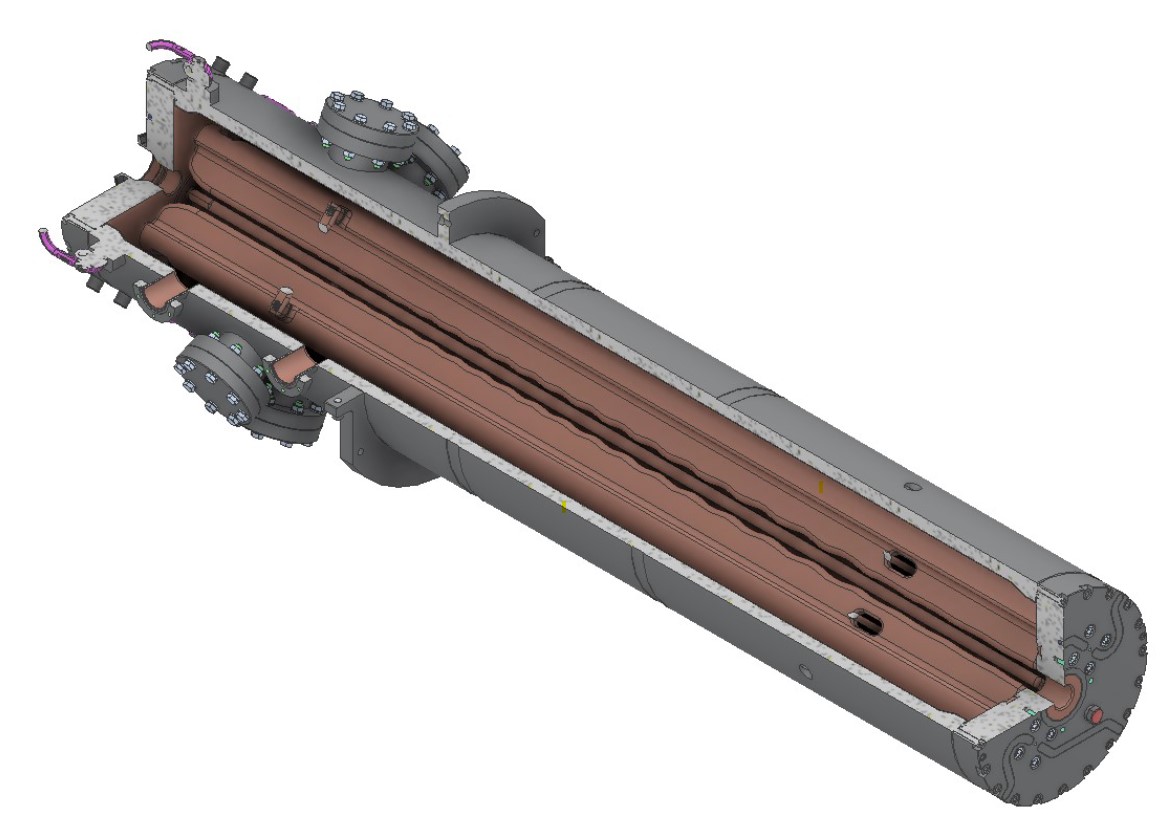}
\caption{Cut view of the IsoDAR RFQ. In the split-coaxial design, vanes are attached to the front and end plates rather than radially to the mantle. From Ref.~\cite{waites_high_2022}.}
\label{fig:RFQ1}
\end{figure}

The \htp beam coming from the LEBT will be injected into a Radio Frequency Quadrupole (RFQ) that is partially embedded in the cyclotron yoke. A schematic of the RFQ and cyclotron can be seen
in Fig.~\ref{cycloISO} where only about a third of the RFQ is visible above the top surface of
the cyclotron. We show a cut view of the RFQ in Fig.~\ref{fig:RFQ1}. Of note are the small diameter
of the RFQ (27.6~cm) while operating at a low resonant frequency of 32.8~MHz. This is achieved
by utilizing the so-called ``split-coaxial'' mode where the vanes are attached longitudinally at the
entrance and exit flanges instead of radially at the mantle. The main operating parameters of the
RFQ are listed in Table~\ref{tab:RFQ_Params}.
The RFQ accelerates only moderately from 7.5~keV/amu to 35~keV/amu and acts mainly as a very
efficient buncher. This leads to a conservatively estimated power consumption of $<6$~kW.

\begin{table}[b!]
    \centering
        \caption{A summary of the design parameters for the RFQ.}
        \label{tab:RFQ_Params}
    \begin{tabular}{||l | c||}
\hline\hline
Parameter & Nominal Value\\
\hline
Diameter & 27.6~cm \\
Frequency & 32.8~MHz\\
Tuning range & $\pm 270$~kHz \\
Q simulated & 2800 \\
Shunt impedance R\textsubscript{p} & 4.9~k$\Omega$/m \\
Input Energy & 7.5~keV/amu\\
Output Energy & 35~keV/amu\\
Duty factor & 100\% (CW) \\
Power & $<6$~kW \\
Cooling & DI Water channels in tank and vanes\\
\hline\hline
    \end{tabular}
\end{table}

\subsubsection{Technical Design}
The technical design was presented in 
Refs.~\cite{ratzingerTechnicalDesignReport2020,winklehner_high-current_2021,holtermann_technical_2021} and the optimization
of the RF loop coupler in Ref.~\cite{sangroula_design_2021}. The thermal properties and cooling considerations were presented in Ref.~\cite{koser_thermal_2021}. An optimization scheme using machine learning was presented in Ref.~\cite{koser_input_2022}.

\begin{figure}[!t]
\centering
\includegraphics[width=0.7\linewidth]{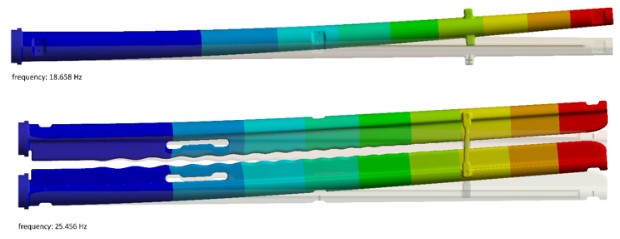}
\caption{Modes of the mechanical deformation of the RFQ vanes calculated with COMSOL. The frequencies are 18~Hz and 25~Hz. From Ref.~\cite{holtermann_technical_2021}.}
\label{fig:RFQ_MechModes}
\end{figure}

In terms of mechanical stability, our analyses show two quarter wave modes which are associated with the choice of a split-coaxial resonator. The addition of bridges between opposite vanes (see Fig.~\ref{fig:RFQ_MechModes}) significantly reduces
detuning due to the Lorentz force with mechanical frequencies 
around 18~Hz and 25~Hz.

\subsubsection{Beam Dynamics}
Beam dynamics studies were performed of the full 3D particle distribution resulting from the LEBT simulations going through the RFQ, including the effects of the entrance gap and asymmetry (due to the split coaxial design) and the exit gap and asymmetry. This was performed by the subcontracted company Bevatech GmbH in Germany, who is currently also manufacturing the RFQ. The well-established tracking code TraceWin was used~\cite{tracewin}.
Fine tuning of the phase of the exit gap (i.e. tuning the length of a drift preceding the exit gap), leads to final output beam parameters listed in Table~\ref{tab:RFQ_BeamOut}.

\begin{table}[bt!]
    \centering
        \caption{RFQ output beam parameters.}
        \label{tab:RFQ_BeamOut}
    \begin{tabular}{||l | c||}
\hline\hline
Parameter & Nominal Value\\
\hline
$\epsilon_x$ (RMS, norm.) & 0.54~mm-mrad \\
$\epsilon_y$ (RMS, norm.) & 0.52~mm-mrad \\
$\epsilon_z$ (RMS, norm.) & 0.114~MeV-deg \\
Phase width (RMS) & 40\degree\\
\hline\hline
    \end{tabular}
\end{table}

\subsubsection{Risks and Mitigation}
\noindent\textbf{Risk: New RFQ Technology}. Direct axial injection into
a compact cyclotron has never been demonstrated. It is possible
that there are limitations that have not been thought of, that will
prove to be show-stoppers. 

\noindent\emph{\textbf{Mitigation:} A prototype machine is currently being built and
will be thoroughly tested in the coming year. As a fallback, conventional LEBTs
using a multi-harmonic buncher are well understood and reliable. However,
they are much less efficient than RFQs.
The focus in this case will shift to increasing ion source performance
further to allow for higher beam losses during injection.}

\noindent\textbf{Risk: RFQ Power Estimate} The RFQ costs depend strongly 
on the RF power necessary to drive the rods/vanes, the length of 
the RFQ, the engineering complexity involved in machining the 
vanes/rods and designing the vacuum vessel.
     
\noindent\emph{\textbf{Mitigation:} We have worked with RF engineers with intimate 
knowledge of RFQ design. The upcoming prototype will demonstrate the
power requirements.}

\noindent\textbf{Risk: Beam spread between RFQ exit and spiral inflector.}
The simulations showed that the 
beam spreads transversally and longitudinally after exiting the RFQ.
Reality might prove more unforgiving than computer simulations and we may incur higher beam losses than anticipated.

\noindent\emph{\textbf{Mitigation:} Our high-fidelity simulations suggest that the beam spread
is manageable. The prototype will allow measuring this precisely. Additional focusing
elements can be inserted between RFQ and spiral inflector if necessary. More aggressive
collimation is possible as well, at the expense of requiring higher current from the source.}

\section{Design of the Cyclotron \label{sec:cyclotron}}

\begin{figure}[!t]
\centering
\includegraphics[width=0.65\linewidth]{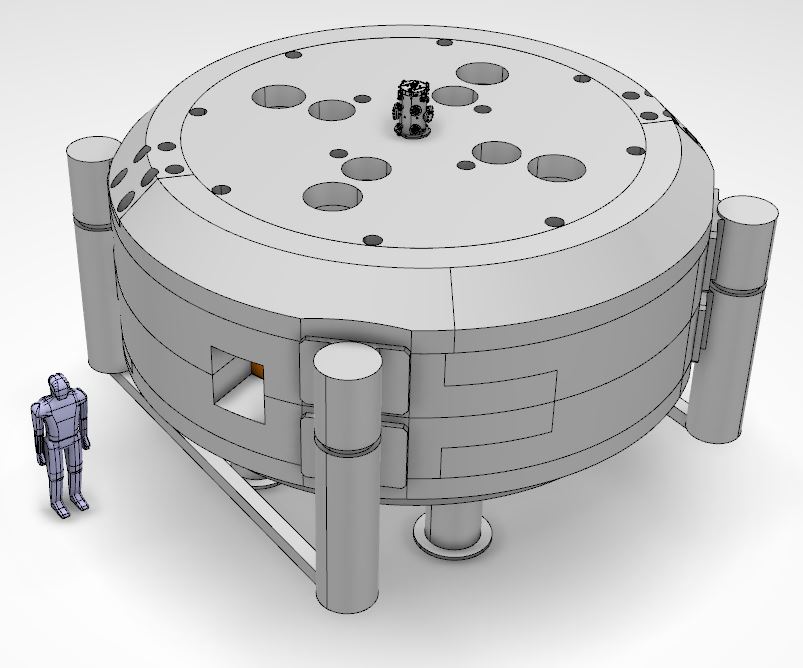}
\caption{Isometric view of the cyclotron.}
\label{cycloISO}
\end{figure}

In this section, we describe the details of the cyclotron at the level of a preliminary design report. This section includes details of the design of the cyclotron, the spiral inflector that transmits ions from the RFQ into the cyclotron, and the extraction system.
This section also characterizes the latest simulated performance of these systems.



\subsection{Magnet Yoke}


The magnet design presented here is physically realistic, but is necessarily preliminary. We do not anticipate the broad design of the magnet to change in future. As we finalize the details of beam extraction small modifications to the magnets, especially at the fringes, may accordingly be modified. 

The engineering design of the IsoDAR magnet benefits from the experience gained by IBA in the fabrication of its C400 cyclotron 
, which is slightly larger, at 7~m outer diameter~\cite{jongenCompactSuperconductingCyclotron2010b}.


\subsubsection{Design and Fabrication of the Cyclotron Magnet}

\begin{figure}[tb]
\begin{minipage}[c]{0.48\linewidth}
\includegraphics[width=\linewidth]{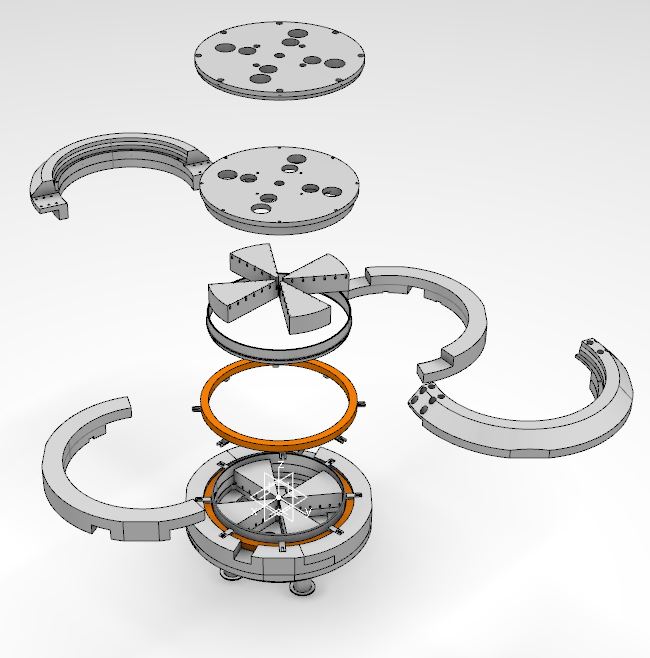}
\caption{HCHC-60 cyclotron upper steel exploded.}
\label{exploded-top}
\end{minipage}
\hfill
\begin{minipage}[c]{0.48\linewidth}
\includegraphics[width=\linewidth]{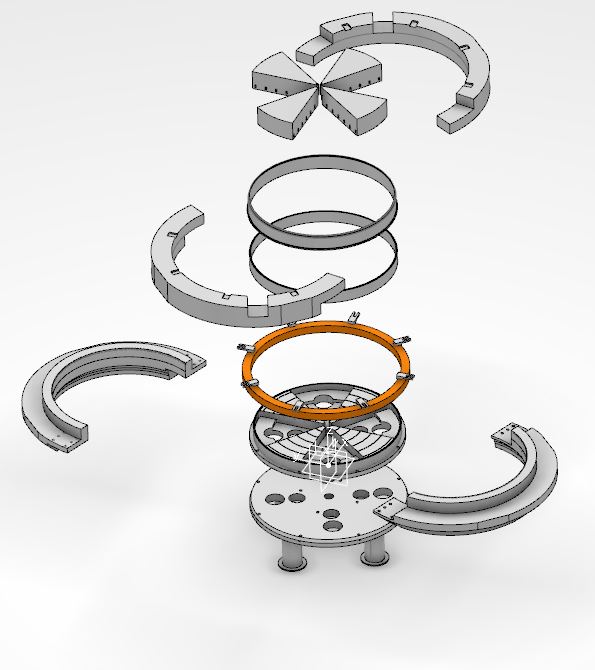}
\caption{HCHC-60 cyclotron lower steel exploded.}
\label{exploded-bottom}
\end{minipage}
\vspace{0.2in}
\end{figure}

As with the C400, the IsoDAR cyclotron will be divided into smaller parts to facilitate transport to the site. The magnet is split radially and will be assembled on site. The vacuum chamber is split into 3 self-centering parts along its height. The vacuum tightness is ensured by 2 axial O-rings at each interface except for the interface between the upper and lower chambers where 2 radial O-rings are used. An exploded visualization of the upper and lower parts is shown in Figs.~\ref{exploded-top} and \ref{exploded-bottom}.
The material used for the magnet parts is cyclotron steel (the exact composition shall be determined after a detailed physics study), the vacuum chambers will be made of stainless steel.

\begin{figure}[H]
    \centering
    \includegraphics[width=0.9\linewidth]{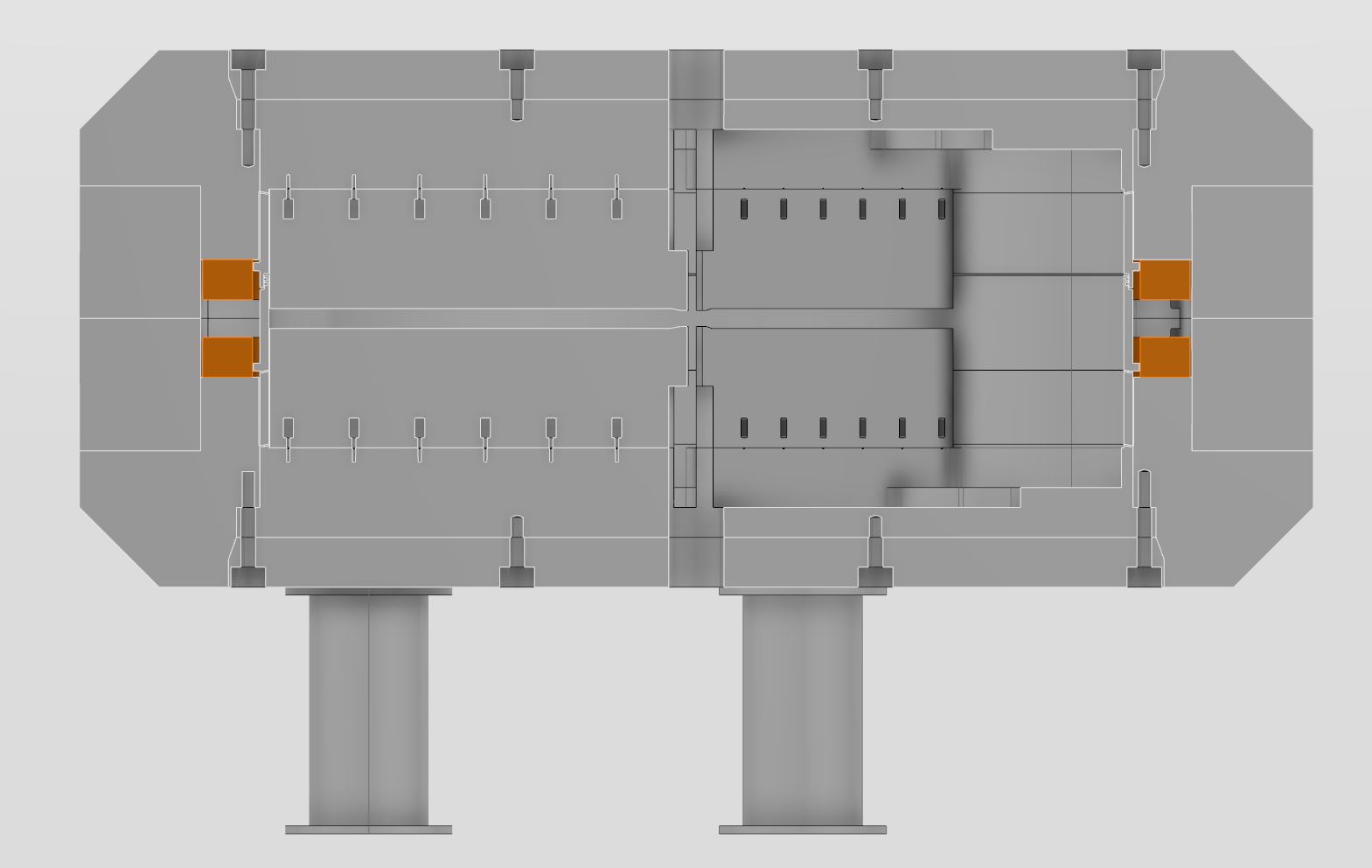}
    \caption{Cyclotron cut view}
    \label{Cyclo cut view}
\end{figure}

\begin{figure}[H]
    \centering
    \includegraphics[width=1\linewidth]{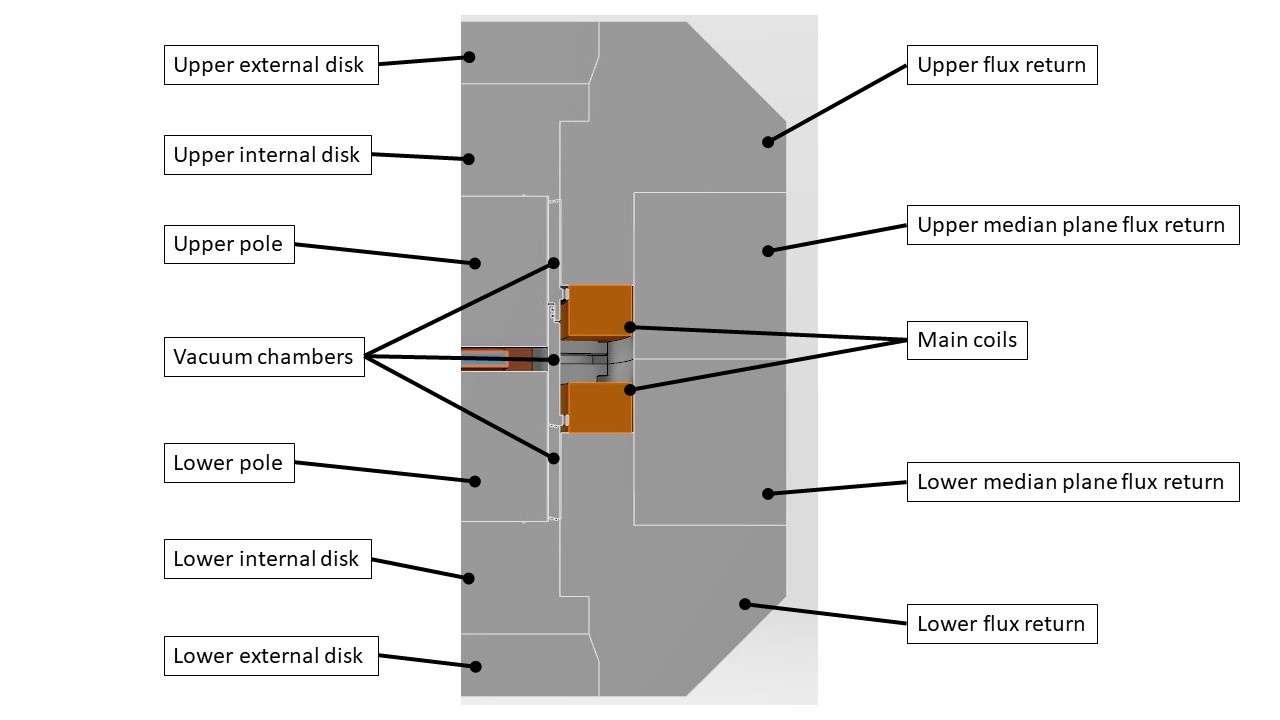}
    \caption{HCHC-60 cyclotron, detailed cut view calling out the major components of the steel, vacuum chamber, and coils.}
    \label{Cyclo detailed cut view}
\end{figure}

A cut-section view of the cyclotron is shown in Fig.~\ref{Cyclo cut view}. The external and vacuum disks are connected to each other as well as to the flux return using specific bolts (such as the Superbolt from Nord-Lock). The poles are split into several parts and an easily removable part should be added in the next iteration to ease the mapping and possibly improve the magnetic performance. A detailed view of the cyclotron components is shown in Fig.~\ref{Cyclo detailed cut view} and additional details about the yoke parts can be found in Table~\ref{Yoke part properties}

\begin{table}[tb]
    \centering
    \caption{Steel yoke, individual parts masses and dimensions.}
    \label{Yoke part properties}
    \begin{tabular}{|l|c|c|c|}
    \hline
    Name & Mass (t) & Dimensions (mm) \\
    \hline\hline
    Upper and lower external disk & 30  & \diameter$4700\times250$\\
    \hline
    Upper and lower internal disk & 36  & \diameter$4620\times463$ \\
    \hline
    Upper and lower flux return (1 and 2)   & 36 & $3100\times1950\times1015$\\
    \hline
    Upper and lower median plane flux return (1 and 2) & 29  & $3100\times1950\times705$\\
    \hline
    Upper and lower pole  & 7 & $2013\times1290\times600$ \\
    \hline
    \end{tabular}
\end{table}

\subsection{Simulated cyclotron magnetic performance}

In this section we describe the latest modifications applied to the cyclotron design regarding its magnetic properties. We will compare these results with the model from the Conceptual Design Report (CDR) from \cite{abs_isodarkamland_2015} to ensure its theoretical magnetic performance remains within specifications.

\begin{figure}[t!]
\begin{minipage}[c]{0.48\linewidth}
\includegraphics[width=\linewidth]{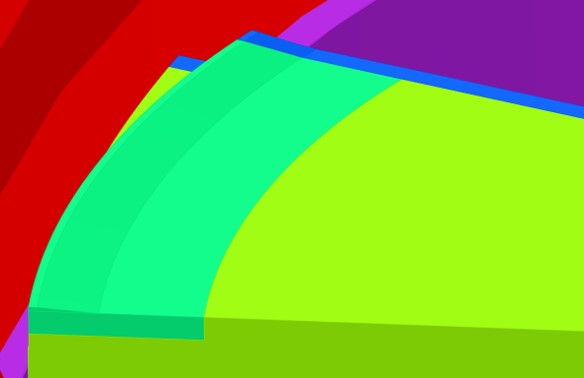}
\end{minipage}
\hfill
\begin{minipage}[c]{0.48\linewidth}
\includegraphics[width=\linewidth]{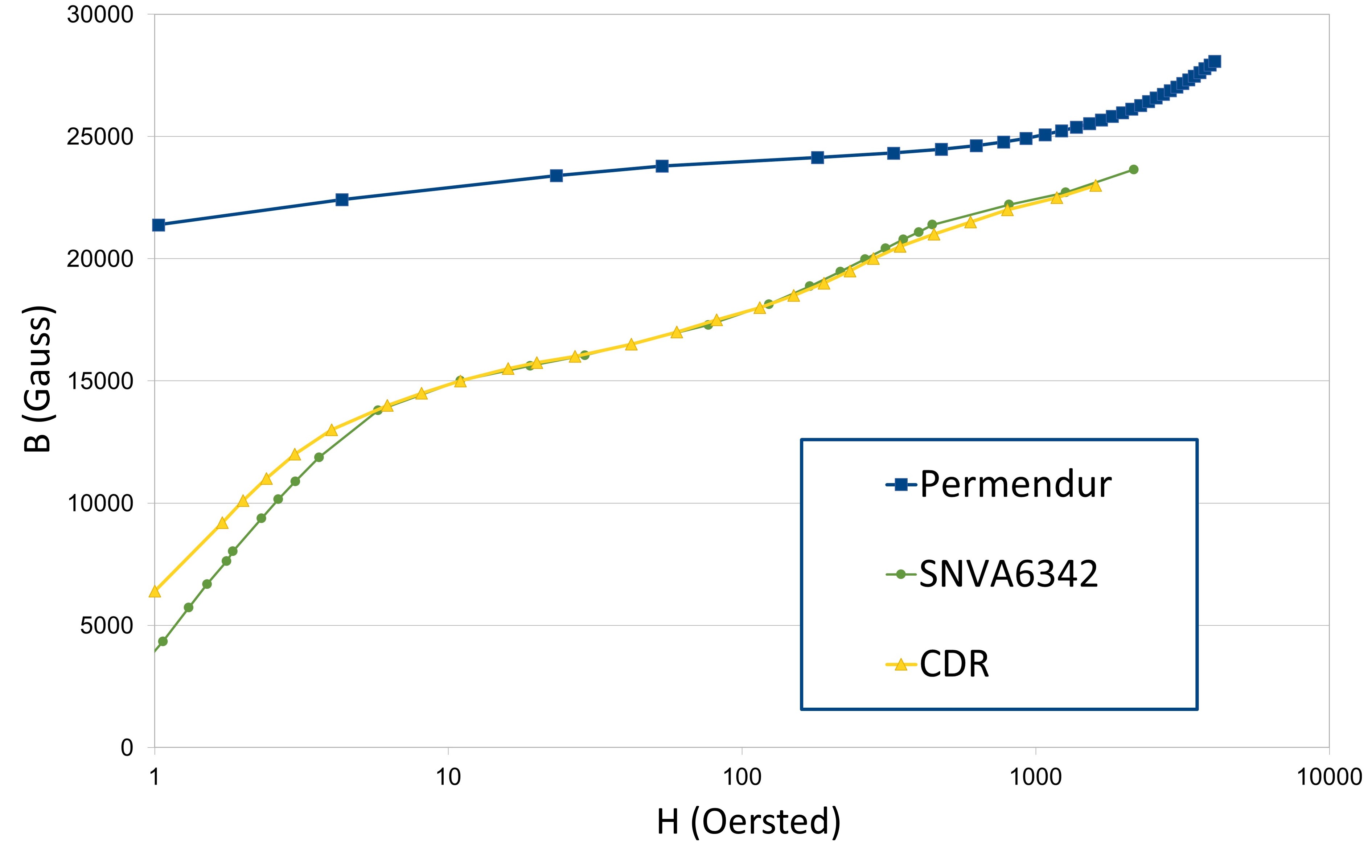}
\end{minipage}
\vspace{0.2in}
\caption{Left: Detailed view of the outer radius of the pole; only half of one pole is shown. In red is the coil, in light green is the pole made of SNVA6342 steel. The turquoise part is a piece 30~mm in height and 200~mm in width made of permendur. In blue we see the pole edge made of SNVA6342 steel, which can be shimmed to isochronize the magnetic field. 
Right: BH curves for the two steel types used: Permendur and SNVA6342. As reference, the BH curve used for the cyclotron in the CDR is also shown. }
\label{fig:pole_and_bhCurves}
\end{figure}

With the latest design of the vacuum chamber, see Section~\ref{sec:vacuumchamber}, the pole radius was slightly reduced. To keep the magnetic field isochronous up to extraction at 120 MeV, the outer radius required small modifications, see left image in Fig.~\ref{fig:pole_and_bhCurves}. First, the curvature was modified to follow the curvature of the last orbit. This enhances the total flux in the region where needed. Second, the top layer of the pole at its outer radius will be made of permendur. This alloy has better magnetic properties at high field strengths. Using permendur will enhance the magnetic flux. Finally, the pole height is increased by a slanted ridge at its extremity. This reduction of the pole gap should have no effect on the transmission and can be tuned to improve the dB/dr for the last orbits. 

The steel used for the cyclotron yoke and poles is the SNVA6342, also used, for example, in the C400 cyclotron (see Ref.~\cite{bib:c400_iron}). Its BH-curve is shown in Fig.\ref{fig:pole_and_bhCurves}. Here, we also see the better performance of the permendur alloy. 

\subsubsection{Isochronization}

\begin{figure}[tb]
\begin{minipage}[c]{0.48\linewidth}
\includegraphics[width=\linewidth]{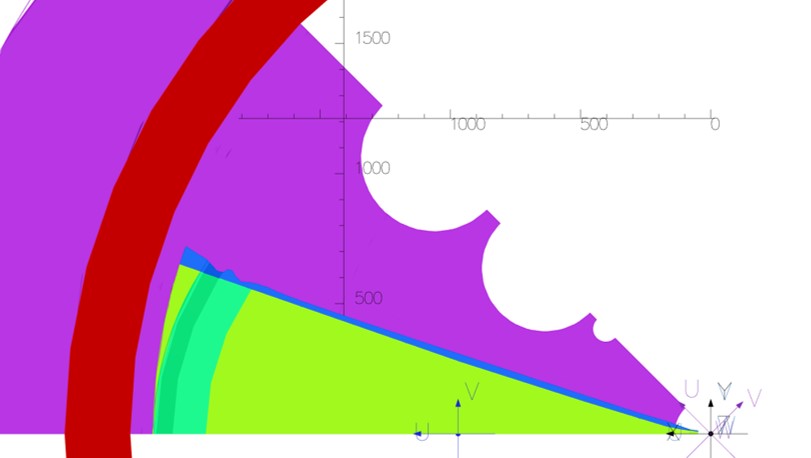}
\end{minipage}
\hfill
\begin{minipage}[c]{0.5\linewidth}
\includegraphics[width=\linewidth]{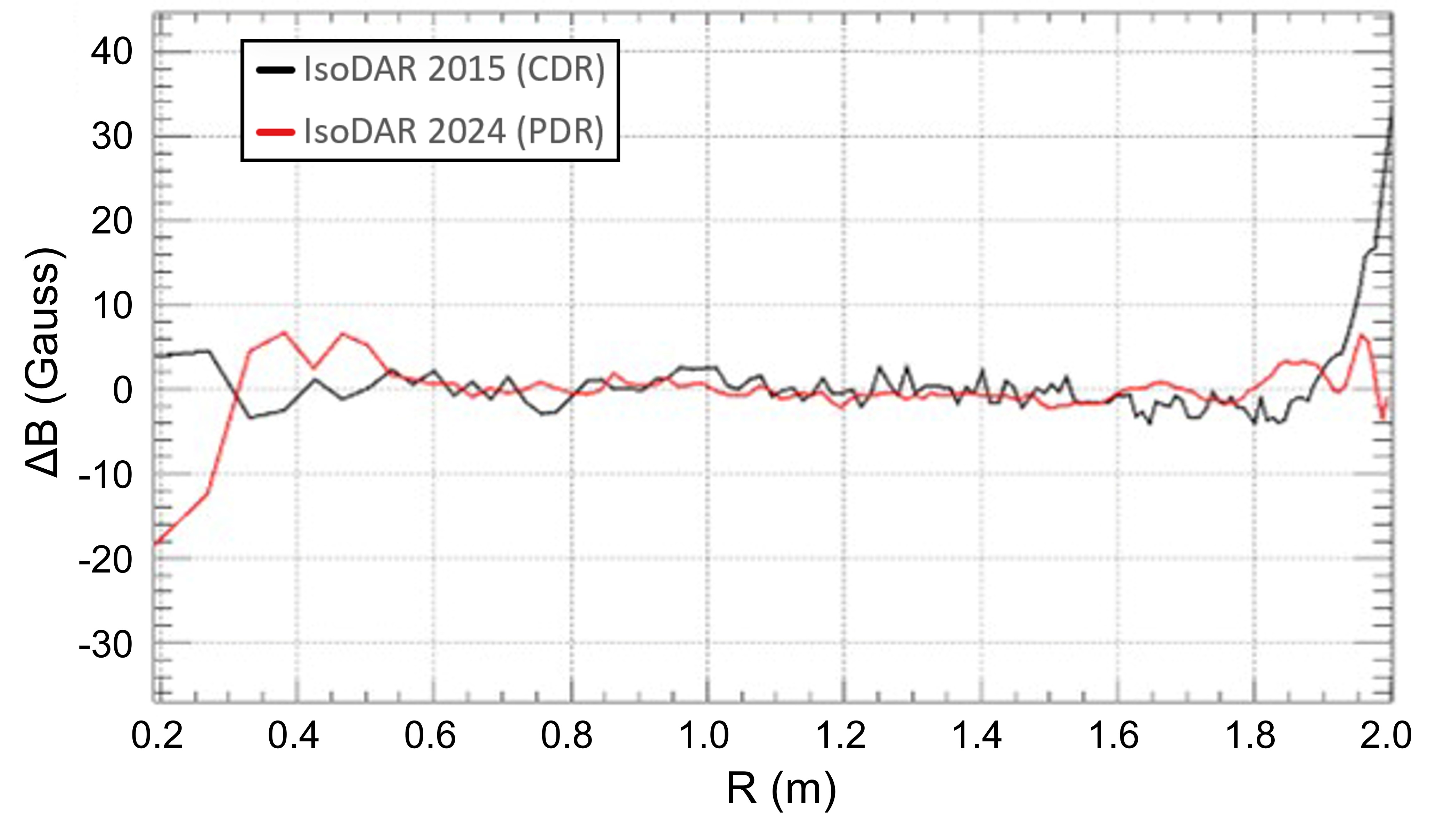}
\end{minipage}
\vspace{0.2in}
\caption{Left: Pole edge of isochronized model.
Right: The $\Delta$B (Gauss) as a function of radius (m), for the CDR model (in black) and the newest model (in red). Note that a positive dB means there is too little field at specific radius, a negative $\Delta$B implies there is too much field. The RF frequency for the CDR model is at 32.77~MHz, for the newest model it is 32.82~MHz.}
\label{fig:pole_edge_isochronized}
\end{figure}

The edge of the pole can be shimmed to isochronize the magnetic field map. In Fig.~\ref{fig:pole_edge_isochronized}, the latest results are given. The left image shows the shimming applied, the right image shows the offset in magnetic field, as required for a given RF frequency. For comparison, the offset (expressed as dB in Gauss) for the model of the CDR is also shown. 

\begin{figure}[tb]
    \centering
    \includegraphics[width=0.49\linewidth]{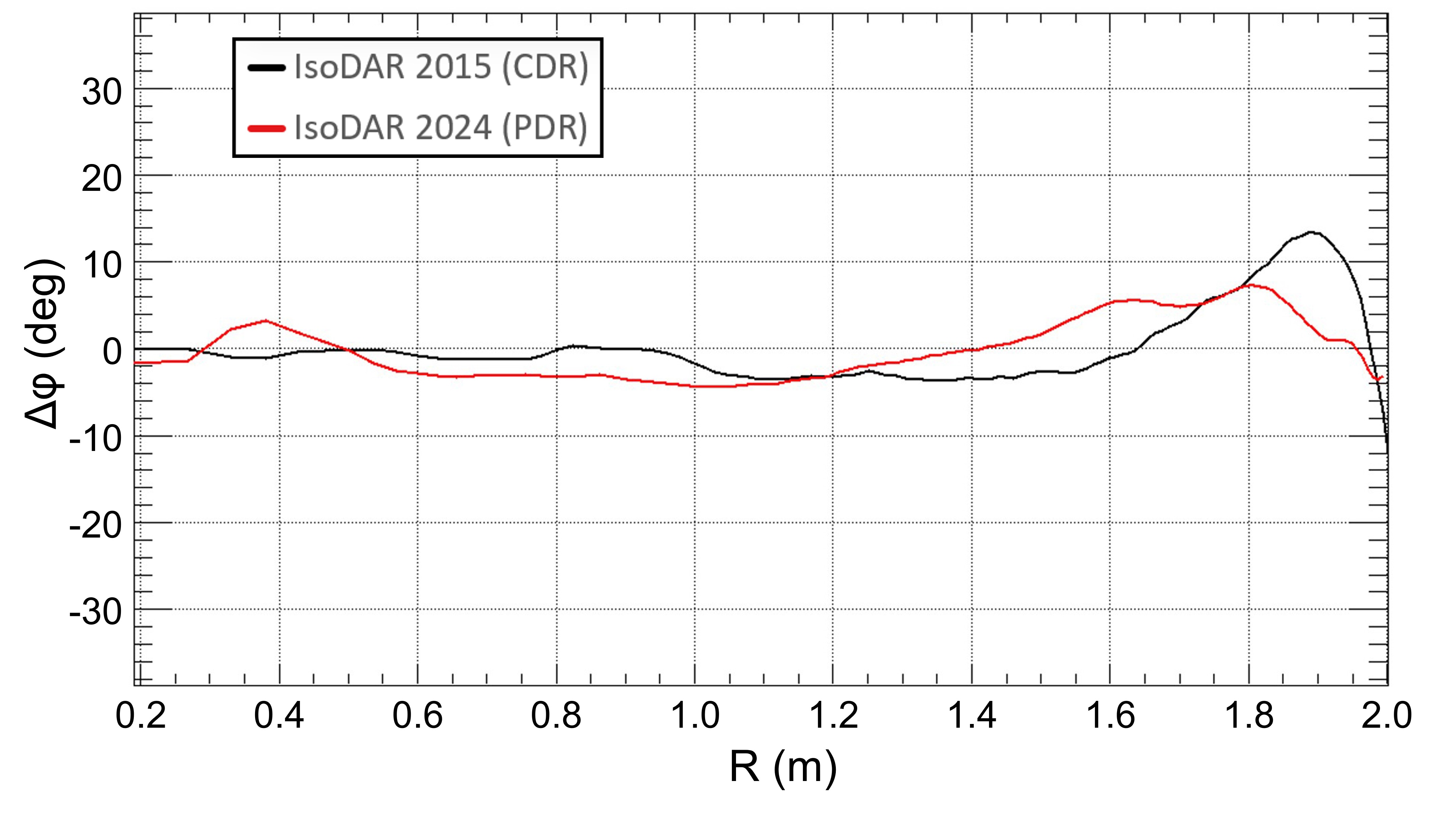}
    \includegraphics[width=0.49\linewidth]{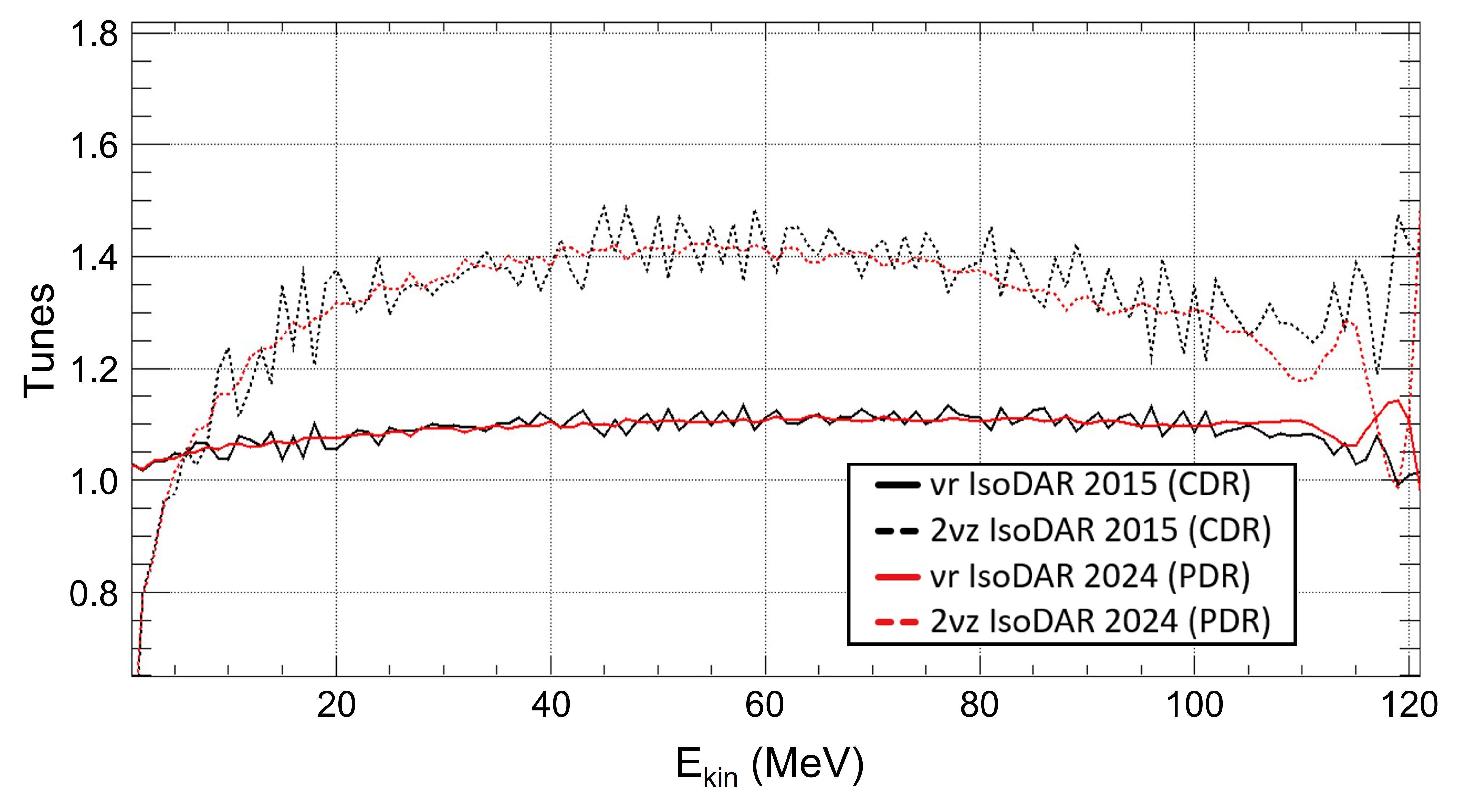}
    \caption{Left: phase slip as a function of radius. Right: vertical and horizontal tunes as a function of energy (dashed and solid lines respectively). Shown are the results for the CDR model (in black) and the newest model (in red).}
    \label{fig:phaseslip_and_tunes}
\end{figure}

The plot shows that the magnetic field is stable and strong enough up to a radius of 2~m, which is the equivalent of 121~MeV. In Fig.~\ref{fig:phaseslip_and_tunes}, the phase slip can be seen to stay within the $\pm 10 \degree$. The horizontal and vertical tunes are also given in the same figure. Here we see that there is a crossing of the tunes at high radius, invoking resonances which could impact the extraction.

As a conclusion, we can state that a configuration is possible with an isochronous magnetic field for up to 121~MeV. The transition to permendur at high radius, together with an abrupt slanted pole face, requires shimming of the pole edge. In a further study, these last modifications can be made more smooth, as not to have too much of a fluctuation in the dB for the last orbits and thus tunes that do not cross anymore, avoiding possible resonances. 

\subsubsection{Risks and Mitigation}

\textbf{Risk: Manufacturing Imperfections.}
Because no manufacturing process is perfect, we must
measure the local magnetic field developed by the accelerator magnet in an iterative process.
During the mapping iterations, physicists compare measurements results with expected calculated (theoretical) magnetic field map values and propose local magnet iron shape adjustments,
typically on pole tips. Difficulty of the process is a risk.

\noindent\emph{\textbf{Mitigation:} For that reason, pole tips, or at least the edges of the pole tips (boundary valley/hill)
must be dismountable to allow re-machining and multiple tuning iterations. This mapping
process is repeated until the final expected map is obtained, allowing, theoretically, the appropriate
beam acceleration and extraction. In the case of IsoDAR, we can either perform the entire process underground, or
the machine must be dismounted after surface
tests and reinstalled underground. This will require re-measuring the magnetic field after assembly on site, during a second (shorter) mapping campaign.}

\noindent\textbf{Risk:} Mapping Tool. Developing a mapping tool requires care as this tool accuracy will
define performance accuracy of the measured magnet itself.

\noindent\emph{\textbf{Mitigation:} Many precautions are mandatory concerning calibrations, mechanical stiffness, thermal stability and tracking as well as the use of perfect ”non-magnetic” materials (to avoid measures perturbations and unexpected distortions).}

\subsection{Interface with the RFQ} \label{sec:rfq_interface}



\begin{figure}[t!]
    \centering
    \includegraphics[width=0.7\linewidth]{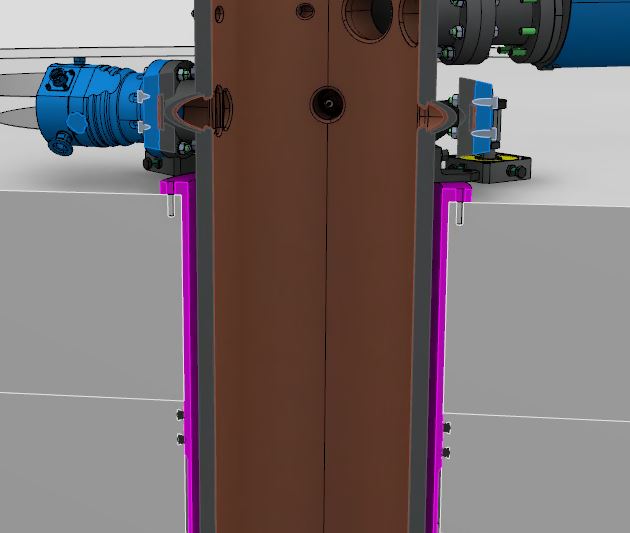}
    \caption{CAD rendering of the RFQ inside the cyclotron yoke. The vacuum sleeve is called out in magenta.}
    \label{RFQ cut section yoke interface}
\end{figure}

The RFQ will sit on top of the cyclotron along its central axis, as shown in Fig.~\ref{cycloISO}. This places the RFQ outside of the cyclotron's vacuum. Therefore, a vacuum sleeve will be used to ensure the integrity of the cyclotron's vacuum. This sleeve will be fixed on top of the upper external disk, and the vacuum sealing will be done radially on the upper internal disk (see Fig.~\ref{RFQ cut section yoke interface} ). The O-rings will have a cross-sectional diameter of 8~mm and an inner diameter of 305~mm.

To allow for adjustment, the RFQ sleeve permits a rotation of 5~mrad around the lower O-ring contact patch (as discussed with the RFQ manufacturer). The RFQ sleeve is as thin as possible to avoid any loss in magnetic field in the central region (hence the O-rings in the yoke parts).

\begin{figure}[t!]
    \centering
    \includegraphics[width=0.8\linewidth]{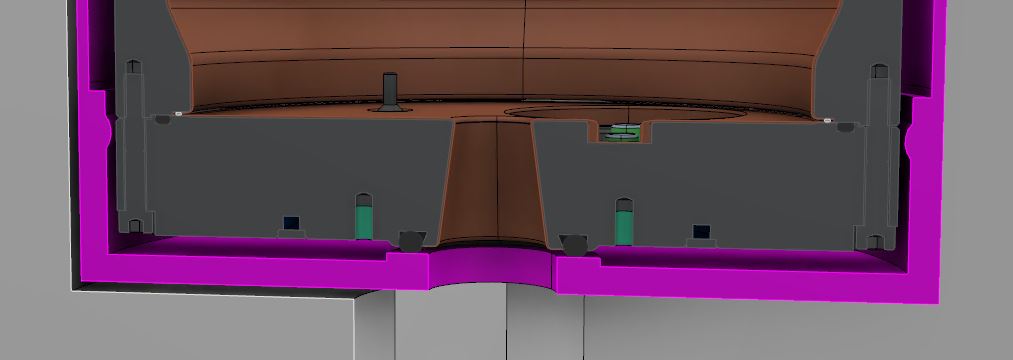}
    \caption{CAD rendering of the RFQ inside the cyclotron yoke. The vacuum sleeve is called out in magenta and the O-ring forming the vacuum interface between RFQ and sleeve is zoomed in.}
    \label{RFQ cut section sleeve interface}
\end{figure}

As shown in Fig.~\ref{RFQ cut section sleeve interface}, vacuum sealing will occur at the base of the RFQ using a thick O-ring (diameter 10~mm) which will self-seal under the RFQ's weight ($\approx3000$~N). The dimensions of the vacuum sleeve are listed in Table~\ref{tab:vac_sleeve_dims}.

\begin{figure}[t!]
    \centering
    \includegraphics[width=0.7\linewidth]{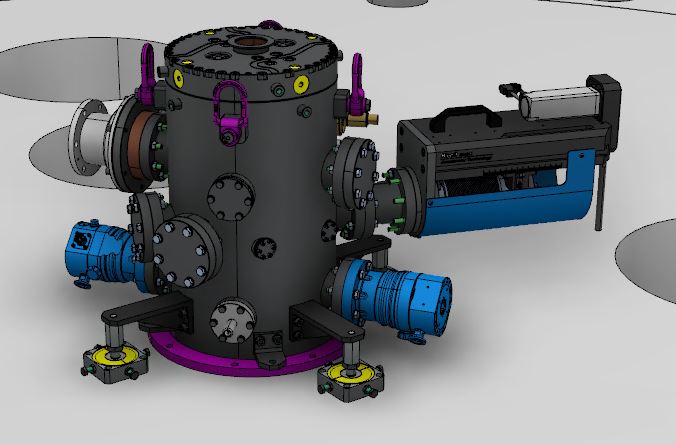}
    \caption{CAD rendering of the top of the RFQ above the cyclotron top surface. The adjustment feet are called out in yellow.}
    \label{RFQ zoom}
\end{figure}

To allow fine-tuning the cyclotron alignment, a set of adjustment feet, shown in Fig.~\ref{RFQ zoom}, are fixed to the cyclotron yoke. By using four groups of set screws and one vertical adjustment nut, the following alignment modifications can be made.
\begin{itemize}
    \item Rotation: $\pm5$\degree
    \item Tilt: $\pm5$~mrad
    \item Lateral movement: $\pm1$~mm 
\end{itemize}
In addition to this fine-tuned alignment, a coarse alignment step is envisioned during the initial assembly and commissioning procedure.

\begin{table}[b!]
    \centering
    \caption{Vacuum Sleeve Dimensions.}
    \label{tab:vac_sleeve_dims}
    \begin{tabular}{|l|c|c|c|}
    \hline
    Name & Value \\
    \hline\hline
    Height & 1030.5~mm\\
    \hline
    Outer diameter & 304.55~mm \\
    \hline
    Inner diameter & 296.55~mm\\
    \hline
    Weight & 18~kg\\
    \hline
    \end{tabular}
\end{table}

\subsection{Central Region} \label{sec:central}

\begin{figure}[!t]
    \centering
    \includegraphics[width=0.5\textwidth]{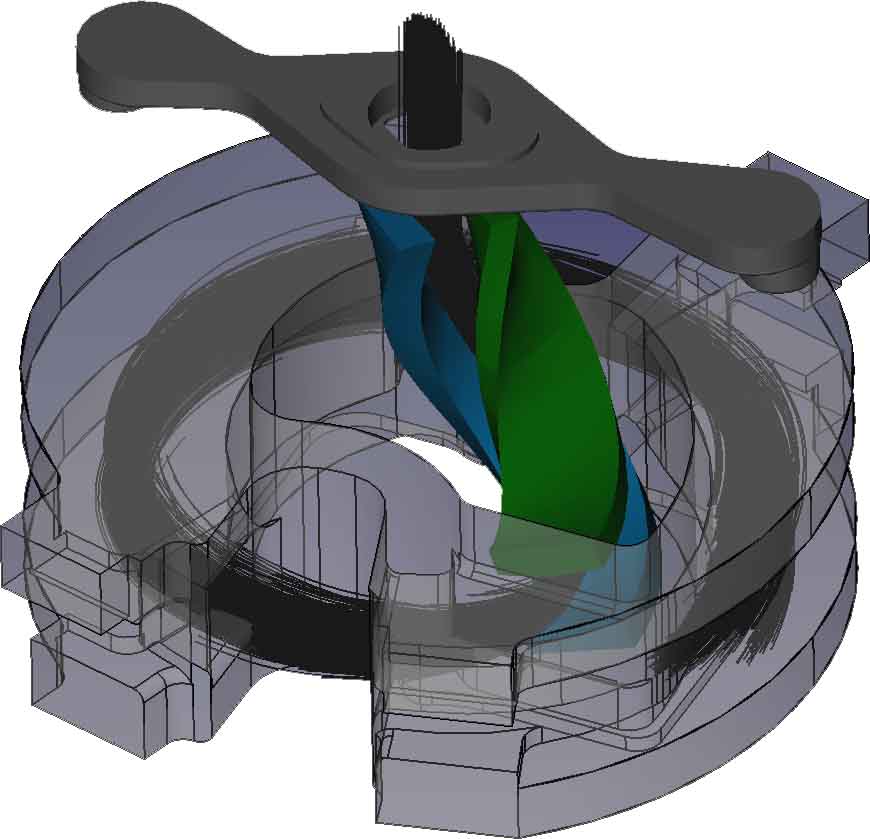}
    \caption{CAD rendering of the spiral inflector and particle trajectories 
             as implemented previously for early tests at Best Cyclotron Systems, 
             Inc.~\cite{alonso_isodar_2015}. 
             The positive and negative electrostatic deflectors are shown in 
             blue and green respectively. The cyclotron magnetic field in this 
             image is directed vertically upwards. Particles enter the spiral 
             inflector via the rectangular grounded collimator (solid gray) and 
             are guided into the cyclotron mid-plane by means of the cyclotron 
             magnetic field and the electrostatic potential between the 
             electrodes. The copper housing (transparent gray) isolates the 
             spiral inflector from the RF fields driving the
             cyclotron. From Ref.~\cite{winklehner_realistic_2017}.}
    \label{fig:bcs_inflector_opal}
\end{figure}

The injection system design is a crucial component that allows us to achieve a high beam current. Our design balances injecting the  \htp molecule at high energy against avoiding the use of complex high-voltage platforms.
Higher energy injection mitigates space charge and reduces 
geometrical (non-normalized) emittance.
An energy of 35~keV/amu was selected. This in turn determines the target output energy of the RFQ. 
As in commercial cyclotrons, an axial injection system based on a Spiral 
Inflector (SI) will be used to bend the beam from the axial direction to the median plane. We have investigated several refinements to a traditional spiral inflector which will be discussed subsequently, but the concept, as implemented successfully in a previous iteration of this work and tested at Best Cyclotron Systems, Inc. (BCSI), is visualized in Fig.~\ref{fig:bcs_inflector_opal} using trajectories calculated with OPAL~\cite{adelmann_opal_2019}.

The exit point of the spiral inflector is also the starting point of 
acceleration in the Central Region (CR) of the cyclotron. First turn acceleration is then achieved by the dee tips extending into recesses in the
spiral inflector housing. The shapes of the dee tips and the spiral inflector housing have been designed to guide the particles from the spiral inflector exit to the acceleration region while providing the necessary energy gain and beam focusing.

The central region and the spiral inflector must be designed carefully 
and together because they have strong interplay. This procedure has already 
undergone one past iteration in the summers of 2013 and 2014 at BCSI. In the following sections, we will discuss improvements to this design which will be incorporated into the preliminary design for the full IsoDAR machine.

\subsubsection{Spiral Inflector Overview \& Design Considerations}

The axial injection of the ionized beam into a cyclotron is 
realized using an electrostatic device called a spiral inflector, which consists of two curved electrode deflectors (see Fig.~\ref{fig:bcs_inflector_opal}). 
The electrostatic potential between these electrodes is able to bend the beam 90\degree from the axial line to the median plane of the cyclotron. The helical trajectory of the beam is determined both by the shape of the electrodes, the electrode potentials, and by the magnetic field of the cyclotron as the beam is bent into the median plane. In order to isolate the inflector region from external fields as much as possible, grounded entrance and exit apertures and a grounded copper housing are also included. 

An ideal spiral inflector would do nothing more than rotate the beam by 90\degree, introducing no additional beam divergence or losses and seamlessly merging with initial accelerating in the central region. Further, in order to achieve this degree of bending in the small space of the central region, $\mathcal{O}(kV)$ potentials are required, so it is important to minimize or even eliminate any \htp impact on the electrodes to prevent sparking. 

It is also important to consider the impact of a realistic magnetic field when designing our spiral inflector. This is particularly important for the IsoDAR inflector because of its relatively large size. Unlike other spiral inflectors that have been constructed, we will require a larger gap and plate size due to the non-negligible space charge and higher rigidity of \htp. This larger volume means that variations in the magnetic field along the trajectory are non-negligible. Accordingly, all following work uses a full 3D magnetic field map generated using OPERA. 

\subsubsection{Designing the Inflector}

Ultimately, an additional iteration of production and empirical testing of the final IsoDAR inflector will be performed. To optimize the inflector beforehand, we use a numerical optimization procedure to inform the best theoretical parameters.

For an inflector in a uniform magnetic field, analytic solutions to particle tracking exist for the central trajectory. However when 3D magnetic variations and/or off center trajectories are considered a numerical approach is required. 

An initial estimate based on an analytical solution for a central track serves as a numerical starting point, after which the mesh generating software Gmsh \cite{GMSHarticle} is used to produce surface representations which inform subsequent electrostatic analysis. Python's BEMPP API \cite{bempp} is then used to perform boundary analysis.

In short, this procedure enables a geometric model of our inflector to yield a full 3D electrostatic. Particles can then be tracked with high precision using the Boris method \cite{boris}. 

A major strength of this procedure and choice of software is speed, with a full simulation iteration taking only $\mathcal{O}$(hours). Prior work \cite{weigelspiral} comparing this process to commercial software such as COMSOL demonstrates that the differences in field calculations are negligible. 

\subsubsection{Geometric Refinements}

In most spiral inflectors, the two electrodes would be rectangular prisms if flattened out. They remain parallel to each other with locally flat surfaces while spiraling. Hereafter we will call this a ``Traditional'' spiral inflector. A traditional inflector can effectively transfer particles, but for off center paths these inflectors can introduce beam divergences in energy and position which were not present in the initial beam. 

We address this by adding several refinements to prior inflector iterations.
\begin{itemize}
\item Beam focusing between the RFQ and spiral inflector
\item Introducing V-shape
\item Introducing plate angling
\item Modifying the terminal end to a wedge
\end{itemize}

The addition of intermittant beam focusing ensures that the \htp entering the spiral inflector are as near to the central trajectory as possible. This is accomplished by the addition of electrostatic quadrupoles in the gap between the RFQ and inflector entrance (see also Fig.~\ref{fig:MainCycloOverview}). Using the output of the RFQ simulation as input to the quadrupoles, geometry and potentials were optimized such that the tightest possible input beam is produced.

The remaining three modifications: V-shape, plate angling, and the terminal wedge were motivated by work done by Barnard et al. \cite{barnard_inflector}. These all have the effect of introducing quadrupole moments to the field of the inflector itself, which can provide mid-track focusing to minimize beam dispersion in flight.

\begin{figure}[htb]
    \centering
    \begin{subfigure}[b]{0.3\textwidth}
        \centering
        \includegraphics[width=\textwidth]{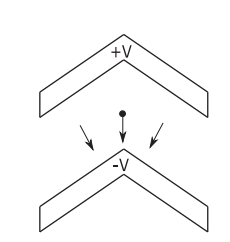}
        \caption{V-Shape Illustration}
        \label{fig:vshapeelectrodes}
    \end{subfigure}
    \hfill
    \begin{subfigure}[b]{0.3\textwidth}
        \centering
        \includegraphics[width=\textwidth]{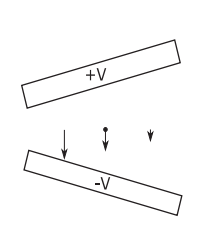}
        \caption{Angling Illustration}
        \label{fig:anglingelectrodes}
    \end{subfigure}
    \hfill
    \begin{subfigure}[b]{0.3\textwidth}
        \centering
        \includegraphics[width=\textwidth]{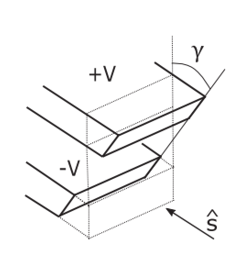}
        \caption{Gamma Wedge Illustration}
        \label{fig:gammawedgeelectrodes}
    \end{subfigure}
    \caption{Illustrations of different focusing concepts. From Ref.~\cite{barnard_inflector}.}
\end{figure}

The V-shape introduces a concave v-shape to the anode and a convex v-shape to the cathode when viewed along the long axis of the electrode (see Fig.~\ref{fig:vshapeelectrodes}). Plate angling reorients the plate angle relative to each other along the course of the inflector. In other words the plates are parallel only in the center with a positive angle between the plates initially linearly transitioning to an equal and opposite negative angle by the end (see Fig.~\ref{fig:anglingelectrodes}). The terminal gamma wedge cut modifies the end of the anode into a triangular rather than flat profile at the end and shortens the cathode. This assists with the fringe fields at the end which can produce vertical spreading. (see Fig.~\ref{fig:gammawedgeelectrodes}).

These modifications and their associated focusing fields interplay, but can be independently varied alongside other parameters like electrode potential, electrode separation, etc. We are actively exploring different techniques to handle this high-dimensional optimization problem. Monte Carlo sampling is typically well suited to this sort of optimization, but studies in which only one parameter is varied at a time have illustrated that optimization extrema can be relatively narrow, necessitating a significant, if not intractable, amount of computation time. 

As we are at the edge of practical optimization using Monte Carlo, this leads us to consider other potential alternatives. Surrogate models using a fully connected neural network capable of learning and interpolating underlying features have been previously explored \cite{EdelenMachingLearning, KoserMachineLearning, villarrealMachineLearning}. Initial iterations have proved promising and have informed the parameters for our current best inflector performance. As this work matures, it will be detailed in upcoming publications.

\subsubsection{Characterization of Spiral Inflector}

\begin{figure}[H]
    \centering
    \includegraphics[width=0.5\linewidth]{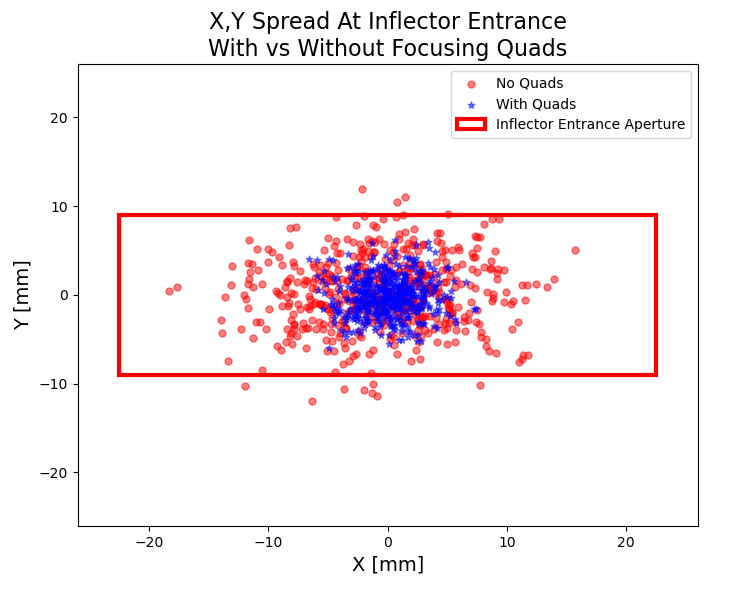}
    \caption{Illustration of beam tightness at the entrance of the spiral inflector before and after the addition of focusing quadrupoles between the RFQ and inflector.}
    \label{fig:quadfocusbeforeafter}
\end{figure}

The result of adding optimized focusing quadrupoles on the beam distribution entering the spiral inflector is illustrated in Fig.~\ref{fig:quadfocusbeforeafter}.

\begin{figure}[htb]
    \centering
    \begin{subfigure}[b]{0.5\textwidth}
        \centering
        \includegraphics[width=\textwidth]{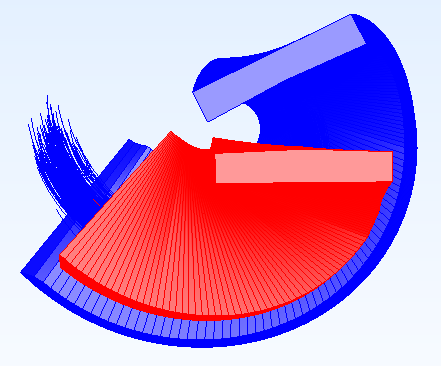}
    \end{subfigure}
    \hfill
    \begin{subfigure}[b]{0.3\textwidth}
        \centering
        \includegraphics[width=\textwidth]{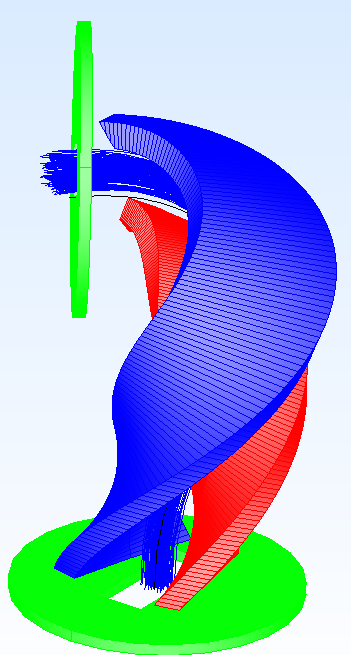}
    \end{subfigure}
    \hfill
    \caption{CAD renderings of the optimized spiral inflector. Note the v-shape to the electrodes as well as the non-parallel angle seen in the bottom view with the apertures removed, and the angled wedge visible in the top view.}
    \label{fig:NewInflectorView}
\end{figure}

The optimized spiral inflector, with \htp tagged through, is illustrated in Fig.~\ref{fig:NewInflectorView}.

The efficiency of this modified inflector is significantly improved. Efficiency must be thought of from two fronts. On the one hand we wish to transmit as much current from the RFQ into the cyclotron as possible.Particles which are not successfully transmitted must terminate somewhere. Termination on grounded apertures is acceptable while termination on the HV electrodes will result in sparking and can have significantly impact on running. It is therefore also vital to specifically consider the fraction of ions that terminate on an electrode.

We thus consider two similar cases. Both have identical optimized spiral inflectors. The difference is whether or not a grounded collimator is placed at the entrance to further narrow the entering beam beyond what the quadrupoles have provided. This collimator is a circular aperture which, for this study, was chosen to be as large as possible while guaranteeing that less than one percent of \htp impact the electrodes. The resulting efficiencies and \htp-electrode collisions are given in Table~\ref{tab:inflector_effic_comparison}.

\begin{table}[h]
    \centering
        \caption{Comparison of efficiency and ion-electrode impact for the optimized versus traditional spiral inflector.}
    \begin{tabular}{|c||c|c|}
        \hline
         & \textbf{Efficiency} & \textbf{Electrode Impact \%} \\ 
        \hline\hline
        \textbf{Traditional Inflector} & 74\% & 17\% \\ 
        \hline
        \textbf{Optimized Inflector} & 97\% & 1\% \\ 
        \hline
        \textbf{Traditional Inflector (Coll.)}& 39\% & $<1$\% \\ 
        \hline
        \textbf{Optimized Inflector (Coll.)}  & 93\% & $<1$\% \\ 
        \hline
    \end{tabular}
    \label{tab:inflector_effic_comparison}
\end{table}

\begin{figure}[htb]
    \centering
    \begin{subfigure}[b]{0.49\textwidth}
        \centering
        \includegraphics[width=\textwidth]{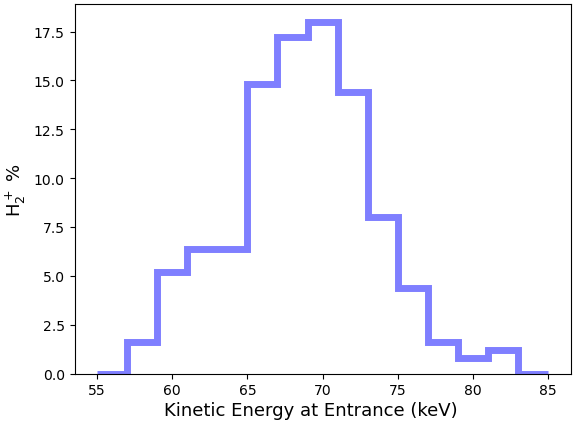}
    \end{subfigure}
    \hfill
    \begin{subfigure}[b]{0.49\textwidth}
        \centering
        \includegraphics[width=\textwidth]{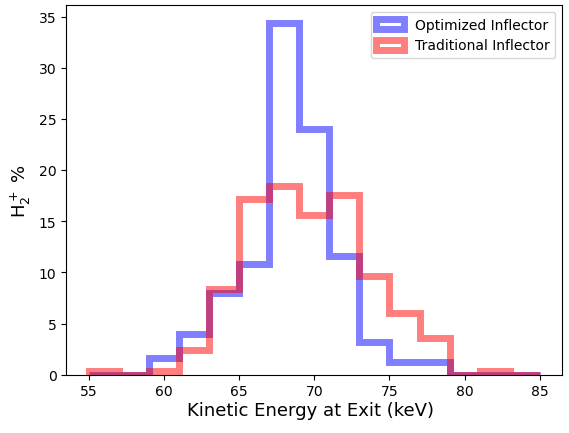}
    \end{subfigure}
    \hfill
    \caption{Kinetic energy distribution of the \htp before the inflector entrance versus at the inflector exit.}
    \label{fig:InflectorKEComparison}
\end{figure}

We also need to consider the impact of the new inflector on the quality of the beam which is successfully transmitted. These changes are characterized in Table~\ref{tab:inflector_beampar_comparison}. Detailed histograms showing the kinetic energy distribution of the \htp at the entrance and exit of the inflector are shown in Fig.~\ref{fig:InflectorKEComparison}.

\begin{table}[tbh]
    \centering
    \caption{Comparison of transmitted beam parameters at the inflector exit.}
    \begin{tabular}{|c||c|c|c|c|c|}
        \hline
         & \textbf{$\sigma_x$} (mm) & 
            \textbf{$\sigma_y$} (mm) & 
            \textbf{$\sigma_z$} (mm) & 
            \textbf{$\sigma_r$} (mm) & 
            \textbf{$\sigma_{KE}$} (keV) \\ 
        \hline\hline
        \textbf{Traditional Inflector} & 3.0 & 3.9 & 10.1 & 2.3 & 4.0 \\ 
        \hline
        \textbf{Optimized Inflector} & 4.9 & 2.7 & 6.4 & 2.4 & 3.1 \\ 
        \hline
    \end{tabular}
    \label{tab:inflector_beampar_comparison}
\end{table}

\subsubsection{Orbit Matching and Beam Centering}
\begin{figure}[htb]
    \centering
    \includegraphics[width=0.75\textwidth]{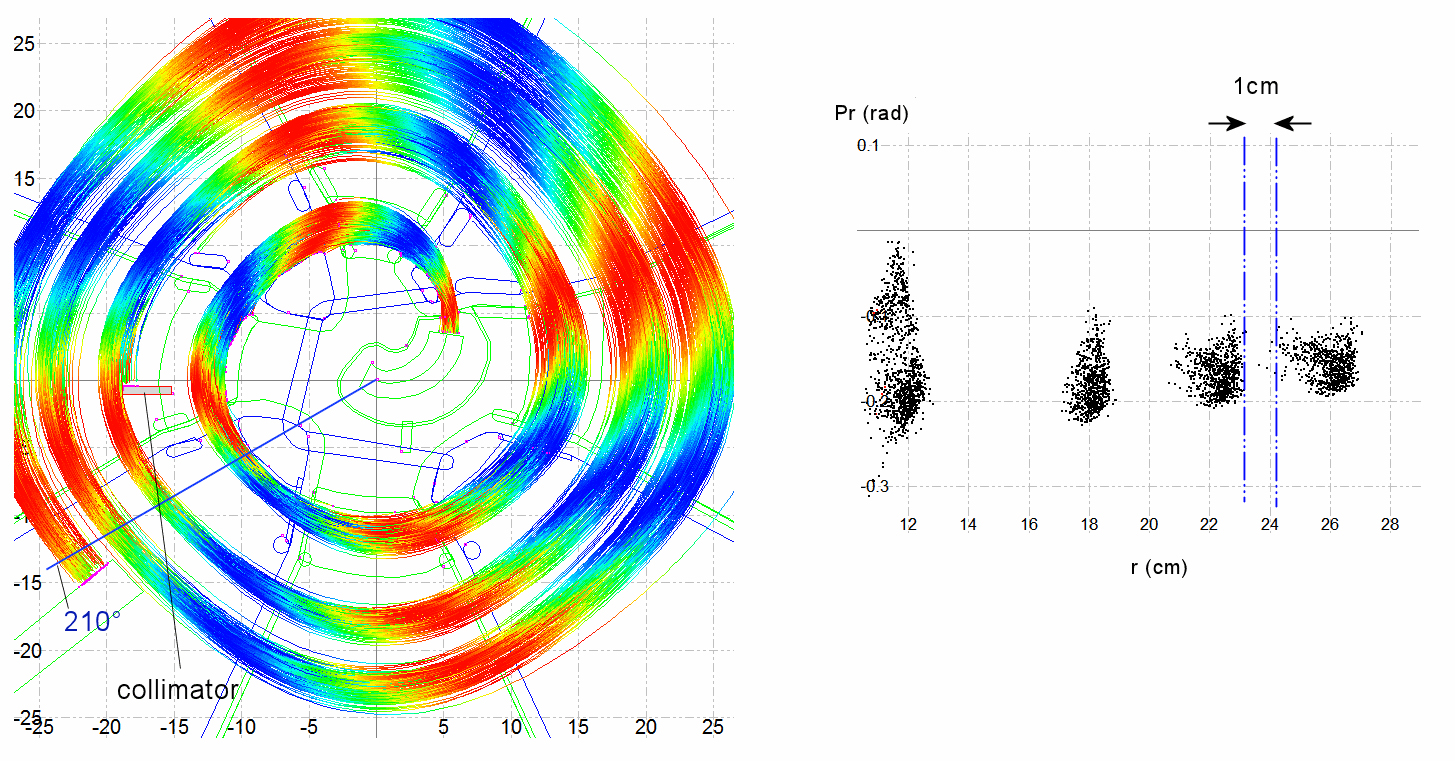}
    \caption{Beam tracking result from the AIMA central region study. Left: beam centering
    and acceleration on top of a view of the electrodes. Colors correspond to prticle phase. Right: 1~cm turn separation in the fourth turn with a single collimator. From~\cite{aima:central}.}
    \label{fig:AIMA}
\end{figure}
Once the beam emerges from the spiral inflector, it must be properly matched in terms of
orbit radius, radial momentum, and phase. A specially designed shape of the inner ends of the RF cavities performs this duty. The angles and widths of the accelerating gaps of each cavity are modified in an optimization procedure during the first two turns to place the beam onto a static equilibrium orbit appropriate for the respective energy. The center of a full 
integrated orbit after these two turns should then align well with the cyclotron center.

The spiral inflector end can be tilted (often this parameter is called $k'$) to aid in the matching and thus becomes part of the optimization procedure.

A preliminary design study of the IsoDAR central region was performed by the company AIMA in France~\cite{aima:central}. The AIMA central region model is shown in Fig.~\ref{fig:AIMA}. This central region was designed with a slightly different magnetic pole shape and with its own matched spiral inflector, using the ouput beam from an early RFQ beam tracking and single electrostatic quadrupole transverse refocusing element. It demonstrated that the RFQ-generated beam could be injected and vertically focused. By using a single collimator (removing $\approx 42$\% of the beam), an edge-to-edge turn separation of 1~cm was achieved in the fourth turn (approximately 1~MeV/amu).

The RMS beam size and energy spread was significantly worse than what we see in our optimized
inflector. Furthermore, the AIMA study did not consider space charge, which is fundamental to establishing vortex motion (see below). Somewhat counterintuitively, including space-charge will
improve the results further; as will our new optimized spiral inflector.

In the next step, we will modify the central region and optimized spiral inflector to match
well and track particles through the finalized geometry with OPAL using the PIC method to demonstrate a stable vortex.

For this preliminary design, we use a Gaussian beam with size and emittance informed by the AIMA study for further simulations of acceleration to 60~MeV/amu.

\subsubsection{Low Energy Collimation}
\begin{figure}[t!]
    \centering
    \includegraphics[width=0.65\textwidth]{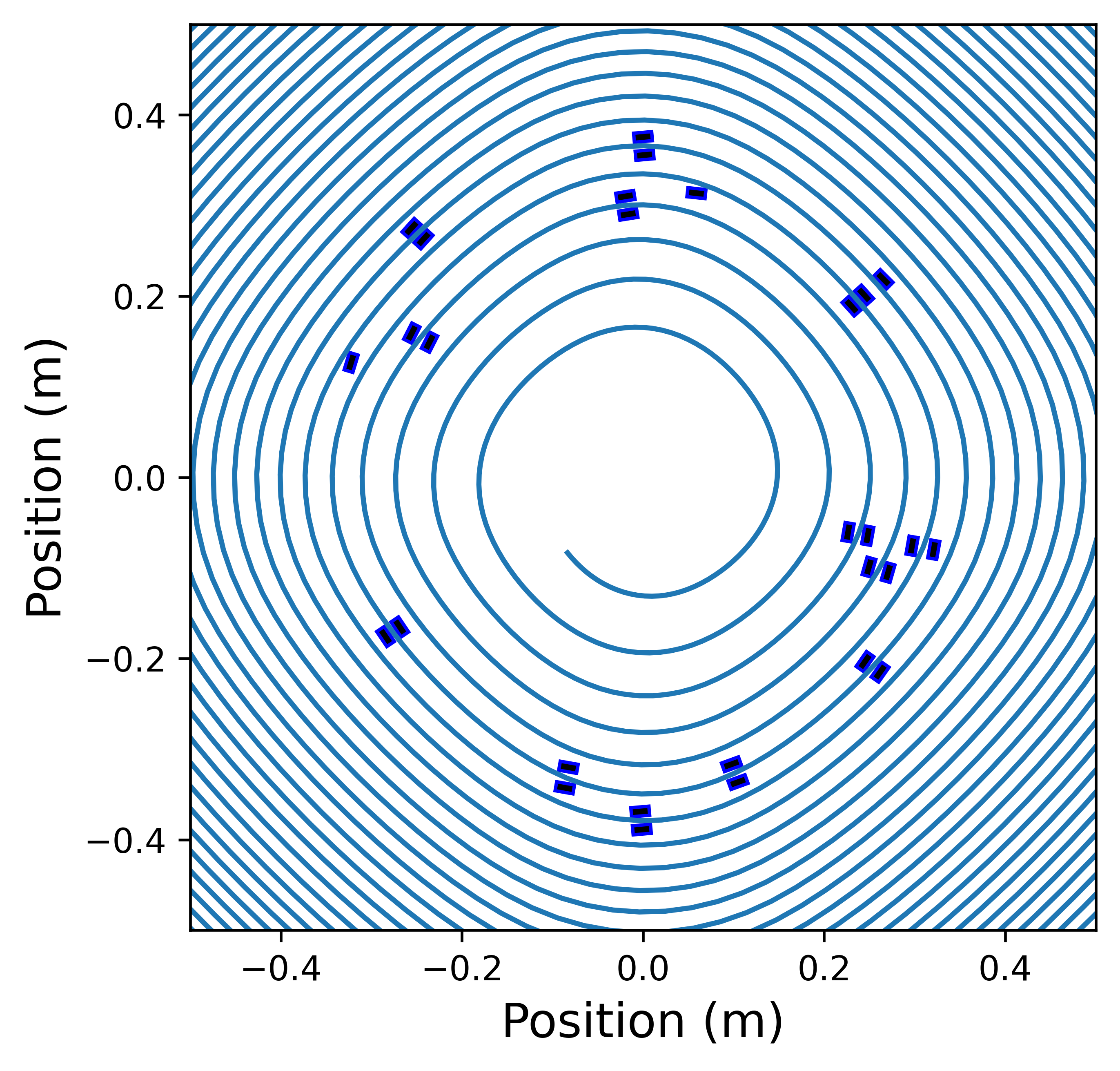}
\caption{Placement of 16 collimators in the first seven turns. All collimators are $10\times20$~mm.}
    \label{fig:CollimatorPlacement}
\end{figure}

\begin{figure}[t!]
    \centering
    \includegraphics[width=0.75\textwidth]{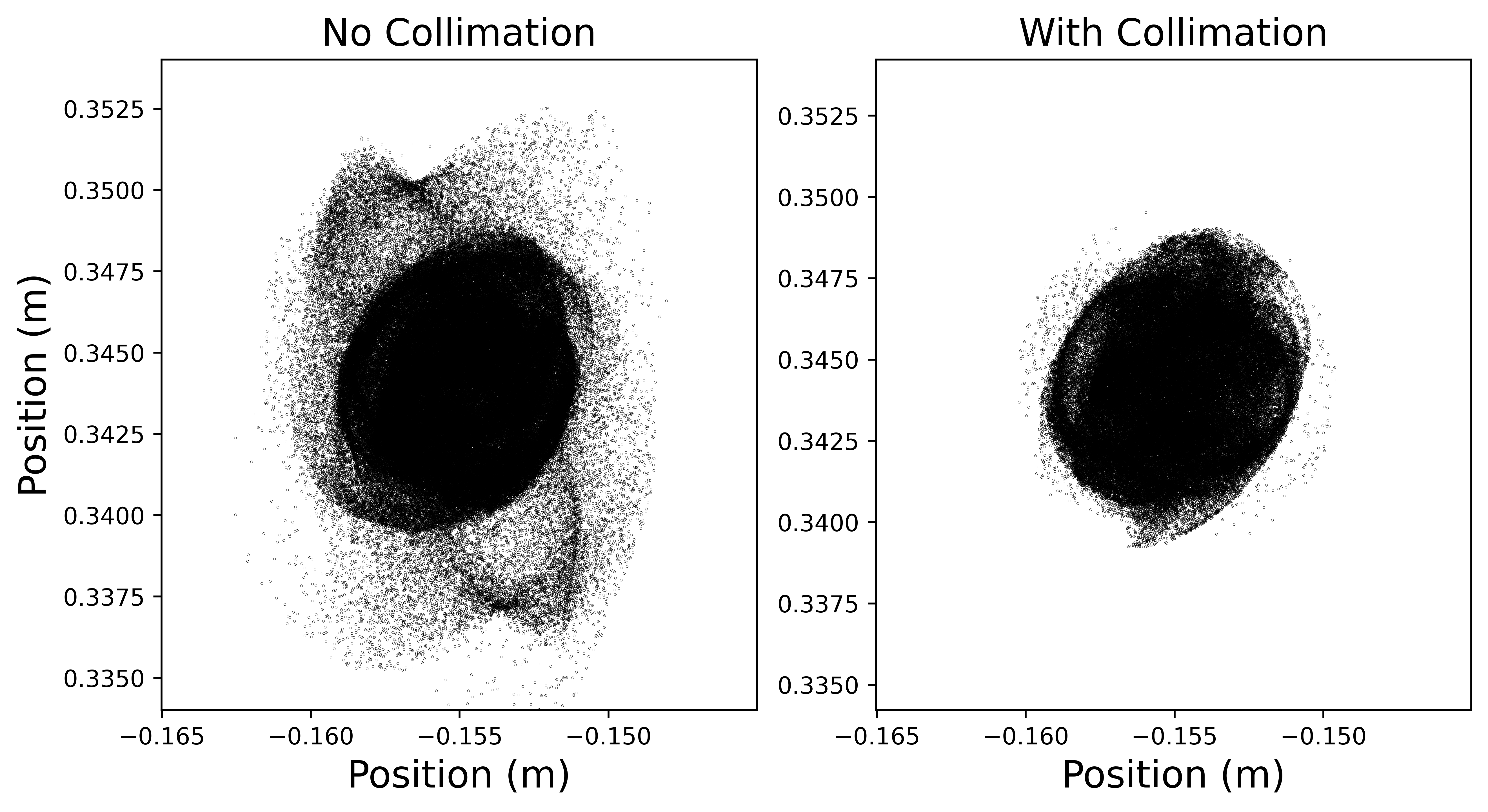}
    \caption{Snapshots of the bunch in radial-longitudinal coordinates (aka local frame of the bunch) demonstrating the reduction of beam halo by collimators.}
    \label{fig:BeamShapeImprovement}
\end{figure}

Following the injection, there is a significant beam spread in the first few turns of the cyclotron that we need to manage. We do this by placing a series of collimators around the beam core in the early turns. We must do all of our collimation in the first 7 turns of the primary acceleration, corresponding approximately to a 2 MeV/amu limit, to avoid activating our collimators. The placement of the collimators is shown in Fig. ~\ref{fig:CollimatorPlacement}. The reduction in RMS is shown in Fig.~\ref{fig:RMS_Collimation}. In total, we have 16 collimators that remove 36.4\% of the beam. The improvement in beam shape is shown in Fig.~\ref{fig:BeamShapeImprovement}. With time, the halo slowly redevelops but to a lesser extent than it would without collimation.

It is necessary to make the beam as circular as possible following the low energy collimators so that the beam is well-matched to ``vortex motion.'' Vortex motion stabilizes the beam and significantly reduces beam growth over time. It arises from the interaction between the self-field of the beam and the external magnetic field of the cyclotron. Vortex motion is discussed in depth in Ref.~\cite{baumgarten:vortex1}.

\subsection{Risks and Mitigation}
\textbf{Risk: High voltage on the spiral inflector electrodes.}
The high voltage on the spiral inflector electrodes can lead to arcing from one electrode to the other or to ground,  creating unstable injection conditions.

\noindent\emph{\textbf{Mitigation:} During the BCS tests we were able to run the spiral inflector electrodes as high as ±13.5 kV without beam using conventional polishing and cleaning methods for the electrodes.
Electro-polishing and conditioning in an inert gas environment will improve this even more.
Beam striking the electrodes could potentially induce arcs through electron emission. This can
be avoided by presenting a well-defined beam to the spiral inflector to assure close to 100\%
transmission. We have and will continue to emphasize minimizing or completely eliminating ion impacts in simulo with these tools. Finally, we will set up machine protection algorithms that can break a spark and restart acceleration in a controlled manner to 
minimize downtimes due to sparking.}

\subsection{Vacuum Chamber\label{sec:vacuumchamber}} 

\begin{figure}[t]
    \centering
    \includegraphics[width=1\linewidth]{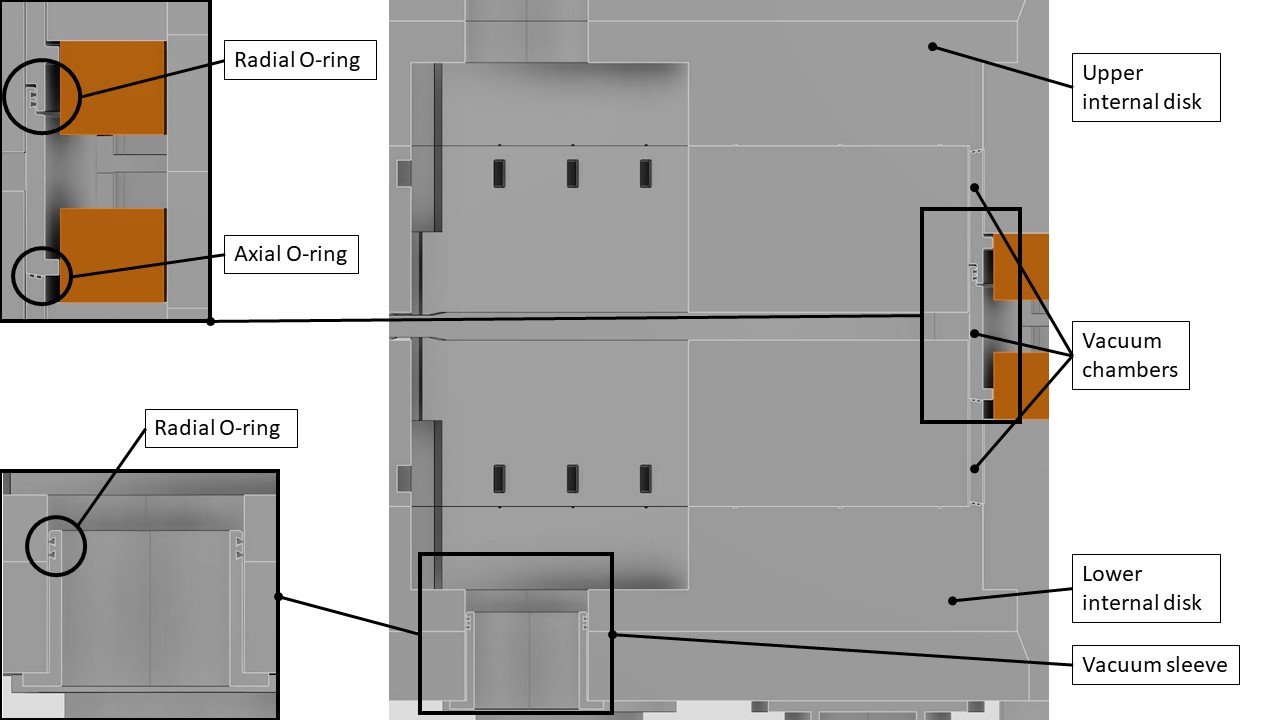}
    \caption{Cross-section view of a CAD rendering of the HCHC-60 cyclotron, calling out  vacuum chamber details.}
    \label{fig:vac_chamber}
\end{figure}

The vacuum enclosure comprises five parts: the upper and lower internal disk and 3 vacuum chambers. Fig.~\ref{fig:vac_chamber} shows the vacuum enclosure in detail and the dimensions are in Table~\ref{Vacuum part properties}. The vacuum sealing is done using 2 metallic O-rings per interface (e.g., from Omniseal~\cite{PrecisionPolymerMetal}) except for the radial interface where Polymer O-rings could be used. The interface for the accessories will be done using a sleeve design and radial O-ring, described in Fig.~\ref{fig:vac_chamber}.

\begin{table}[b]
    \centering
    \caption{Vacuum part properties}
    \label{Vacuum part properties}
    \begin{tabular}{|l|c|c|c|}
    \hline
    Name & Mass (Metric Ton) & Dimensions (mm) \\
    \hline\hline
    Upper and lower internal disk & 36  & \diameter4620*463 \\
    \hline
    Upper vacuum chamber   & 2.3 & \diameter4452*483\\
    \hline
    Lower vacuum chamber - median plane & 2.3  &  \diameter4452*483\\
    \hline
    Lower vacuum chamber - internal disk  & 1.8 & \diameter4386*380 \\
    \hline
    \end{tabular}
\end{table}

\subsubsection{Risks and Mitigation}

\textbf{Risk: Achieving the required vacuum level in the presence of the beam.}
The calculated required vacuum is $5\cdot10^{-8}$ mbar. Before the beam is turned on, the pumping plan provided will reach this with little difficulty, assuming there are no vacuum leaks. However, introduction of beam, with gas-desorption from surfaces hit by stray beam particles, could raise the pressure above the acceptable level.

\noindent\emph{\textbf{Mitigation:} Beam losses from other sources will be minimized. In addition, options for further
surface cleaning and treatment, as well as the possible addition of more pumping capacity, should be developed.}

\subsection{Coil configuration}

The CDR anticipated a slightly shorter coil than we now anticipate using. Specifically, the coil height has been increased from 200 to 240 mm. Its radial width remains at 250 mm. The total current used in the calculations is 180,850 amp-turns in each of the 180 half coils. The apparent current density is thus 3.01 A/mm$^2$, i.e. $5\% $ lower than the 3.17 A/mm$^2$ foreseen in the CDR. For now, no extraction channels are anticipated in the return yoke. If they are added, an increase in main coil current will be required

\begin{figure}[tb]
    \centering
    \includegraphics[width=0.7\linewidth]{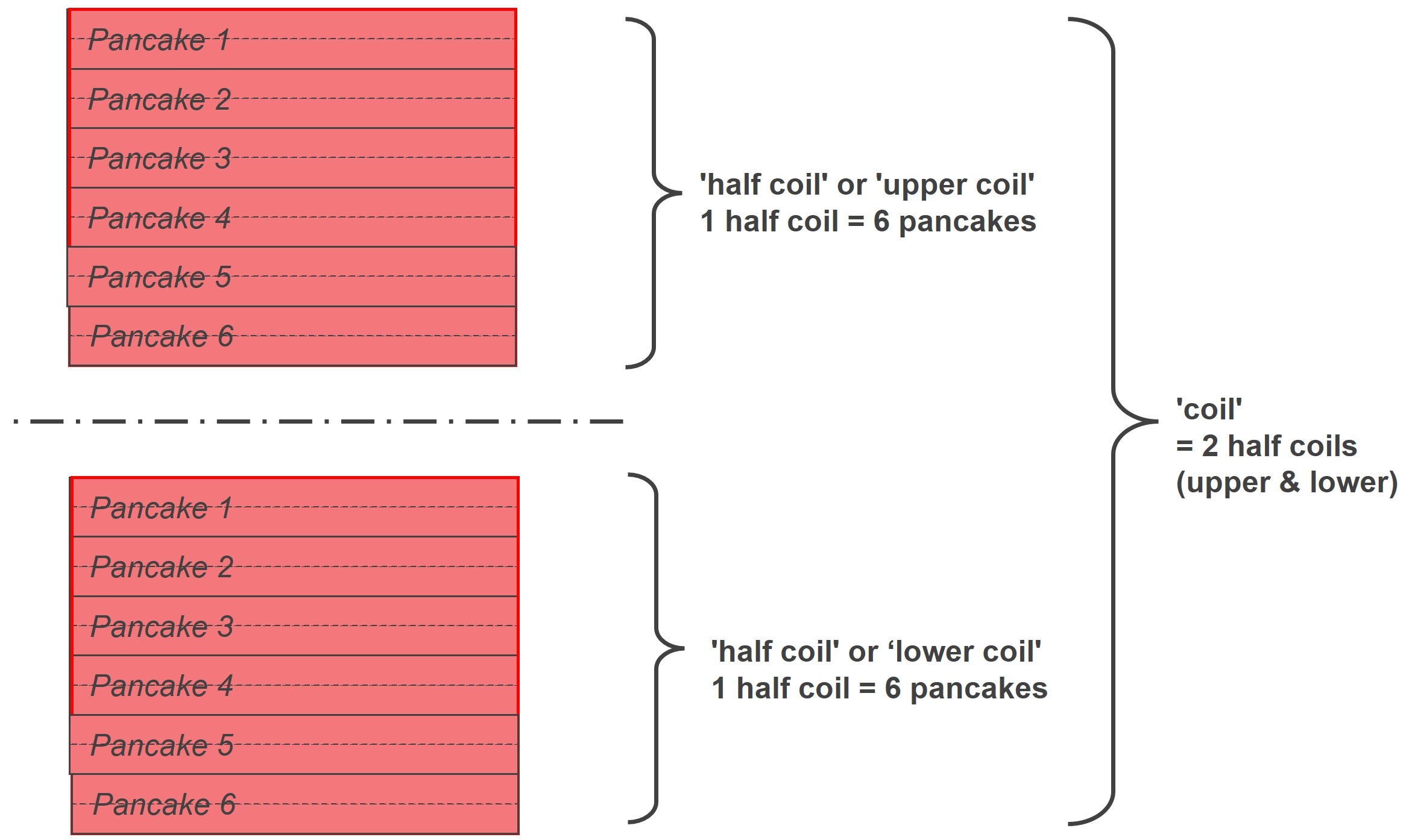}
    \caption{Left: Schematic view of a cross-section of the main coil, with six pancakes per half coil. The parameters of these coils are given in Table~\ref{tab:PancakeDescription}.}
       \label{fig:IBA_coil_config}
\end{figure}

In Fig.~\ref{fig:IBA_coil_config} a schematic view of the coil is given. We show a possible configuration of winding and cooling in Table~\ref{tab:PancakeDescription}. Although this configuration does not exactly sum to a cross-section of 240 by 250 mm$^2$, it shows that a cooling configuration is feasible. 

Working in pancakes gives the possibility of manufacturing the pancakes outside the cave and stacking them inside. The increase in height from 200 to 240 mm is the equivalent of adding an extra pancake. 

\begin{table}[h]
\caption{Proposed coil configuration for a single pancake. The HCHC-60 will require 12 pancakes.}
\label{tab:PancakeDescription}
\centering
\renewcommand{\arraystretch}{1.25}
\begin{tabular}{ll}
\hline
\textbf{Parameter}  & \textbf{Value} \\
\hline \hline
\# of axial turns & 2 \\
\# of radial turns & 12 \\
Wire size & $20.5\times20.5$~mm$^2$ \\
Cooling hole diameter & 15~mm \\
Pancake width & 251~mm \\
Copper weight & 708~kg \\
Current & $1256\times24$~Ampere-turns \\
Power & 45~kW \\
Water flow & 21.5~L/min \\
Pressure drop & 12.3~kg/cm$^2$ \\
\hline
\end{tabular}
\end{table}

\subsection{Cyclotron RF System}
The RF system for the IsoDAR cyclotron produces the primary acceleration of the ion beam and can be divided into the following subsystems:

\begin{itemize}
\item The RF cavities which accelerate the beam and provide large inter-turn
separation. The cavities are wedge-shaped and are located in the magnet valleys. There are 4 such cavities. The inner part is often called a {\it dee} whereas the outer part is
called a {\it dummy-dee}.
\item The RF tuning system that will compensate for the thermal drift of the
  cavities and maintain a stable resonating frequency.
\item The RF amplifiers that provide the necessary power to create the required RF electric field inside the cavity.
\item The RF couplers that inject the RF power from the amplifiers into the cavities
\item The Low Level RF system (LLRF) that controls and regulates the RF amplitude on the dee and drives the tuning system.
\end{itemize}

\begin{table}[!b]
	\caption{Important parameters of the RF cavities. These are per cavity. The HCHC-60 will require four RF cavities.}
    \label{tab:rf_design}
	\centering
    \renewcommand{\arraystretch}{1.25}
		\begin{tabular}{ll}
            \hline
		    \textbf{RF System Component} & \textbf{Design Value} \\
            \hline \hline
			Resonance Frequency & 32.8~MHz\\
			Dee voltage in the central region & $65$~kV \\
			Dee voltage at extraction radius & $230$~kV\\
			Dee radial extension & 2~m \\
			Acceleration gap angle & 42$^\circ$\\
			Cavity height & 1812~mm \\
			Number of dee stems & 4 per dee (2 up, 2 down)\\
			Number of dees & 4\\
			Acceleration harmonic & 4th\\
			Power dissipated per cavity & 113~kW\\
			Cavity Q-factor & 9620 \\
			\hline
		\end{tabular} 
\end{table}
We consider these in the Sections below. The numerical parameters of the RF system are listed in Table~\ref{tab:rf_design}.

\subsection{RF Cavities (Dees, Stems and Liners)}

The RF cavities of the cyclotron have been designed using CST microwave 
studio \cite{cst:microwave_studio} in order to define the shape of the dee
and the dee stems. We optimized our designs to reach the right resonance frequency, satisfy the dee voltage Law, and minimize power dissipation. Ultimately, this improves the quality factor (Q-factor).

\begin{figure}[tb]
    \centering
    \includegraphics[width=1\linewidth]{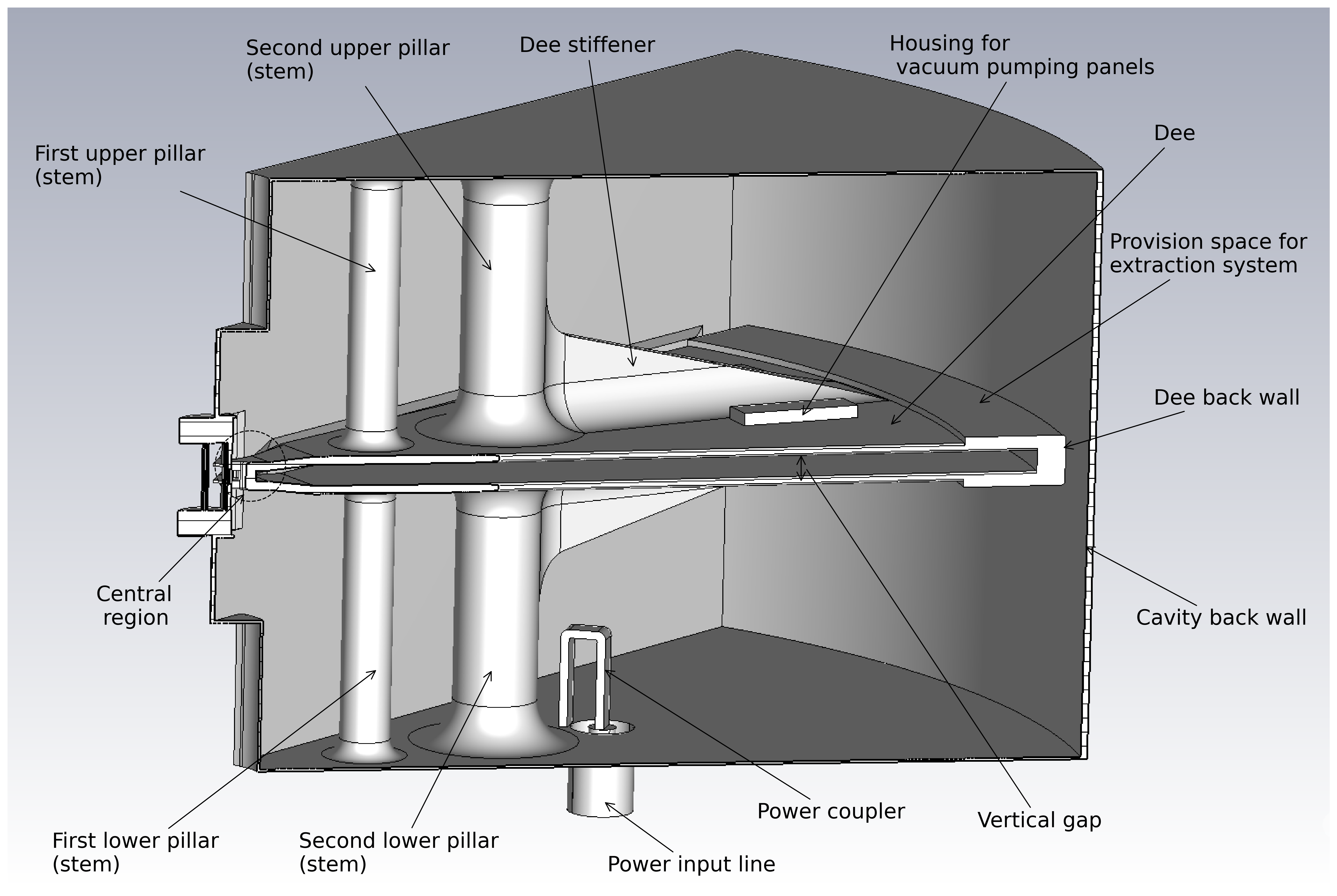}
    \caption{Section view of the HCHC-60 double-stem, double-gap RF cavity,
             calling out the most important internal features.}
    \label{fig:RF_items_legend1}
\end{figure}

\begin{figure}[tb]
    \centering
    \includegraphics[width=1\linewidth]{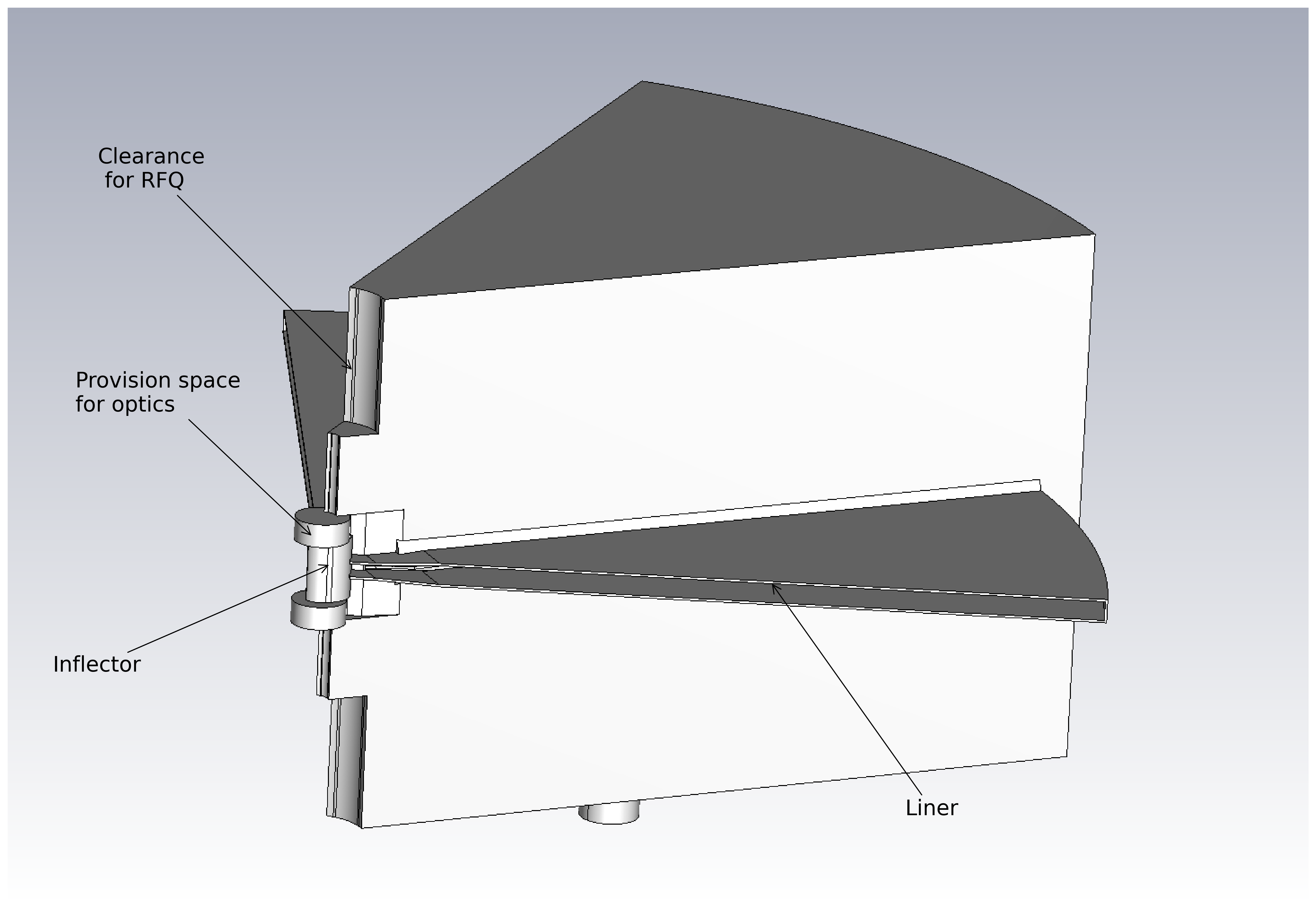}
    \caption{CAD rendering of the HCHC-60 double-stem, double-gap RF cavity,
             calling out the most important external features.}
    \label{fig:RF_items_legend2}
\end{figure}

The dees and cavity walls are all made of OFHC copper, 
which has good electrical conductivity and good solderability.
The dees are fabricated out of a 20~mm thick copper plate and are made
rigid by a supporting arm.
The four round stems support the dee and provide the right inductance in order to produce 
the required resonance frequency. Their size and location produce the desired Dee 
voltage versus radius profile. Each dee will be equipped with vacuum pumping panels to 
efficiently pump the median plane.
One dee is equipped with one electrostatic deflector for
extraction. This fact has not been taken into account in detail in the RF simulations to 
date, but some room for housing the deflector has been provisioned at the back of the dee. 
Figures \ref{fig:RF_items_legend1} and \ref{fig:RF_items_legend2} present the main elements composing the 
RF cavities.


\subsubsection{Integration of preliminary central region in RF Cavities }

In order to compute a more accurate voltage radial dependency law, we have integrated the central region design proposed by AIMA~\cite{aima:central} into the RF cavity modeling, several technical choices were made to ensure the effective implementation of this innovative configuration:
\begin{itemize}
\item The central region design was trimmed at 150~mm from center. 
\item	Both the dee and dummy-dee original central region was removed by trimming the cavities at 250~mm from center.
\item	Sections were cut straight, and a linear transition (loft) technique was used to achieve the connections between the trimmed components.
\end{itemize}

\begin{figure}[tbh]
    \centering
    \includegraphics[width=0.75\linewidth]{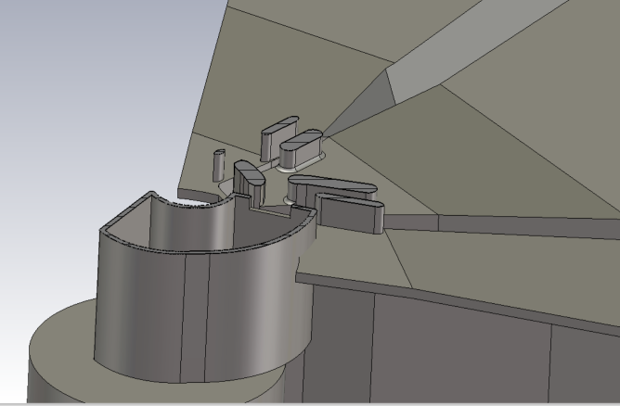}
    \caption{CAD rendering of the dee--close-up of the central region, showing the spiral inflector housing and field-shaping posts.}
    \label{fig:RF_1_central_region_close_view}
\end{figure}

\begin{figure}[tbh]
    \centering
    \includegraphics[width=0.75\linewidth]{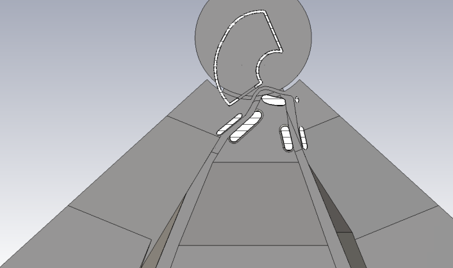}
    \caption{CAD rendering of the dee--top view, showing the spiral inflector housing and field-shaping posts.}
    \label{fig:RF_2_lofted_transition_close_view}
\end{figure}

\begin{figure}[tbh]
    \centering
    \includegraphics[width=0.75\linewidth]{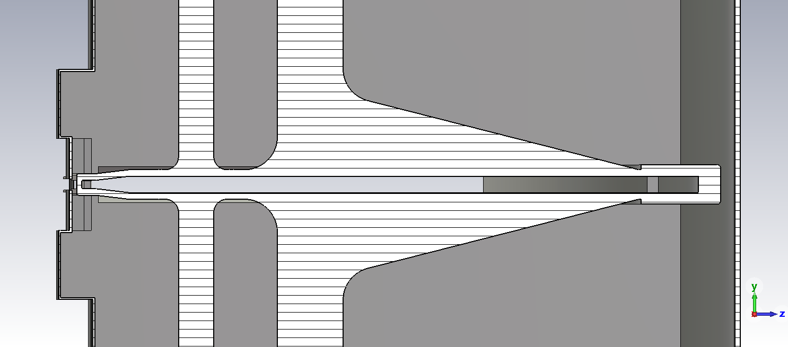}
    \caption{CAD rendering of the dee--side view, showing the radial evolution of the gap.}
    \label{fig:RF_3_vertical_gap_evolution_close_view}
\end{figure}

Figure~\ref{fig:RF_1_central_region_close_view} shows the final integration of the central region. The lofted transitions can be better seen from the top view in Fig.~\ref{fig:RF_2_lofted_transition_close_view}. They leads to an acceptable linear vertical gap opening from 26 mm at a radius of 150~mm, to 50~mm at a radius of 250~mm. The section view  of the model, presented in Fig.~\ref{fig:RF_3_vertical_gap_evolution_close_view}, shows the radial evolution of the vertical gap. 
In addition, adequate clearance provisions were incorporated, specifically to fit the inflector and optics coil. Symmetry with regard to the median plane symmetry was maintained.
Power bridges between dees in the central regions were removed since we recommend to power the 4 cavities independently.

\subsubsection{RFQ Integration in RF Cavities}

\begin{figure}[tb]
    \centering
    \includegraphics[width=0.75\linewidth]{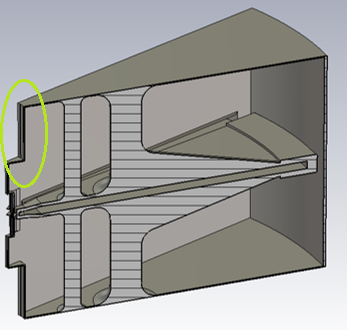}
    \caption{CAD rendering of the dee, showing the clearance for RFQ.}
    \label{fig:RF_4_RFQ_clearance}
\end{figure}

Some clearance provision has been set for the integration of the RFQ, as seen in 
Fig.~\ref{fig:RF_4_RFQ_clearance}:
\begin{itemize}
\item Diameter 280~mm
\item Down to 35~cm of median plane
\end{itemize}

While the RFQ will only be in the upper location, we have decided to keep median plane symmetry, for computation, manufacturing, and performances purposes. 
We have observed that the RFQ clearance has a low impact on cavity frequency and performances, and the output of the RFQ could be lowered more without impacting the cavity behavior.

\subsubsection{RFQ getter pumps in RF Cavities }

\begin{figure}[tb]
    \centering
    \includegraphics[width=0.75\linewidth]{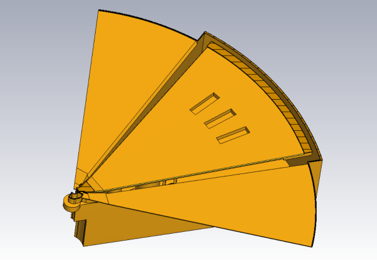}
    \caption{CAD rendering of the deer, showing the provisional getter housings (rectangular recesses).}
    \label{fig:RF_5_getter_housings}
\end{figure}

 Incorporating getter pumps into the dees allows us to improve our vacuum quality. By integrating three housing for SAES UHV1400 getter 
pumps~\cite{ModulesUHV} per dee, we aim to ensure highly efficient and reliable 
vacuum performance within the cyclotron system. The use of six of those getter pumps per cavity facilitates the effective removal of residual gases, since achieving and maintaining ultra-high vacuum conditions essential for optimal particle acceleration. Figure~\ref{fig:RF_5_getter_housings} shows acceptable locations for three getter pump housings. Furthermore, our preliminary assessments suggest that it may be feasible to scale up the configuration to include up to two sets of five getter pumps per cavity. The very low electrical field of the locations of the housings will allow the getter pumps to function safely.

This early integration of the housings of the getters in the dees is mandatory to assess their impact on the cavity frequency, and on surfaces currents.
However, to finalize this integration, comprehensive vacuum computations will be needed. These computations will enable us to determine the ideal amount of getter pumps, ensuring a properly engineered vacuum system that underpins the overall success and efficiency of the cyclotron.

\subsubsection{RF Cavities design optimization}

\paragraph{Performance requirement target}

We have optimized the cavity geometry to ensure the following requirements:
\begin{itemize}
\item RF frequency of accelerating mode: 32.8~MHz
\item Reference accelerating voltages: 65~kV at 80~mm (capture), 230~kV at 2000~mm (extraction)
\end{itemize}

The initial target value for the outer radius voltage was initially set at 250~kV, but from extensive prior experience we expect such high voltage will be very difficult to maintain in cyclotron accelerator gaps of this size. It has been decided to reduce the target to 230~kV which remains a challenging, albeit manageable, value.

\paragraph{Parametric optimization of the design}

\begin{figure}[tb]
    \centering
    \includegraphics[width=0.75\linewidth]{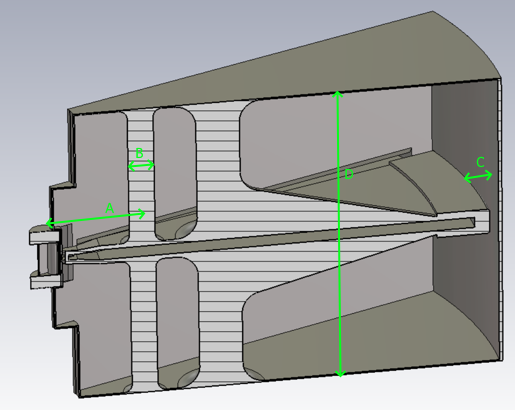}
    \caption{CAD rendering of the dee, overlayed with an indication of the design optimization parameters.}
    \label{fig:RF_6_optim_parameters}
\end{figure}

We have intensively used parameters while building the CST model, to allow fast design exploration. We have retained four parameters to play with, for optimizing the cavity behavior. The parameters, shown in Fig.~\ref{fig:RF_6_optim_parameters}, are defined as a deviation from the early cavity design values.

\begin{itemize}
\item 	A: 1st pillar radial position
\item 	B: 1st pillar diameter
\item 	C: Deflector wall distance
\item 	D: Deflector height
\end{itemize}

We did some initial exploration of the parameter space, but we did the final optimization using the built-in CST optimizer.

\begin{table}[tbh]
	\caption{RF cavity optimization goals definition and achieved values.}
    \label{tab:goals_rf_design}
	\centering
    \renewcommand{\arraystretch}{1.25}
    \begin{tabular}{llll}
    \hline
    \textbf{Goal description} & \textbf{Target}  & \textbf{Final   Value} & \\
    \hline \hline
    Ratio between voltage at   80~mm and 2000~mm  & 0.28 & 0.283 \\
    Mode 1 frequency   & 32.8~MHz        & 32.805~MHz      \\       
    \hline
    \end{tabular}
\end{table}

\begin{table}[tbh]
	\caption{RF cavities optimized design parameters.}
    \label{tab:optim_rf_design}
	\centering
    \renewcommand{\arraystretch}{1.25}
    \begin{tabular}{lll}
        \hline
         \textbf{Design parameter}              &  \textbf{Value (mm)}                                  \\
        \hline\hline
        First pillar radial position (A)       & 444.6    \\
        First pillar diameter (B) & 117.7 \\
        Distance between dee and cavity back wall (C)   & 42  \\
        Total cavity height (D) & 1412 \\
        \hline
    \end{tabular}
\end{table}

We selected the trust region algorithm and have defined scaled goals that would ensure convergence toward the required frequency and voltages. The trust region algorithm converged in a few tens of iterations. The definition of the optimization goals, and their target and reached values are contained in Table~\ref{tab:goals_rf_design}. Table~\ref{tab:optim_rf_design} shows the optimal values reached by the optimizer.

\subsubsection{RF cavity computation results} \label{RF cavity computation results}
\mbox{}

\begin{figure}[tb]
    \centering
    \includegraphics[width=0.6\linewidth]{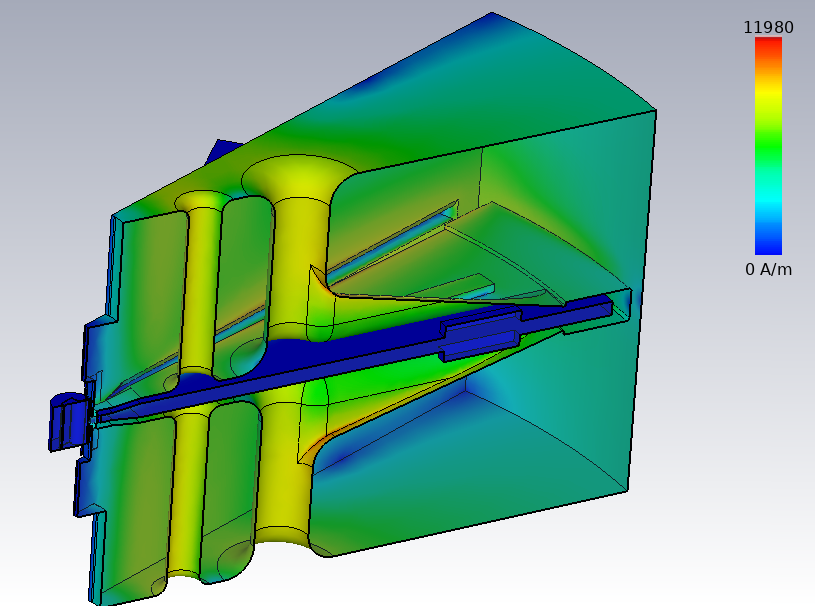}
    \caption{RF cavity: Surface currents, calculated with CST.}
    \label{fig:RF_11_surface_current}
\end{figure}

\begin{figure}[tb]
    \centering
    \includegraphics[width=0.6\linewidth]{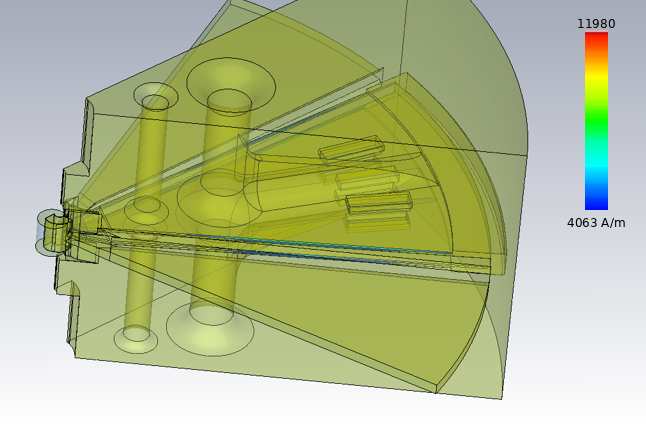}
    \caption{RF cavity: Surface current generating more than 10~W/cm$^{2}$}
    \label{fig:RF_12_surface_current_above_10W_m}
\end{figure}

The RF cavity computation employing the eigen mode solver on the optimized geometry yielded the following performance and characteristics of the cyclotron system:

\begin{itemize}
    \item \textbf{Frequency:} The RF cavity operates at a frequency of 32.805~MHz, ensuring compatibility with the designed acceleration and the RFQ integration.
    \item \textbf{Voltage Ratio (center to outer radius):} The reached voltage ratio is 0.283, indicating that computed voltages at center would be 65.2~kV for a chosen voltage of 230~kV at the outer radius (2~m).
    \item \textbf{CST computed quality factor (Q):} 10586
    \item \textbf{Estimated quality factor (Q):} 7281
    \item \textbf{Stored Energy (for max RF of 230~kV at 2~m):} 4.06~J
    \item \textbf{CST Computed Losses:} The computed losses per cavity using CST simulations are 79~kW.
    \item \textbf{Estimated Power Dissipation:} By experience, senior RF designers at IBA state that the CST eigenmode simulations underestimate the actual losses by up to 30\%. If we increase the CST computed losses (79~kW) by this percentage, the actual estimated power dissipation per cavity becomes 113~kW.
    \item \textbf{Estimated Power Requirement:} To effectively drive the cyclotron, each cavity will require an estimated power input of at least 263~kW, which include power dissipation and beam loading.
    \item \textbf{Maximal Surface Current:} The RMS surface current reaches a maximum of 17052.6~A/m in some reduced locations. Which leads to a maximum surface power of 43.5~W/cm$^2$. This maximal surface power is localized on the dee edges. Careful attention will be required for thermal management in these regions. For the rest of the surfaces, the thermal dissipation is well below 10~W/cm$^2$. See Figs.~\ref{fig:RF_11_surface_current} and \ref{fig:RF_12_surface_current_above_10W_m}.
    While matching the requirement, these computation results provide valuable information for next steps of the cyclotron's RF cavity design including consideration in managing thermal effects.
\end{itemize}

\begin{figure}[tb]
    \centering
    \includegraphics[width=0.6\linewidth]{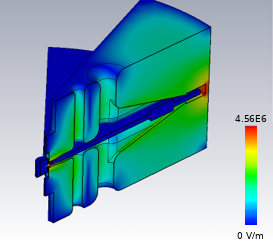}
    \caption{RF cavity: E-field vertical cross section, calculated with CST.}
    \label{fig:RF_7_E_field_Vertical_section}
\end{figure}

\begin{figure}[tb]
    \centering
    \includegraphics[width=0.6\linewidth]{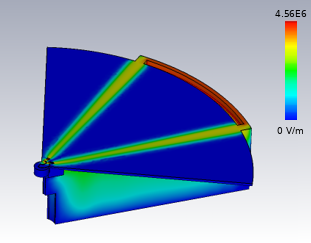}
    \caption{RF cavity: E-field horizontal cross section, calculated with CST.}
    \label{fig:RF_8_E_field_Horizontal_section}
\end{figure}

\begin{figure}[tb]
    \centering
    \includegraphics[width=0.6\linewidth]{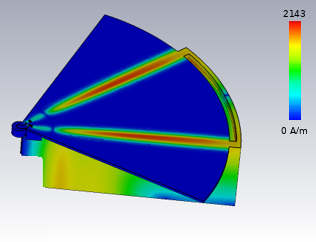}
    \caption{RF cavity: H-field horizontal cross section, calculated with CST.}
    \label{fig:RF_9_H_field_Horizontal_section}
\end{figure}

\begin{figure}
    \centering
    \includegraphics[width=0.6\linewidth]{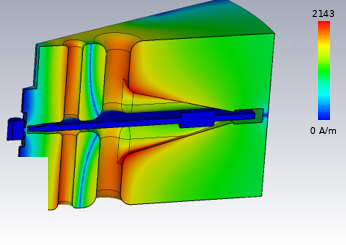}
    \caption{RF cavity: H-field vertical cross section, calculated with CST.}
    \label{fig:RF_10_H_field_vertical_section}
\end{figure}

Some illustrations of the electrical and magnetic fields computed by CST are given for reference in Figs.~\ref{fig:RF_7_E_field_Vertical_section}, \ref{fig:RF_8_E_field_Horizontal_section}, \ref{fig:RF_9_H_field_Horizontal_section} and \ref{fig:RF_10_H_field_vertical_section}.

\paragraph{Detailed power analysis}

We will now detail the post treatment of the CST results that lead to the results presented in Section~\ref{RF cavity computation results}. The CST eigenmode solver results are computed for a given reference energy in the system (about 1~J). They must then be scaled accordingly, to reach the expected voltage at a reference location. In our case, the reference location, is the orbit at 2~m, and the desired voltage is 230~kV. 

The ratio of the desired voltage to the reference voltage at 2~m is 2.013. We can scale the currents and voltages using this factor. We can also scale the energies, power and losses using this factor squared (formulae \ref{RF_eq1}, \ref{RF_eq2} and \ref{RF_eq3}). We thus have:

\begin{equation}
\label{RF_eq1}
E_{stored} = E_{stored,ref} \frac{V_{target}^2}{V_{ref}^2} 
\end{equation}

\begin{equation}
\label{RF_eq2}
P_{loss} = P_{loss,ref}{ V_{target}^2 \over V_{ref}^2}
\end{equation}

\begin{equation}
\label{RF_eq3}I{_s} = I_{s,ref}{ V_{target} \over V_{ref}}\end{equation}

Here, $E_{stored}$ and $E_{stored,ref}$ are the actual stored anergy and the reference
stored energy from the 1~J CST solution. $P_{loss}$ and $P_{loss,ref}$ are the actual power loss and the reference power loss from the 1~J CST solution. $V_{target}$ is the desired cavity accelerating voltage and $V_{ref}$ is the voltage obtained from the 1~J CST solution.

\begin{table}[tb]
\caption{Optimized cavities, functional characteristics.}
	\centering
    \renewcommand{\arraystretch}{1.25}       
    \begin{tabular}{llll}
    \hline
     & \textbf{CST reference} & \textbf{Scaled value} &  \\
    \hline\hline
    Stored energy (J) & 1 & 4.06 &  \\
    Voltage at on the 2m orbit (kV) & 114.2 & 230 &  \\
    Voltage at center (kV) & 32.4 & 65.2 &  \\
    Maximal voltage at outer edge of the dee (kV) & 124.6 & 250.9 &  \\
    Computed losses (kW) & 19.48 & 78.9 &  \\
    Maximal surface current (A/m) & 8470 & 17052.6 \\
    \hline
    \end{tabular}
\end{table}

CST provides the maximal surface current $I_{s}$ in the cavity which isn’t a directly useful value, but we can compute from it the maximal surface power dissipation $P_s$, which will be useful for designing  the cooling system sizing and layout. The physical properties needed for this determination are included in Table~\ref{tab:CavityProperties}. We first need to compute the surface resistance $R_s$ whose formula,~\ref{RF_eq4}, uses the conductivity 
$\sigma$ and the RF skin depth $\delta$ (see formulae \ref{RF_eq4}, \ref{RF_eq5} and \ref{RF_eq6}). we end up with a maximal surface dissipation Power $43.5$~W/cm$^2$. This value, despite being high, only concerns the very small surface of the dee edges, and should not be considered as an issue for the design of the cooling system. 

\begin{equation}
\label{RF_eq4}
R_{s} =  {1 \over \sigma\cdot\delta} \end{equation}

\begin{equation}
\label{RF_eq5}\delta = \sqrt{2\over {\sigma\cdot\mu\cdot \omega}}\end{equation}

\begin{equation}
\label{RF_eq6}P_{s} = R_{s}  I_{s}^2\end{equation}

Here, $\omega = 2\pi f$ is the angular frequency and $\mu = \mu_r\cdot\mu_0$ the product of relative permeability and vacuum permeability.

\begin{table}[tb]
\centering
\caption{Cavity properties for computing surface dissipation}
\begin{tabular}{cc}
\hline
Property & Value \\
\hline
\hline
Copper magnetic permeability $\mu$ & $0.99991 \times 4\pi \times 10^{-7}$~H/m \\
Copper conductivity $\sigma$  & $5.8 \times 10^{7}$~S/m \\
Angular Frequency $\omega$ & $2\pi \times 32.805 \times 10^{6}$~rad/s \\
Skin Depth $\delta$ & $11.5$~$\mu$m \\
\hline
\end{tabular}
\label{tab:CavityProperties}
\end{table}

\begin{figure}[tb]
    \centering
    \includegraphics[width=0.8\linewidth]{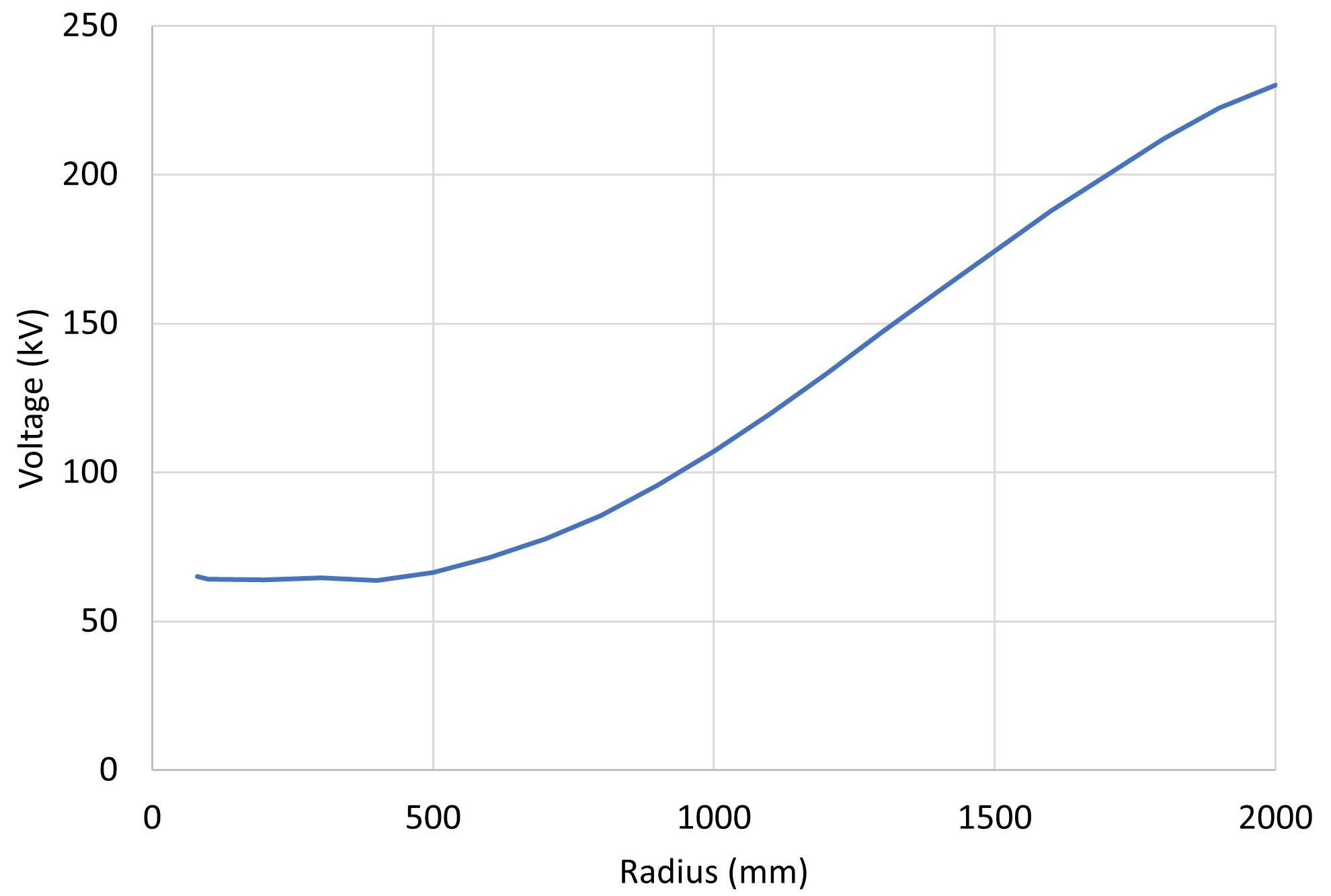}
    \caption{Radial dependency of the potential difference (``accelerating voltage'') across the accelerating gaps.}
    \label{fig:RF_13_voltage_radial_dependency}
\end{figure}

\paragraph{Radial voltage dependency}
The voltages at different radii have been calculated by integrating the electrical field along circular arcs crossing the acceleration gaps, and reference curves crossing central gap and the dee outer edge gap. The transit time factor has not been considered. The computed voltages have then been scaled to match a target voltage of 230~kV at 2~m.
The resulting absolute and normalized voltage radial dependency are given in Fig.~\ref{fig:RF_13_voltage_radial_dependency} and Table~\ref{tab:rad_dep}.

\begin{table}[tb]
    \caption{RF voltage radial dependency.}
    \label{tab:rad_dep}
	\centering
    \renewcommand{\arraystretch}{1.25}     
    \begin{tabular}{cc}
        \hline
        \textbf{Radius (mm)} &\textbf{ Voltage (kV)} \\
        \hline
        \hline
        Capture (80) & 65.17         \\
        100 & 64.07     \\
        200 & 63.95        \\
        300 & 64.60        \\
        400 & 63.76        \\
        500 & 66.37        \\
        600 & 71.30        \\
        700 & 77.60        \\
        800 & 85.50        \\
        900 & 95.67        \\
        1000 & 107.09      \\
        1100 & 119.72      \\
        1200 & 133.04      \\
        1300 & 147.22      \\
        1400 & 160.77      \\
        1500 & 174.31      \\
        1600 & 187.84      \\
        1700 & 199.73      \\
        1800 & 211.94      \\
        1900 & 222.49      \\
        Extraction (2000) & 230.00      \\
        back of dee  & 250.91\\
        \hline
    \end{tabular}
\end{table}

\subsubsection{RF Tuning System}

The tuning system is a movable plate that corrects the resonance frequency drift mainly caused by RF heating.
The system here is based on four separated tuning plates (one per cavity) and located in the median plane. These plates are located  at the back of the dee and are controlled via a ball screw that is driven by a motor. This system is similar to that of many commercial cyclotrons.

\subsubsection{RF Amplifiers}

Traditionally a large part of the cyclotron RF budget is dedicated to the RF power source.
This is especially true for this cyclotron, as it requires a high dee voltage and accelerates 600~kW of beam power. This
means that the total RF power exceeds 1 MW.   Here we assume
1200~kW of RF power, which corresponds to a 50$\%$ efficiency of RF power to beam power.  
Since the four cavities are driven separately we are speaking about 4 distinct amplifier chains of 300~kW each.

We plan to use vacuum tubes (triodes or tetrodes) to produce the RF power. We have considered using LDMOS (laterally diffused metal-oxide semiconductor) amplifiers given the recent improvements in power density, reliability, and cost. However, considering the harsh environment in question, vacuum tubes will perform more reliably and will also be better able to accommodate the rapid variations in load due to the beam temporal patterns.

\subsubsection{RF Lines and Couplers}

The RF power is fed to the cavities through EIA 6 1/8" solid copper
coax line. There are basically two ways to couple RF to the cavities:

\begin{itemize}
\item Capacitively in the median plane
\item Inductively in the bottom (or top) of the cavity liner.
\end{itemize}

Both of these systems are capable of providing the necessary 300~kW power to each cavity. Capacitive coupling is less prone to multipactoring because the coupling antenna could be easily DC biased. But, our expertise in high power coupler design advocates for an inductive solution. The design of inductive coupler is easier to water-cool, and there is a wide area where H-field is high enough for efficient coupling behind the second pillar of the cavity.  

\begin{figure}[tb]
    \centering
    \includegraphics[width=0.75\linewidth]{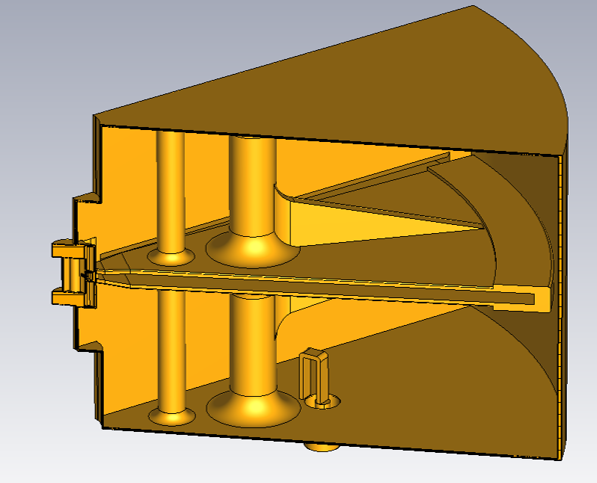}
    \caption{CAD rendering of the RF cavity showing the position and size of the loop coupler.}
    \label{fig:RF_14_coupler}
\end{figure}

We have initiated a preliminary design computation of such an inductance coupler, feeding the RF power from a EIA 6 1/8" port. The suggested design is a loop made out of a $20\times60$~mm rectangular section bar, offering room for water-cooling. See Fig.~\ref{fig:RF_14_coupler}. The adaptation tuning will be performed by rotation of the loop around the vertical axis. The design has been optimized with a default angle of 45\degree to the cavity axis, in order to allow a maximal tuning range.

\begin{figure}[tb]
    \centering
    \includegraphics[width=0.6\linewidth]{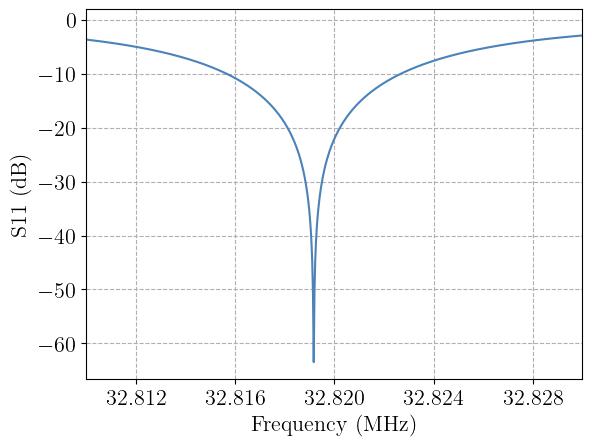}
    \caption{RF loop coupler reflected power ($S_{11}$) showing excellent coupling at -60~dB. }
    \label{fig:RF_15_coupler_port_impedance}
\end{figure}

The optimization of the coupler loop, for a cavity loaded with 150 kW of beam leads to an height of 235 mm, which provides a good matching (see Fig.~\ref{fig:RF_15_coupler_port_impedance}) and a $S_{11}$ computed scattering parameter smaller than -60~dB.

\subsubsection{LLRF Electronics}

The LLRF (Low Level RF) controls the RF amplitude and phase of the signal sent to the amplifiers. It also controls the tuning
mechanism in order to maintain the resonant frequencies in the cavities.
The amplitude and phase control is based on feed-back signals coming from the cavities and amplifiers.
In order to avoid multipactoring, the LLRF manages also a pulsing mode that provides for an easy start-up and
reduces the power dissipated in the amplifiers. One LLRF system per cavity is planned.

An additional electronic element is needed to keep a stable phase difference between the different cavities.

\subsubsection{Risks and Mitigation}
\textbf{Risk: Design and Manufacturing Process.} The Q factor and eigenfrequency depend on
the manufacturing process of the RF Cavity, e.g., surface quality and material purity. Furthermore, there is a risk of multipactoring during run-time as electrons can be accelerated back an forth in the RF field.

\noindent\emph{\textbf{Mitigation:} A 30\% margin was built into the RF power estimates, based on operational experience by IBA. An appropriate tuning range is built into the cavity design. A full scale RF cavity will be built and tested before the start of the construction of the other three cavities. Options of cavity coating exist in case of multipactoring, e.g., Aquadag.}

\subsection{Cyclotron Particle Simulations}

It is necessary to perform an in-depth particle simulation of the primary acceleration of the beam to ensure that the beam reaches energy, maintains quality, and is possible to extract. We do this using OPAL ~\cite{adelmann_opal_2019} which is a particle accelerator simulation suite developed at the Paul Scherrer Institute that accounts for space-charge effects. It is based on particle-in-cell (PIC) calculations that track a ``bunch" of particles through time-varying magnetic and electric fields. The temporal integration within OPAL is carried out using a fourth-order Runge-Kutta method. There is a mode within OPAL, OPAL-cycl, which is designed to simulate cyclotrons. This makes it well-suited for the primary acceleration phase of the IsoDAR cyclotron.  Conveniently, OPAL-cycl has built-in objects called probes, that can provide beam diagnostics, and collimators, which can provide halo reduction.

\subsubsection{Fieldmaps}
A critical aspect of these simulations is that they accurately simulate the electric and magnetic fields within the cyclotron. We performed these simulations using the latest updates to our 1D RF model discussed previously. Specifically, the field is derived from the voltages provided in Table~\ref{tab:rad_dep}. The radially increasing voltage leads to a roughly constant turn separation, which is important for ease of extraction. We conducted these simulations using a 2D median plane magnetic field that is derived from a physically realistic cyclotron model produced by IBA. Of note, this magnetic field map does not include a $\nu_{r} = 1$ resonance, which would increase the inter-turn separation in the last few turns at the price of reduced beam quality. This magnetic field is significantly more realistic than our previous work because the cyclotron model allots space for the injection components.

\subsubsection{Probes}
Probes are a built-in OPAL object that records whenever a particle passes through the region defined by the probe. These particles are simply recorded, and the trajectories of the particles are not affected by their intersection with the probe. This is helpful for determining the “extractability” of the beam. The probe collects information about the position and the momenta of the passing particles. By placing a probe at an azimuth of 25 degrees, the angle at which our electrostatic channel starts, we are able to determine the overlap between sequential turns of the beam path. This is necessary because we want to minimize this overlap for ease of extraction. The overlap between beams provides a lower bound for the power that will be deposited on the electrostatic septum, so we aim to minimize this value. The data from the probe is displayed in Fig.~\ref{fig:ProbeData}.

\clearpage
\subsubsection{Simulation Results}
\begin{figure}[t!]
    \centering
    \includegraphics[width=0.8\textwidth]{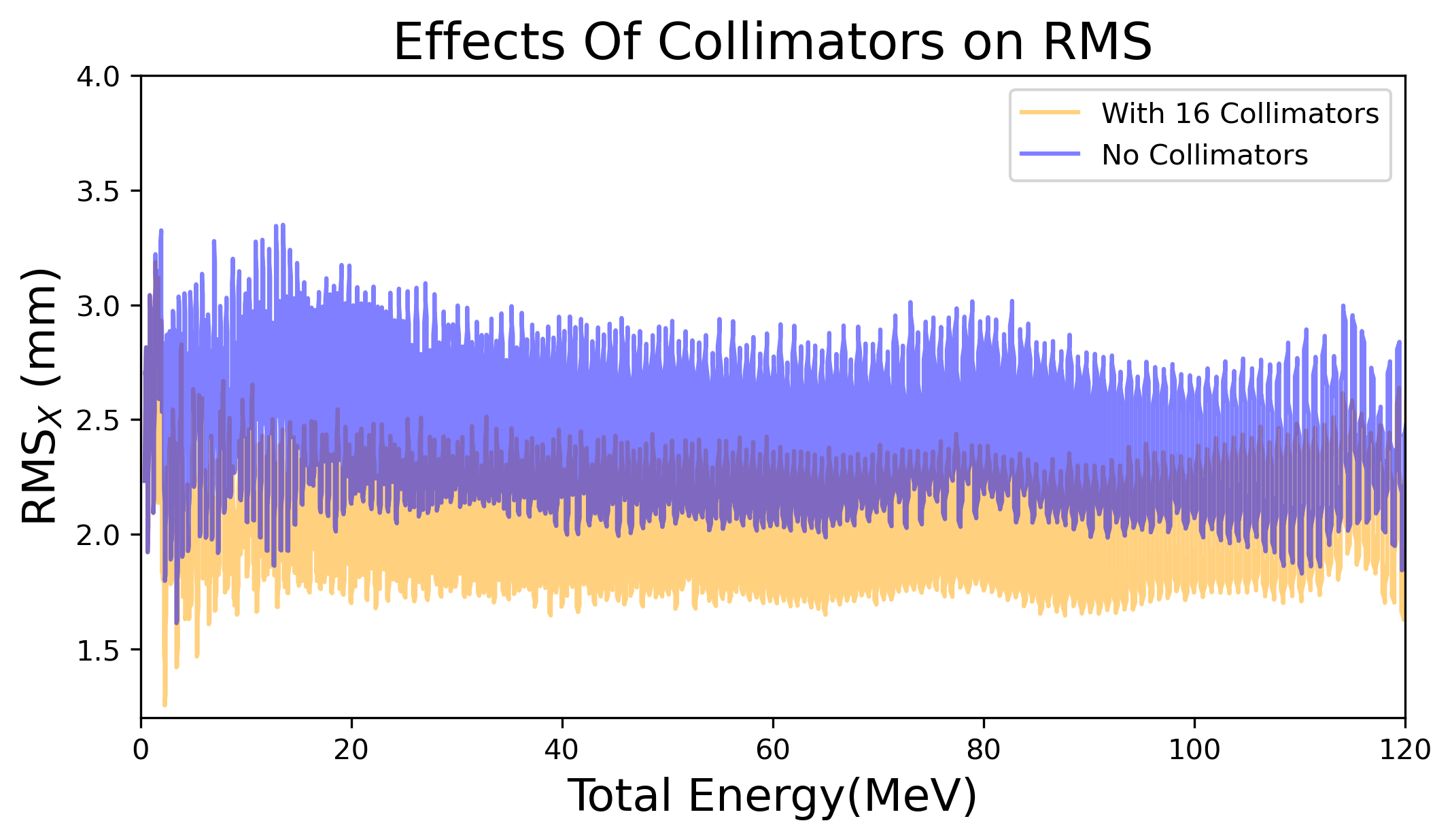}
    \caption{Rms beam size in the radial direction as a function of beam energy. We observe a much reduced radial spread in the collimated beam.}
    \label{fig:RMS_Collimation}
\end{figure}

\begin{figure}[b!]
    \centering
    \includegraphics[width=0.8\textwidth]{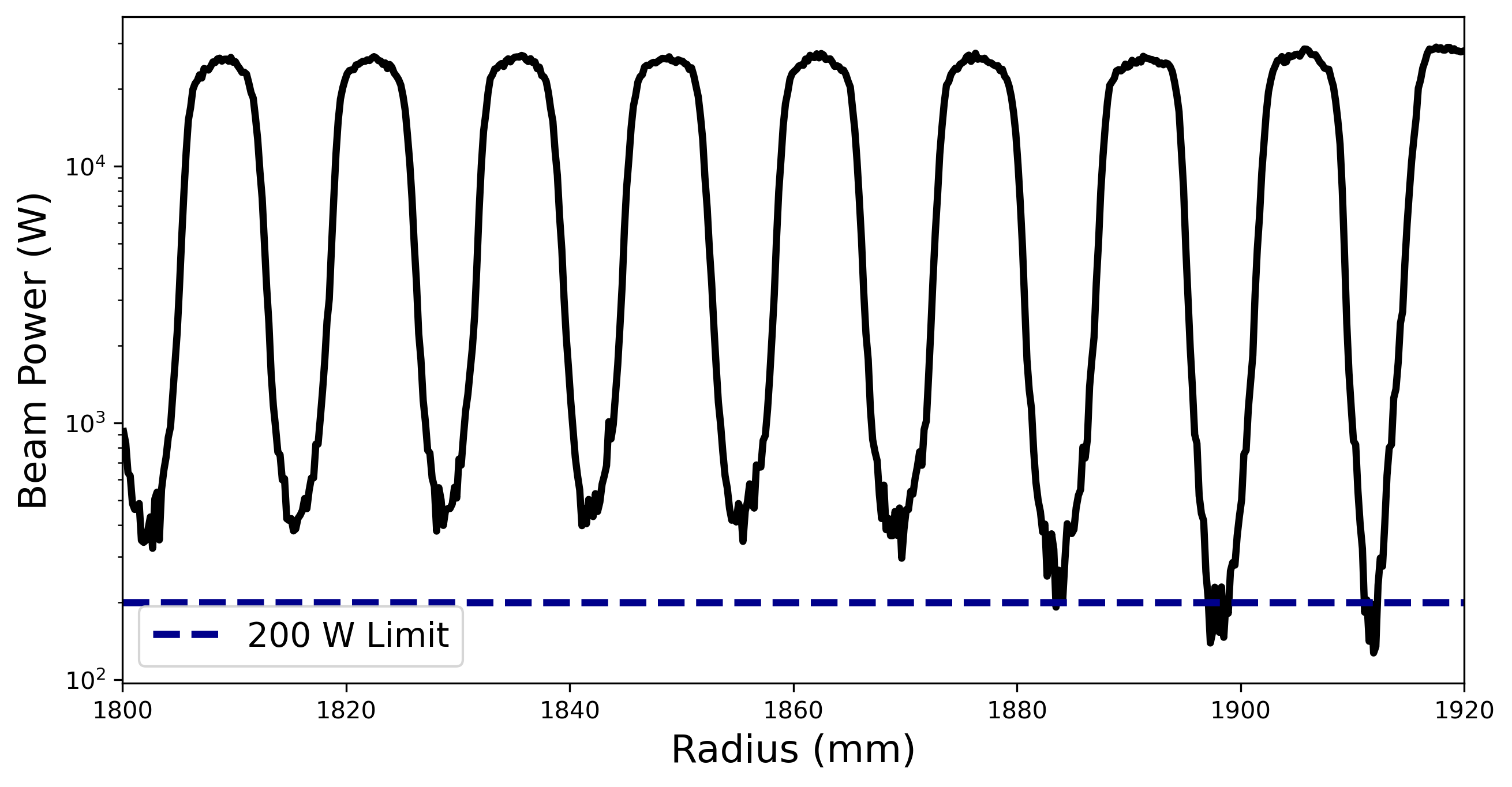}
    \caption{Beam power versus radius on a radial probe inserted at 25\degree azimuth, binned in 0.5 mm wide bins to account for the thickness of the electrostatic septum.
    The final two turns exhibit low enough inter-turn power to extract the beam.}
    \label{fig:ProbeData}
\end{figure}

\begin{figure}[h]
    \centering
    \includegraphics[width=0.8\textwidth]{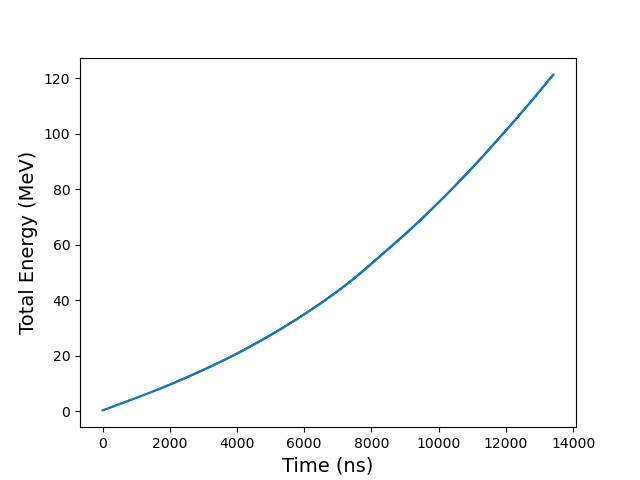}
    \caption{Beam energy versus time during acceleration in the HCHC-60. 120~MeV (60~MeV/amu) is reached smoothly after 13.5~$\mu$s.}
    \label{EnergyVsTime}
\end{figure}

Using the field maps as discussed earlier and the set of 15 collimators as described in the previous sections. We are able to smoothly accelerate the beam up to 60~MeV/amu as shown in Fig.~\ref{EnergyVsTime}. Additionally, there is minimal growth in beam size in the collimated beam as shown in Fig.~\ref{fig:RMS_Collimation}. Additionally, a probe at an azimuth of 25\degree sets a lower limit for extraction at 127~W as seen in Fig.~\ref{fig:ProbeData}.

\subsubsection{Electrostatic Extraction Channels}
\begin{figure}[tbh]
    \centering
    \includegraphics[width=0.9\textwidth]{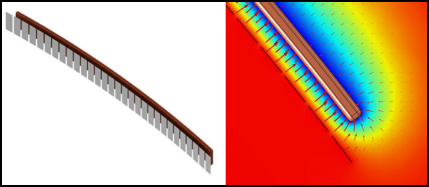}
    \caption{CAD rendering of electrostatic septum. Inventor model (left). Visualization of electric field from COMSOL (right). From Ref.~\cite{winklehner_order--magnitude_2022}.}
    \label{fig:SeptumImage}
\end{figure}

The first step in our extraction is the construction of a radially directed electric field that peels the last turn away from the second to last turn. This is done by placing a series of thin (0.2~mm) grounded electrodes between the last two turns and a larger puller electrode that is at high negative potential outside of the last turn. This generates a well-shielded electric field in the region between the puller and septum. A rendering of the electrostatic septum is shown in Fig.~\ref{fig:SeptumImage}. Two such electrostatic channels increase separation between the last two turns. 

\begin{figure}[tbh]
    \centering
    \includegraphics[width=0.9\textwidth]{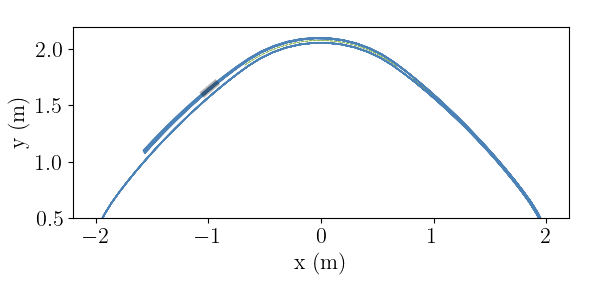}
    \caption{100 randomly selected particle trajectories out of the simulation sample of 200,000 particles as they pass through the electrostatic channels (shown in green). Also shown is the location of the first magnetic extraction element in gray.}
    \label{fig:SeptumParticles}
\end{figure}

We determine the exact location of each element of the extraction channel iteratively. The first step is to run a Python script on a pre-existing OPAL simulation of the last three turns that determines the initial positions of the septum and the puller. We chose to place the septum evenly between the last two turns. A text file is generated that allows for a model of the septum to be built in Autodesk Inventor. With this geometry, we use COMSOL to determine the electric field and add it to the pre-existing RF field of the cyclotron in OPAL. We model the physical dimensions of the septum as COLLIMATOR objects in OPAL. This process is repeated twice at which point further iteration does not result in any improvements in ESC performance. Following this we were able to reduce our losses on the first septum to 158~W which is below our 200~W goal and we lose 7~W on the second septum. In the future, we are considering reintroducing a resonance into the magnetic field which would increase the inter-turn separation at extraction and greatly reduce this number. The end result of the septum optimization process is shown in Fig.~\ref{fig:SeptumParticles}.

\subsubsection{Magnetic Extraction Channels}
\begin{figure}[t!]
    \centering
    \includegraphics[width=0.75\textwidth]{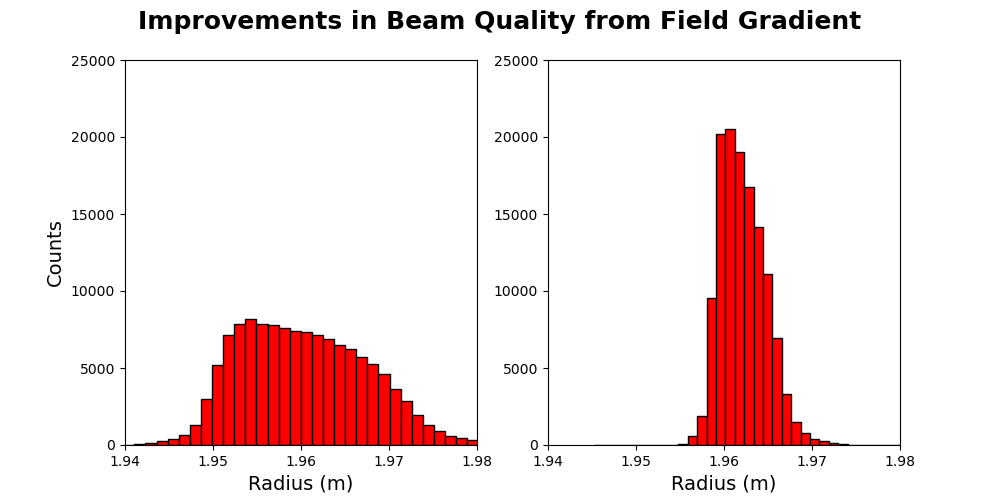}
    \caption{A 1.5~kG/cm gradient after the electrostatic channel dramatically reduces radial beam spread}
    \label{fig:MagneticGradientEffects}
\end{figure}

After the electrostatic extraction channels, the beam will rapidly de-focus as it passes through the radially decreasing magnetic field. In order to avoid this issue, we must include focusing elements that can counteract the effects of the fringe fields of the cyclotron. Due to space constraints, it is not possible to have an active element such as a quadrupole within the cyclotron. Instead, we will use the standard strategy of machining and placing iron bars along the extracted beam path. These bars allow us to locally reshape the magnetic field and are very compact. We model our magnetic extraction off of the MSU cyclotron~\cite{MSU_Cyclotron} which generates a strong radially increasing magnetic field that serves to refocus the beam after the ESC. By using the gradients that are produced at MSU as a reference point, we conservatively set a gradient of 1.5~kG/cm which is much less than the 5~kG/cm that they were able to achieve. We directly modified our magnetic field to mimic these elements. It is visible in Fig.~\ref{fig:MagneticGradientEffects} that a single relatively weak magnetic channel can improve beam quality. We do not anticipate difficulties implementing a series of such channels to guide and focus our extracted beam because we have significantly increased turn-to-turn separation following the electrostatic channels.

\subsubsection{Risks and Mitigation}

\textbf{Risk: Inadequate turn separation.} If beam bunches are not radially separated by a sufficient
amount, unacceptable beam loss will occur on the thin septum rendering its lifetime to be inadequately short.

\noindent\emph{\textbf{Mitigation:} Understanding beam dynamics sufficiently well that this does not happen, and having
adequate beam diagnostics to be able to tune the beam for maximum turn separation. Further, we could place a narrow carbon foil in front of the septum, dissociating the few \htp ions that would otherwise hit the septum. The resulting protons have lower magnetic rigidity and bend away from the septum before hitting it. They follow a specific trajectory that can guide them outside the cyclotron to be dumped in a controlled way.}

\noindent\textbf{Risk: Placement of the narrow stripping foil.} It is not obvious where the optimum location is for
the thin stripper. It must be such that stripped protons do not strike the septum, but also that halo particles do not miss the stripper and strike the septum.

\noindent\emph{\textbf{Mitigation:} Adequate beam dynamics simulations, including tracing of particle orbits.}

\section{Installation in the Yemilab Setting\label{sec:installation}}

\begin{figure}[b!]
    \centering
    \includegraphics[width=\linewidth]{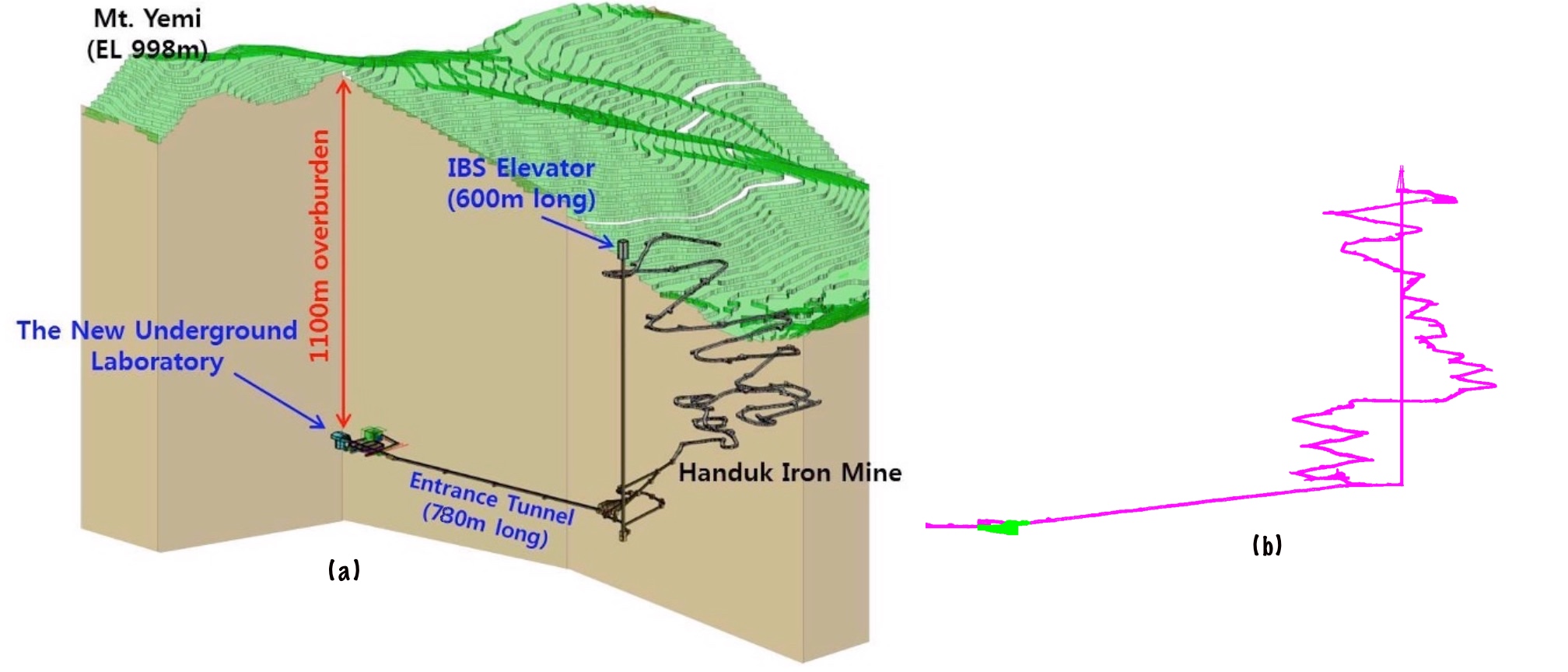}
    \caption{(a) Schematic of Handuk/Yemilab area, (b) Low resolution laser scan of Handuk mine and Yemilab access ramps, green area denotes high resolution scans of IsoDAR deployment area, shown in more detail in the next figure. From Ref.~\cite{kimYemilabNewUnderground2024}.}
    \label{Yemilab environment}
\end{figure}

\begin{figure}[hbt!]
    \centering
    \includegraphics[width=0.9\linewidth]{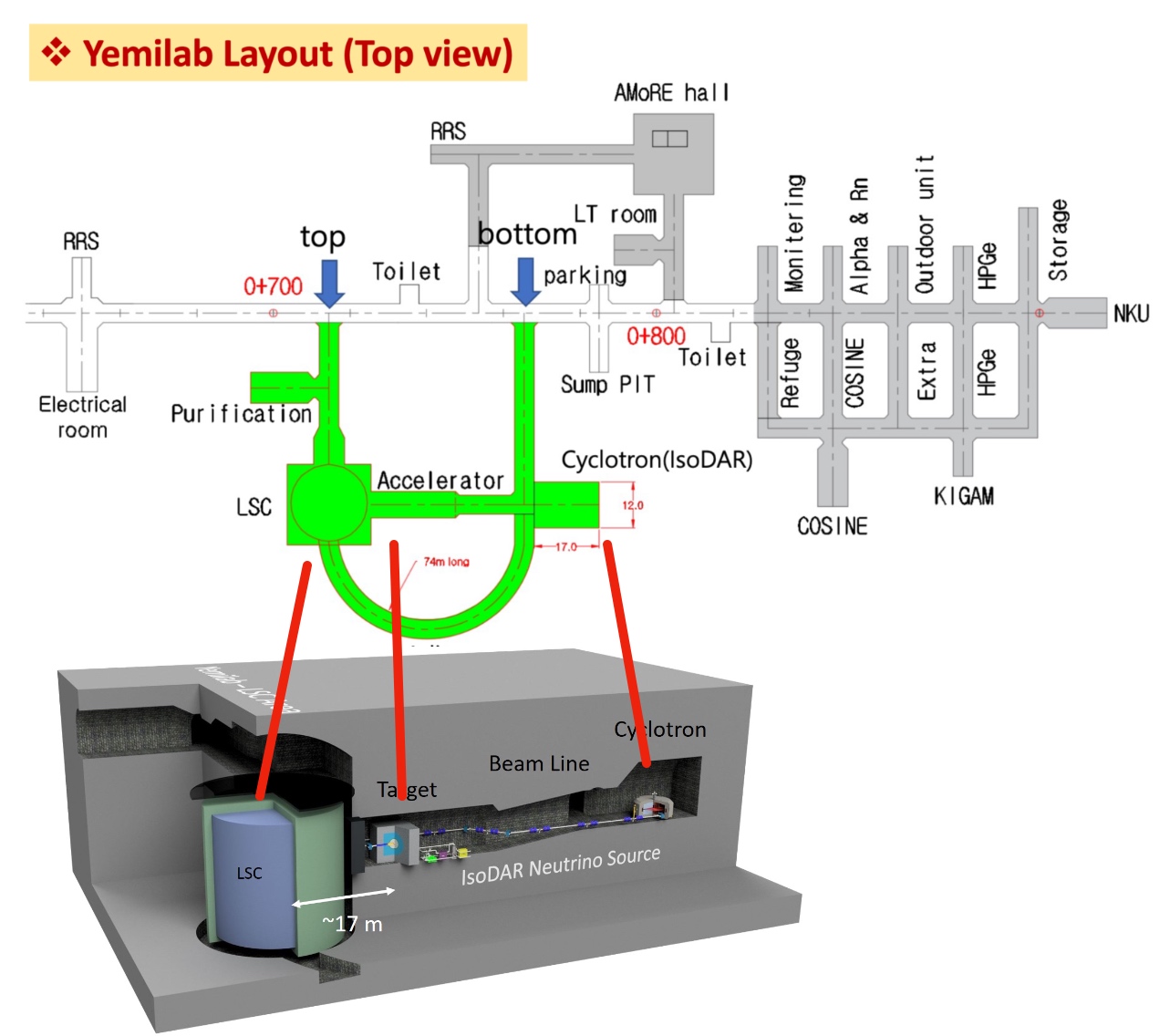}
    \caption{Layout of Yemilab, showing (in green) the area where IsoDAR will 
             be installed. The cyclotron will be assembled in the dedicated 
             cyclotron vault as shown.}
    \label{Yemilab Layout}
\end{figure}

The most likely deployment location for IsoDAR is
at the Yemilab underground laboratory, 1 km below the surface adjacent to the Handuk iron mine in the Republic of Korea.  
Fig.~\ref{Yemilab environment} schematically shows Yemilab, accessed either by a 6~km meter ramp descending 750~m at a 15\% slope from a surface access point into the mine, or by an elevator which descends down a 600~m vertical shaft. Unfortunately, the elevator is suitable only for personnel access. All heavy and bulky equipment must be brought down via the ramp. Fig.~\ref{Yemilab Layout} details the area where the IsoDAR experiment would be installed, with the target only a few meters from the planned 3~kt LSC (Liquid Scintillator Counter).
The following section provides a general description of deployment strategies in an underground setting, pertaining only to the cyclotron. In this document we use Yemilab as the example, noting that most of the issues discussed will arise in an equivalent manner at any alternative installation sites.
A similar approach will be taken for the target system in Volume II of this PDR.

\subsection{Transport of Components}

All of the components of the experiment must be brought to the site from the port-of-entry into the Republic of Korea. The preferred method is to use trucks along the highways. IBA has had prior experience with rail transport and has determined that it is difficult 
to provide adequate shock protection, and heavy pieces can be damaged in transit. As stated above, transport from the surface to the experimental site must be accomplished by way of the mine ramp  depicted in Fig.~\ref{Yemilab environment}. Nominally, its cross section is $5\times5$~m, however we are told that pinch points and sharp bends may occur and might provide challenging areas to transport large, heavy loads through.  
We have received assurance that large mining trucks, 3.5~m wide, can traverse the entire ramp, as this is the way ore was transported to the surface for many years. 

While high-resolution laser scans of the IsoDAR and LSC areas have been performed, 
(cf. Figs.~\ref{fig:cyc_w_cavern1} and \ref{fig:cyc_w_cavern2}),
the resolution of the laser scans of the access ramp areas are not of sufficient resolution to provide enough detailed information about the location or size of the possible constrictions.
Improved scans must be performed as part of final installation planning in the Technical Design Report phase (TDR).

\begin{figure}[tb]
    \centering
    \includegraphics[width=0.6\linewidth]{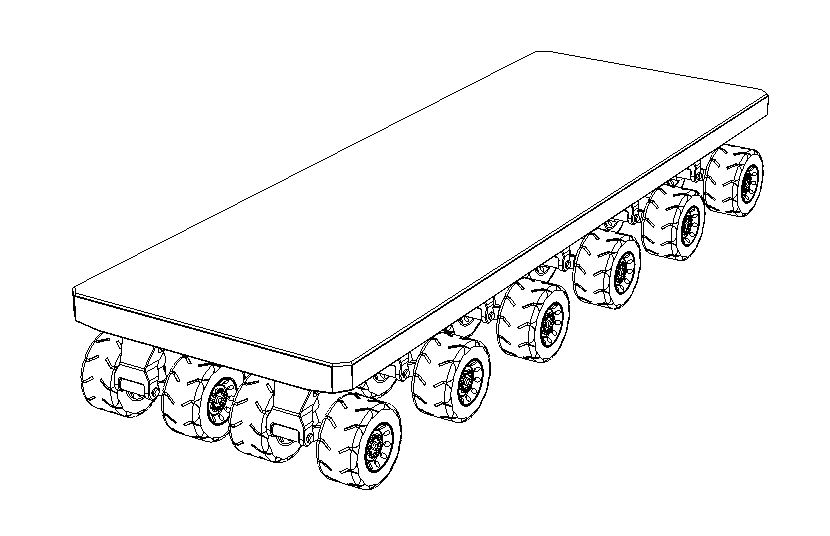}
    \caption{A typical SPMT vehicle, a versatile way to transport heavy loads over irregular terrain.}
    \label{SPMT vehicle}
\end{figure}

As seen in the previous chapter, the iron pieces that make up the cyclotron magnet, although heavy, are not in themselves large, so the pinch points are not likely to be a problem for bringing the magnet steel to the installation site.  The plan is to use an SPMT vehicle (Self-Propelled Modular Transporter) shown in Fig.~\ref{SPMT vehicle} for bringing the magnet pieces from the surface.
These electrically-powered vehicles are flexible in their size, comprising sets of 4 tires, each set  independently steerable,  the vehicle is extremely maneuverable and can easily accommodate the heaviest piece of magnet steel \cite{SPMT}.  The vacuum liner is made mostly of stainless steel pieces, but though large and bulky, the pieces can be brought down and welded in the underground assembly cavern.

\begin{figure}[tb]
    \centering
    \includegraphics[width=0.5\textwidth]{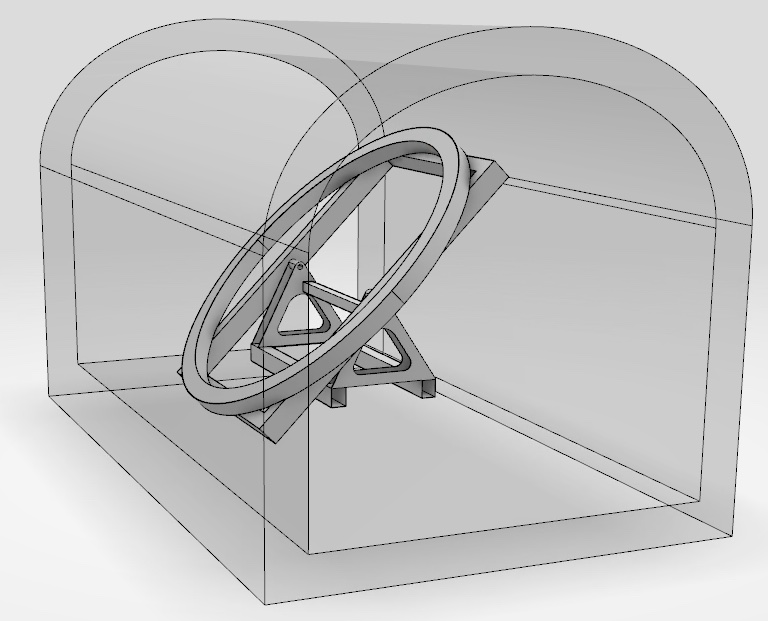}
    \caption{Isometric view of the coil package, on a specially-designed cart, transporting the coil through the access ramp to the Yemilab campus.}
    \label{coil_in_tunnel_1}
\end{figure}

\begin{figure}[tb]
    \centering
    \includegraphics[width=0.8\textwidth]{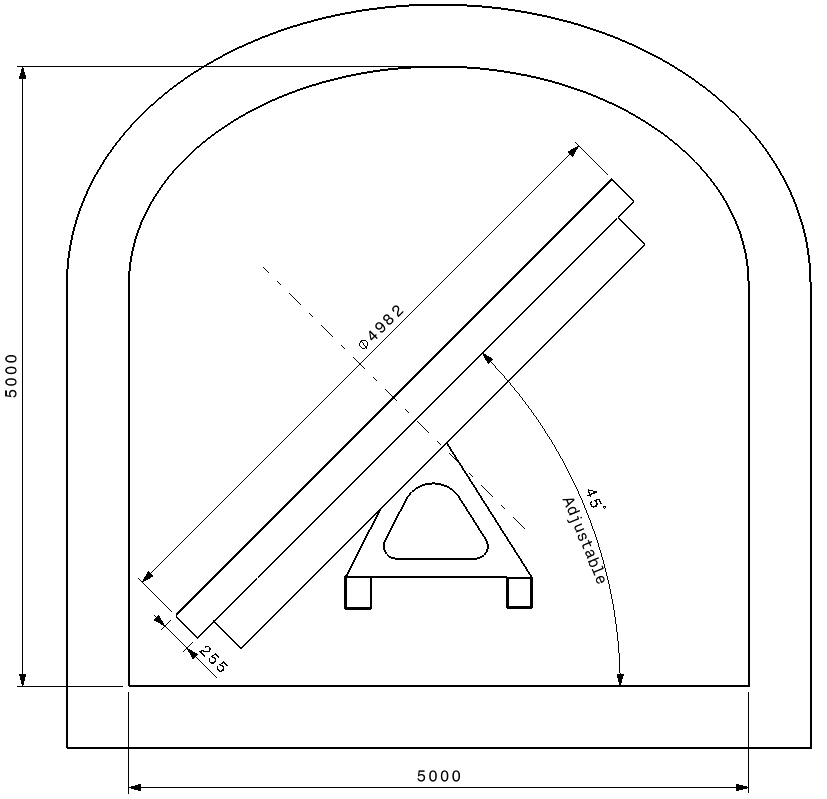}
    \caption{Front view of the coil package in the access tunnel. The ramp is nominally engineered to a 5~m width, and maximum height of 5~m. Dimensions are in mm and degrees.}
    \label{coil_in_tunnel_2}
\end{figure}

The only challenge will be the magnet coils. Each of the two coil packages measures 5 meters across and 0.3~m high, the individual coils inside measure $4.960\times0.252$~m.  Figures~\ref{coil_in_tunnel_1} and \ref{coil_in_tunnel_2} give an approximate view of a coil package and the clearances in the access ramp to the Laboratory (marked as the "Entrance Tunnel" in Fig.~\ref{Yemilab environment} (a).  This particular ramp is well-constructed and engineered, and has a measured width of 5~m.  A special mounting fixture capable of holding the coil packages at an angle should be able to bring these coils through this particular passageway.  There are constrictions even in this ramp:  cable trays and service boxes, however these can be moved if necessary.  Uncertainty remains as to whether this type of conveyance device for the coils can navigate the winding ramp from the surface.

In the unlikely case the individual pancakes do not fit through the constriction points,
the alternative is to wind and pot the coils underground, or make them in
sections.  This has already been studied in chapter~3 of the CDR document~\cite{alonso_isodaryemilab_2021, alonso_isodaryemilab_2022}. The heavy winding machine and fixtures would need to be transported, but could be broken down into parts that will fit through the ramps, then assembled on site.  The biggest challenge would be the ventilation scheme for the potting process.
There is precedence for sectioned coils. For the large TRIUMF cyclotron~\cite{TRIUMF}, coils are separated into six different segments as the outer diameter of the coils is almost 20~m, making the transport of fully assembled coils essentially impossible over conventional roadways.  
The inconvenience for a segmented coil, is that the ends of every turn of every 
section must be spliced, substantially increasing the cost and decreasing the efficiency of operation, as the coil must have as few turns as possible with very high current flowing through each turn.  

\subsection{Assembly underground}

\begin{figure}[tb]
    \centering
    \includegraphics[width=0.7\linewidth]{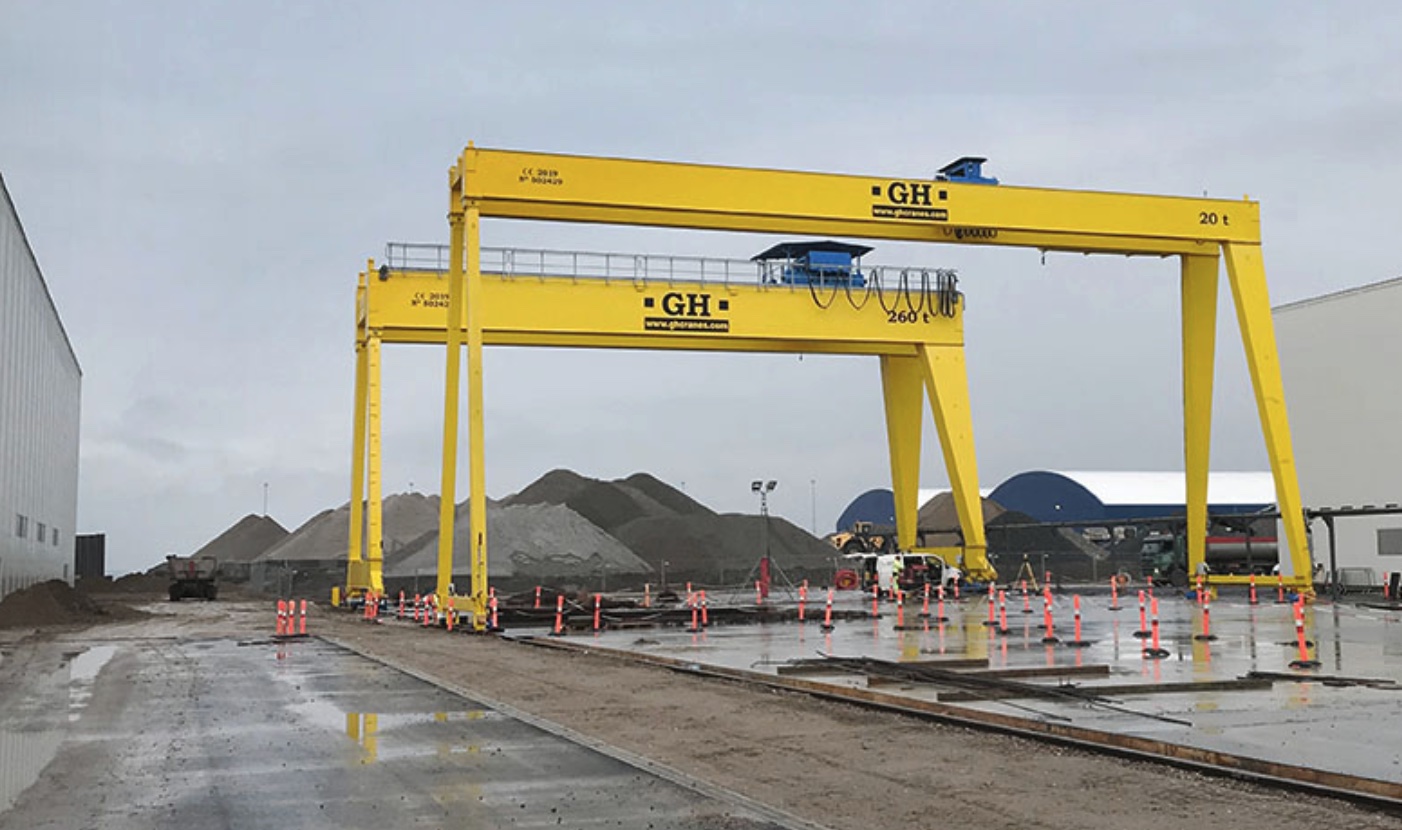}
    \caption{Examples of GH gantry cranes, capacities:  foreground 20~t, background 260 tons. From Ref.~\cite{GHcranes}.}
    \label{crane}
\end{figure}

The cyclotron will be assembled in a hall measuring 12~m wide, 17~m deep and 10~m high.  This hall was excavated to our specifications at the 90\degree bend in the access ramp to the LSC, as seen in Fig.~\ref{Yemilab Layout}.  A flat area, roughly $10\times5$~m at the corner will be the staging area for components transported from the surface.  A bridge crane riding on tracks will be used to rig parts to the cyclotron assembly point.  Conversations with a crane manufacturer, Industrias Electromecánicas GH, S.A.~\cite{GHcranes} gave us assurances that a 40~t capacity crane, riding on rails installed on the cavern floor would be quite practical (see Fig.~\ref{crane}.  It could be configured to fit the dimensions of the cavern, and broken into modular parts that could be transported to the site.  

\begin{figure}[hbt!]
\begin{minipage}[c]{0.48\linewidth}
\centering
\includegraphics[height=2.2in]{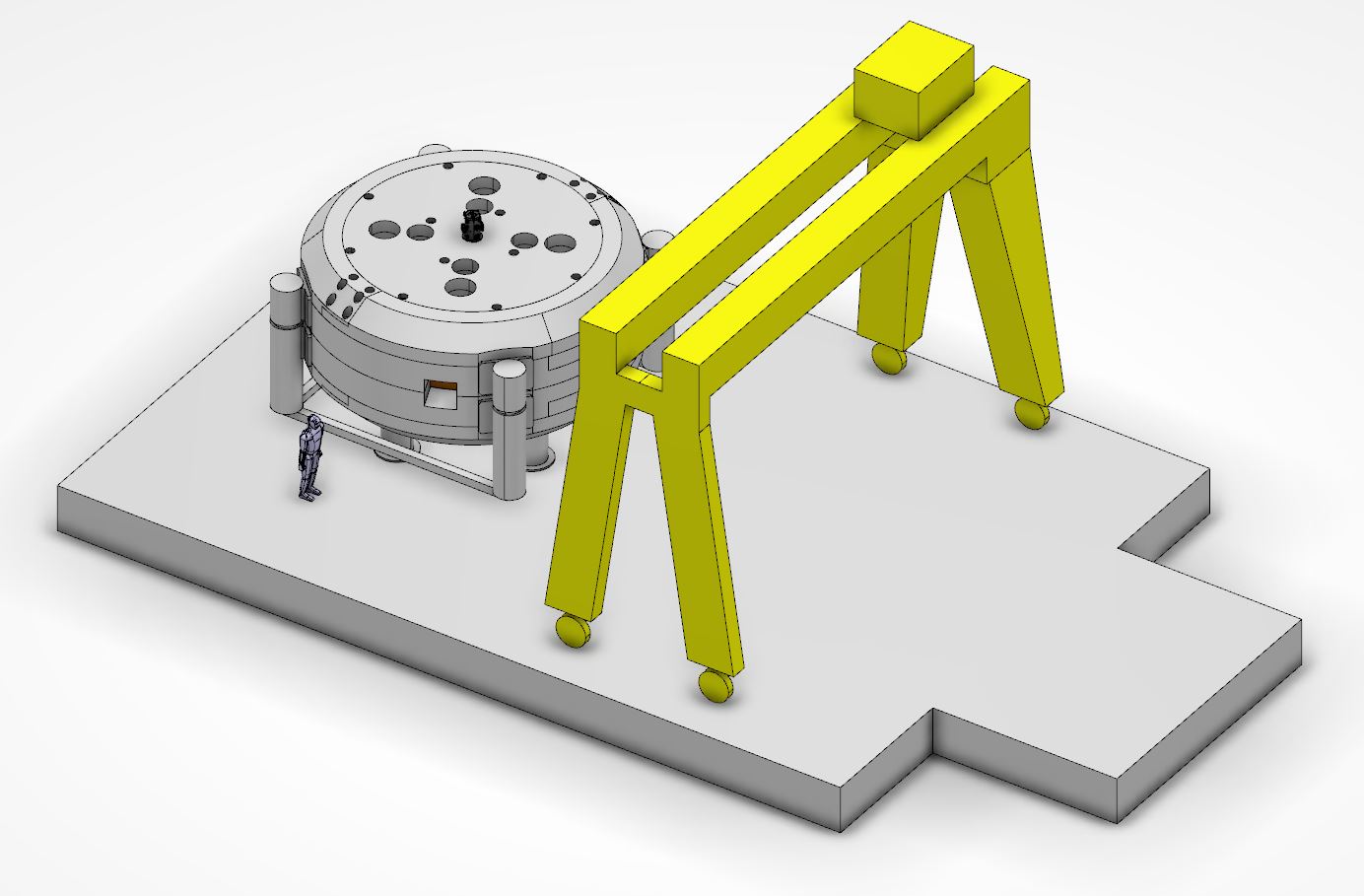}
\caption{Cyclotron and crane on cavern floor.}
\label{fig:cyc_w_cavern0}
\end{minipage}
\hfill
\begin{minipage}[c]{0.48\linewidth}
\centering
\includegraphics[height=2.2in]{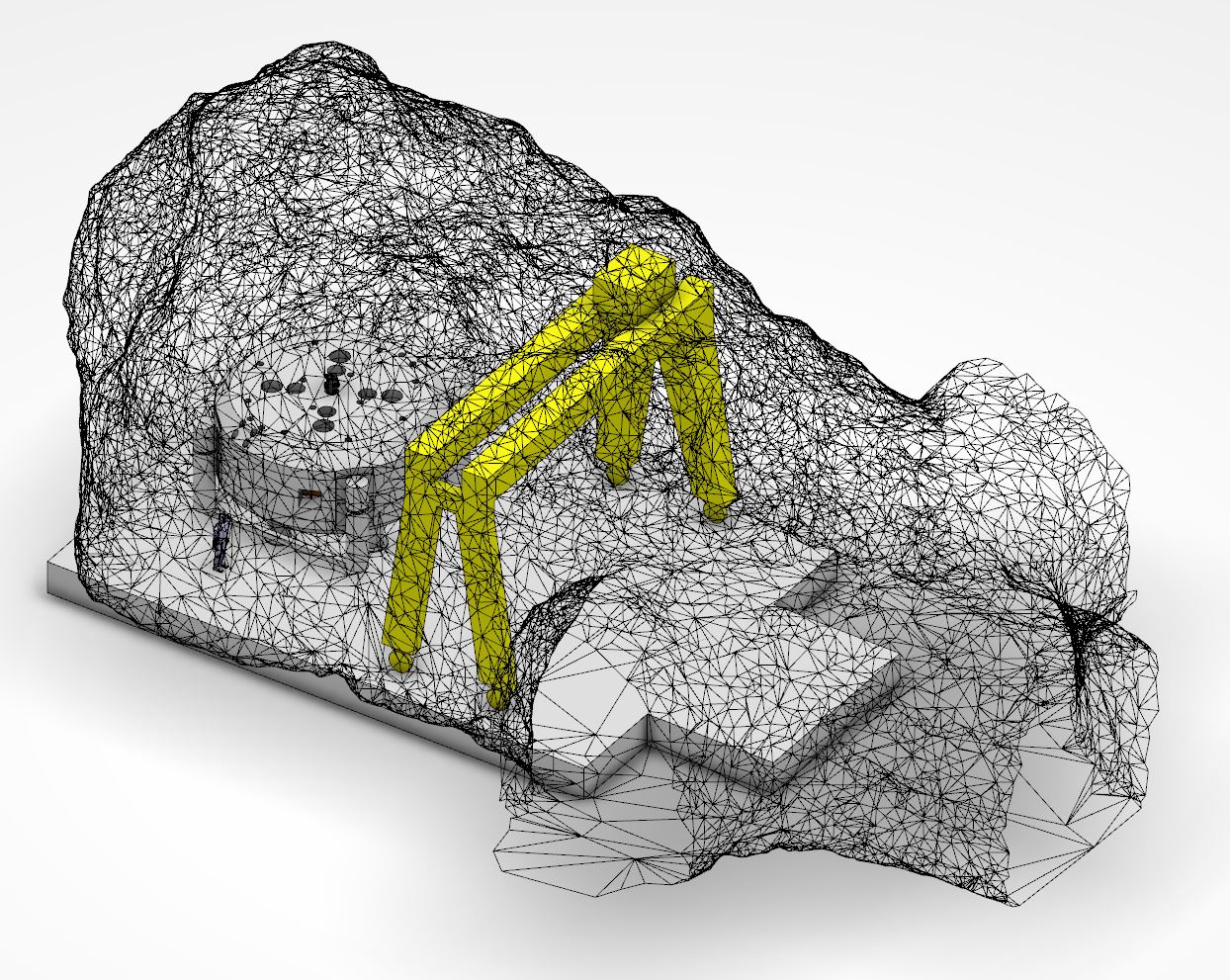}
\caption{Cyclotron and crane inside the cavern.}
\label{fig:cyc_w_cavern1}
\end{minipage}
\vspace{0.2in}
\end{figure}

\begin{figure}[ht]
    \centering
    \includegraphics[width=1.0\linewidth]{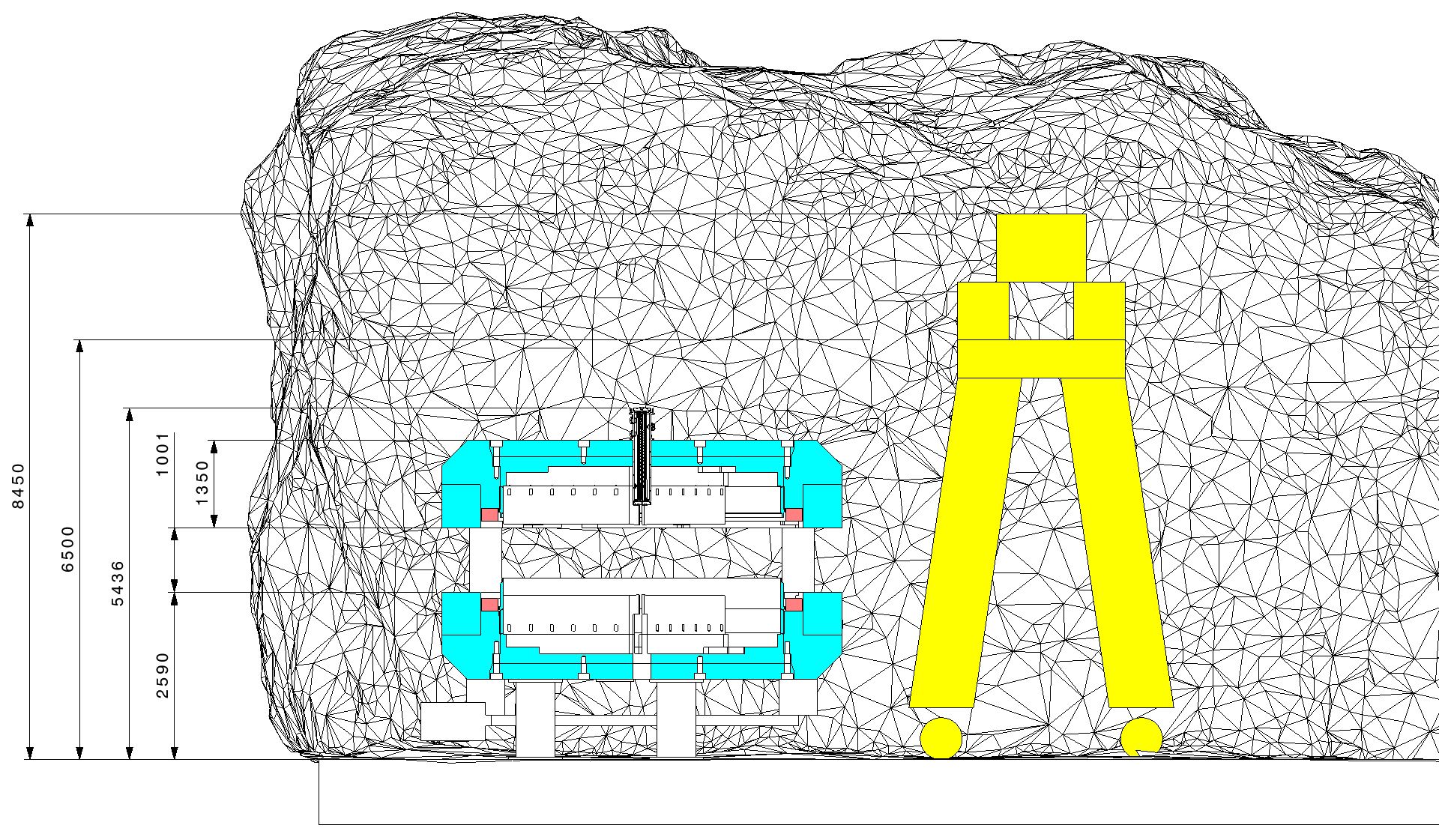}
    \caption{Cyclotron open and crane with cavern mesh - cut-view.}
    \label{fig:cyc_w_cavern2}
\end{figure}

An overview of the cyclotron inside the cavern with the rigging equipment is shown in Figures~\ref{fig:cyc_w_cavern0}~-~\ref{fig:cyc_w_cavern2}. It should be noted that the gantry crane will be dismantled once the heavy parts are rigged. This would increase the clearance above the cyclotron for the additional equipment.  
In addition, the large crane can  be reconfigured and reassembled in the target hall, and used for assembly of the MEBT, target and shielding, and for subsequent maintenance of the target systems.
Note, the crane is not needed for all routine maintenance of the cyclotron, as the top half is raised on jacks, providing access to the central plane and the injection, RF and extraction components.

\clearpage

\subsubsection{Risks and Mitigation}

\textbf{Risk: Coil Size.} 
The cyclotron coils may not fit through all of the constrictions of the mining ramp, even with the most flexible conveyance device, because of irregularities in the ramp shape and size.
     
\noindent\emph{\textbf{Mitigation:} In the case the individual pancakes do not fit through the constriction points, the alternative is to wind and pot the coils underground, or make them in
sections, as discussed above. Both come with drawbacks or additional cost, but are manageable.}

\section{Conclusion}
\subsection{Summary}
In this version 2.0 of the document, we presented the preliminary design of the IsoDAR compact cyclotron with RFQ direct injection, including ion source, low energy beam transport, electrostatic extraction channels and some thoughts on typical magnetic extraction channels. This version presents significant updates over the previously published CDR and also version 0.9 of this PDR. Most notably, we included first information about the upgraded MIST-2 ion source, a much optimized spiral inflector, a realistic cyclotron magnet and RF cavity design and particle-in-cell simulations thereof. Over version 0.9 of this document, we added detailed subsections on the simulations and subsections on risks associated with the different components as well as mitigation strategies. 

The presented design study shows the feasibility of producing a high CW \htp beam current (10~mA of protons through stripping of a 5~mA \htp beam) in a compact cyclotron. It further shows a convincing strategy to segment the cyclotron yoke for transportation into the underground environment to be assembled in situ at the final location. Below, we summarize our designs and findings for the different subsystems (ion source and LEBT, RFQ, cyclotron, underground installation).

\subsubsection{Ion Source and LEBT Summary}
Based on literature and in-depth simulations using well-established codes, we have determined that a filament-driven multicusp ion source presents an attractive solution to producing 10 - 12~mA of DC \htp beam current with high purity and low emittance. The filament lifetime 
is the main drawback. We presented our design and measurements with such an ion source 
using our own MIST-1 prototype and first tests of the upgraded MIST-2 components. Our 
measurements agree extremely well with simulations and, although we have only reached about half of the design current with the MIST-1 prototype, we are confident that we have addressed all the ``lessons learned'' from it in our MIST-2 design and foresee no showstoppers in achieving the needed beam current with it. We estimate the filament lifetime to be about a week of continuous running, which would let us achieve the required 90\% up-time.
As a fallback, flat-field ECR ion sources are a tried-and-tested alternative, albeit with the (manageable) caveats of co-producing significant amounts of protons and having a higher emittance.

The LEBT is a straightforward, highly compact design that includes safety mechanisms,
beam positioning, and beam diagnostics. Our simulations include robustness tests and 
failure-mode simulations showing its capability to handle these. 
The beam is well matched into the RFQ. Parts of the LEBT have been manufactured with the remainder to be delivered at the same time as the RFQ. Experimental verification is thus forthcoming.

\subsubsection{RFQ Summary}
Here and through references to published papers, we presented the technical design of a split-coaxial RFQ of 28~cm diameter, operating at a frequency of 32.8~MHz (matched to the cyclotron frequency). The design includes thermal and mechanical vibration studies and full 3D PIC beam dynamics calculations including all fringe fields, using the LEBT output as input beam.

The RFQ is currently being manufactured by Bevatech, GmbH in Germany. An experimental campaign
has been started to put together ion source, LEBT and RFQ, and ultimately a 1.5~MeV/amu test cyclotron (the HCHC-1.5) for the ultimate demonstration of \htp acceleration, RFQ direct injection, and establishment of vortex motion.

\subsubsection{Cyclotron Design Summary}
We presented a realistic cyclotron mechanical and electrical design consisting of the main magnet, the RF Cavities, and vacuum chamber. The magnetic and electric fields calculated from these models by FEM and BEM were used to perform highly realistic PIC simulations using the OPAL code. 

Simulations merging the RFQ output with the spiral inflector in the central region have now also been performed; completing a simulation chain running seamlessly from initial plasma generation in the ion source into the central region of the cyclotron where acceleration will begin. Simulations of the acceleration beginning with the end of the central region and concluding with extraction at the target energy of 60~MeV/amu have been performed using an input beam that was informed by a standalone central region design. Connecting the latest spiral inflector simulations with a modified version of this central region inside the new magnet model will yield a full start-to-end simulation of the IsoDAR experiment.

This section also discussed several improvements that have been explored to increase the efficiency, safety, and beam quality that emerges from the inflector into the cyclotron. 

Calculations performed with the full 3D model of the RF cavity showed manageable power and cooling. The required radial voltage distribution in the accelerating gaps, and adequate room for vacuum getter pumps.

\subsubsection{Installation Summary}
We described the layout of the Yemilab facility and how to transport the cyclotron steel 
components using SPMTs. In the cyclotron cavern, assembly will be aided by a gantry crane that can be disassembled after cyclotron installation to make room for the peripheral devices.

We identified the main risk as the cyclotron coil size being on the edge of what can be transported through the main access ramp at Yemilab. A careful investigation of possible choke points is needed for the TDR. Possible mitigation is to wind the coil in place or to adopt a split coil design as we described in the CDR.

For the next level of detail, we are working hand in hand with colleagues from Yemilab to 
identify further issues, devise an installation and removal plan, as well as overall integration plans with the facility (power, HVAC, controls, etc.).

\subsection{Outlook}
A future Technical Design Report will have to include a number of additional details, among them a full start-to-end simulation using as few beam dynamics codes as possible, benchmarked against a second set of codes; robustness analysis; a vacuum calculation (most likely using the CERN-developed code MolFlow); specifics of assembly and removal as well as shielding of the cyclotron cavern; RF system details; a commissioning strategy; data acquisition for beam diagnostics; and the cyclotron control system.

We have investigated a number of these topics and published results thereon, but including them here would have gone beyond the level of a PDR. Others are current and ongoing studies
and publications are forthcoming (well before the final TDR, much like we did here).

Most notably, we are in the process of putting together the RFQ-Direct Injection Project (RFQ-DIP), combining the MIST-2 ion source (currently commissioning), the RFQ (currently being machined), and the HCHC-1.5 cyclotron (parts being ordered), as demonstrator of all novel concepts in the HCHC-60 cyclotron design.

\backmatter





\bmhead{Acknowledgements}

This work was primarily supported by NSF grant PHY-2012897. Preliminary work leading up to this was supported by NSF grants PHY-1912764 and PHY-1626069. Work for V1.0 was supported by DOE grants DE-SC0024138 and DE-SC0024914 and NSF grant PHY-2411745.
Winklehner was also supported by the Heising-Simons Foundation.
Winkler thankfully acknowledges the support from the MIT UROP office.

\section*{Declarations}


\begin{itemize}
\item Data availability: Data is available from the authors upon reasonable request.
\end{itemize}

\bibliography{sn-bibliography}


\begin{thebibliography}{72}
\ifx \bisbn   \undefined \def \bisbn  #1{ISBN #1}\fi
\ifx \binits  \undefined \def \binits#1{#1}\fi
\ifx \bauthor  \undefined \def \bauthor#1{#1}\fi
\ifx \batitle  \undefined \def \batitle#1{#1}\fi
\ifx \bjtitle  \undefined \def \bjtitle#1{#1}\fi
\ifx \bvolume  \undefined \def \bvolume#1{\textbf{#1}}\fi
\ifx \byear  \undefined \def \byear#1{#1}\fi
\ifx \bissue  \undefined \def \bissue#1{#1}\fi
\ifx \bfpage  \undefined \def \bfpage#1{#1}\fi
\ifx \blpage  \undefined \def \blpage #1{#1}\fi
\ifx \burl  \undefined \def \burl#1{\textsf{#1}}\fi
\ifx \doiurl  \undefined \def \doiurl#1{\url{https://doi.org/#1}}\fi
\ifx \betal  \undefined \def \betal{\textit{et al.}}\fi
\ifx \binstitute  \undefined \def \binstitute#1{#1}\fi
\ifx \binstitutionaled  \undefined \def \binstitutionaled#1{#1}\fi
\ifx \bctitle  \undefined \def \bctitle#1{#1}\fi
\ifx \beditor  \undefined \def \beditor#1{#1}\fi
\ifx \bpublisher  \undefined \def \bpublisher#1{#1}\fi
\ifx \bbtitle  \undefined \def \bbtitle#1{#1}\fi
\ifx \bedition  \undefined \def \bedition#1{#1}\fi
\ifx \bseriesno  \undefined \def \bseriesno#1{#1}\fi
\ifx \blocation  \undefined \def \blocation#1{#1}\fi
\ifx \bsertitle  \undefined \def \bsertitle#1{#1}\fi
\ifx \bsnm \undefined \def \bsnm#1{#1}\fi
\ifx \bsuffix \undefined \def \bsuffix#1{#1}\fi
\ifx \bparticle \undefined \def \bparticle#1{#1}\fi
\ifx \barticle \undefined \def \barticle#1{#1}\fi
\bibcommenthead
\ifx \bconfdate \undefined \def \bconfdate #1{#1}\fi
\ifx \botherref \undefined \def \botherref #1{#1}\fi
\ifx \url \undefined \def \url#1{\textsf{#1}}\fi
\ifx \bchapter \undefined \def \bchapter#1{#1}\fi
\ifx \bbook \undefined \def \bbook#1{#1}\fi
\ifx \bcomment \undefined \def \bcomment#1{#1}\fi
\ifx \oauthor \undefined \def \oauthor#1{#1}\fi
\ifx \citeauthoryear \undefined \def \citeauthoryear#1{#1}\fi
\ifx \endbibitem  \undefined \def \endbibitem {}\fi
\ifx \bconflocation  \undefined \def \bconflocation#1{#1}\fi
\ifx \arxivurl  \undefined \def \arxivurl#1{\textsf{#1}}\fi
\csname PreBibitemsHook\endcsname

\bibitem[\protect\citeauthoryear{Alonso et~al.}{2022}]{alonso_isodaryemilab_2022}
\begin{barticle}
\bauthor{\bsnm{Alonso}, \binits{J.R.}},
\bauthor{\bsnm{Conrad}, \binits{J.M.}},
\bauthor{\bsnm{Winklehner}, \binits{D.}},
\bauthor{\bsnm{Spitz}, \binits{J.}},
\bauthor{\bsnm{Bartoszek}, \binits{L.}},
\bauthor{\bsnm{Adelmann}, \binits{A.}},
\bauthor{\bsnm{Bang}, \binits{K.M.}},
\bauthor{\bsnm{Barlow}, \binits{R.}},
\bauthor{\bsnm{Bungau}, \binits{A.}},
\bauthor{\bsnm{Calabretta}, \binits{L.}},
\bauthor{\bsnm{Kim}, \binits{Y.D.}},
\bauthor{\bsnm{Mishins}, \binits{D.}},
\bauthor{\bsnm{Park}, \binits{K.S.}},
\bauthor{\bsnm{Seo}, \binits{S.H.}},
\bauthor{\bsnm{Shaevitz}, \binits{M.}},
\bauthor{\bsnm{Voirin}, \binits{E.A.}},
\bauthor{\bsnm{Waites}, \binits{L.H.}}:
\batitle{{IsoDAR}@{Yemilab}: {A} report on the technology, capabilities, and deployment}.
\bjtitle{Journal of Instrumentation}
\bvolume{17}(\bissue{09}),
\bfpage{09042}
(\byear{2022})
\doiurl{10.1088/1748-0221/17/09/P09042} .
\bcomment{Publisher: IOP Publishing}
\end{barticle}
\endbibitem

\bibitem[\protect\citeauthoryear{Seo et~al.}{2023}]{seo_physics_2023}
\begin{botherref}
\oauthor{\bsnm{Seo}, \binits{S.-H.}},
\oauthor{\bsnm{Alonso}, \binits{J.}},
\oauthor{\bsnm{Bakhti}, \binits{P.}},
\oauthor{\bsnm{Conrad}, \binits{J.}},
\oauthor{\bsnm{Dye}, \binits{S.}},
\oauthor{\bsnm{Kim}, \binits{D.}},
\oauthor{\bsnm{Migenda}, \binits{J.}},
\oauthor{\bsnm{Pallavicini}, \binits{M.}},
\oauthor{\bsnm{Park}, \binits{J.-C.}},
\oauthor{\bsnm{Rajaee}, \binits{M.}},
\oauthor{\bsnm{Shaevitz}, \binits{M.}},
\oauthor{\bsnm{Shin}, \binits{S.}},
\oauthor{\bsnm{Spitz}, \binits{J.}},
\oauthor{\bsnm{Winklehner}, \binits{D.}},
\oauthor{\bsnm{Wronka}, \binits{S.}},
\oauthor{\bsnm{Wurm}, \binits{M.}},
\oauthor{\bsnm{Yeh}, \binits{M.}}:
Physics {Potential} of a {Few} {Kiloton} {Scale} {Neutrino} {Detector} at a {Deep} {Underground} {Lab} in {Korea}.
arXiv.
arXiv:2309.13435 [hep-ex, physics:hep-ph]
(2023).
\doiurl{10.48550/arXiv.2309.13435} .
\url{http://arxiv.org/abs/2309.13435}
\end{botherref}
\endbibitem

\bibitem[\protect\citeauthoryear{Abs et~al.}{2015}]{abs_isodarkamland_2015}
\begin{botherref}
\oauthor{\bsnm{Abs}, \binits{M.}},
\oauthor{\bsnm{Adelmann}, \binits{A.}},
\oauthor{\bsnm{Alonso}, \binits{J.R.}},
\oauthor{\bsnm{Axani}, \binits{S.}},
\oauthor{\bsnm{Barletta}, \binits{W.A.}},
\oauthor{\bsnm{Barlow}, \binits{R.}},
\oauthor{\bsnm{Bartoszek}, \binits{L.}},
\oauthor{\bsnm{Bungau}, \binits{A.}},
\oauthor{\bsnm{Calabretta}, \binits{L.}},
\oauthor{\bsnm{Calanna}, \binits{A.}},
\oauthor{\bsnm{Campo}, \binits{D.}},
\oauthor{\bsnm{Castro}, \binits{G.}},
\oauthor{\bsnm{Celona}, \binits{L.}},
\oauthor{\bsnm{Collin}, \binits{G.H.}},
\oauthor{\bsnm{Conrad}, \binits{J.M.}},
\oauthor{\bsnm{Gammino}, \binits{S.}},
\oauthor{\bsnm{Johnson}, \binits{R.}},
\oauthor{\bsnm{Karagiorgi}, \binits{G.}},
\oauthor{\bsnm{Kayser}, \binits{S.}},
\oauthor{\bsnm{Kleeven}, \binits{W.}},
\oauthor{\bsnm{Kolano}, \binits{A.}},
\oauthor{\bsnm{Labrecque}, \binits{F.}},
\oauthor{\bsnm{Loinaz}, \binits{W.A.}},
\oauthor{\bsnm{Minervini}, \binits{J.}},
\oauthor{\bsnm{Moulai}, \binits{M.H.}},
\oauthor{\bsnm{Okuno}, \binits{H.}},
\oauthor{\bsnm{Owen}, \binits{H.}},
\oauthor{\bsnm{Papavassiliou}, \binits{V.}},
\oauthor{\bsnm{Shaevitz}, \binits{M.H.}},
\oauthor{\bsnm{Shimizu}, \binits{I.}},
\oauthor{\bsnm{Shokair}, \binits{T.M.}},
\oauthor{\bsnm{Sorensen}, \binits{K.F.}},
\oauthor{\bsnm{Spitz}, \binits{J.}},
\oauthor{\bsnm{Toups}, \binits{M.}},
\oauthor{\bsnm{Vagins}, \binits{M.}},
\oauthor{\bsnm{Van~Bibber}, \binits{K.}},
\oauthor{\bsnm{Wascko}, \binits{M.O.}},
\oauthor{\bsnm{Winklehner}, \binits{D.}},
\oauthor{\bsnm{Winslow}, \binits{L.A.}},
\oauthor{\bsnm{Yang}, \binits{J.J.}}:
{IsoDAR}@{KamLAND}: {A} {Conceptual} {Design} {Report} for the {Technical} {Facility}.
arXiv: 1511.05130
(2015).
\url{http://arxiv.org/abs/1511.05130}
\end{botherref}
\endbibitem

\bibitem[\protect\citeauthoryear{Stetson et~al.}{1992}]{stetson:vortex}
\begin{bchapter}
\bauthor{\bsnm{Stetson}, \binits{J.}},
\bauthor{\bsnm{Adam}, \binits{S.}},
\bauthor{\bsnm{Humbel}, \binits{M.}},
\bauthor{\bsnm{Joho}, \binits{W.}},
\bauthor{\bsnm{Stammbach}, \binits{T.}}:
\bctitle{The commissioning of {PSI} injector 2 for high intensity, high quality beams}.
In: \bbtitle{13th International Conference on Cyclotrons and Their Applications},
p. \bfpage{4}
(\byear{1992})
\end{bchapter}
\endbibitem

\bibitem[\protect\citeauthoryear{Baumgarten}{2011}]{baumgarten:vortex1}
\begin{barticle}
\bauthor{\bsnm{Baumgarten}, \binits{C.}}:
\batitle{Transverse-longitudinal coupling by space charge in cyclotrons}.
\bjtitle{Physical Review Special Topics-Accelerators and Beams}
\bvolume{14}(\bissue{11}),
\bfpage{114201}
(\byear{2011})
\end{barticle}
\endbibitem

\bibitem[\protect\citeauthoryear{Winklehner et~al.}{2022}]{winklehner_order--magnitude_2022}
\begin{barticle}
\bauthor{\bsnm{Winklehner}, \binits{D.}},
\bauthor{\bsnm{Conrad}, \binits{J.M.}},
\bauthor{\bsnm{Schoen}, \binits{D.}},
\bauthor{\bsnm{Yampolskaya}, \binits{M.}},
\bauthor{\bsnm{Adelmann}, \binits{A.}},
\bauthor{\bsnm{Mayani}, \binits{S.}},
\bauthor{\bsnm{Muralikrishnan}, \binits{S.}}:
\batitle{Order-of-magnitude beam current improvement in compact cyclotrons}.
\bjtitle{New Journal of Physics}
\bvolume{24}(\bissue{2}),
\bfpage{023038}
(\byear{2022})
\doiurl{10.1088/1367-2630/ac5001} .
\bcomment{Publisher: IOP Publishing}
\end{barticle}
\endbibitem

\bibitem[\protect\citeauthoryear{Winklehner et~al.}{2018}]{winklehner_high_2018}
\begin{barticle}
\bauthor{\bsnm{Winklehner}, \binits{D.}},
\bauthor{\bsnm{Bahng}, \binits{J.}},
\bauthor{\bsnm{Calabretta}, \binits{L.}},
\bauthor{\bsnm{Calanna}, \binits{A.}},
\bauthor{\bsnm{Chakrabarti}, \binits{A.}},
\bauthor{\bsnm{Conrad}, \binits{J.}},
\bauthor{\bsnm{D’Agostino}, \binits{G.}},
\bauthor{\bsnm{Dechoudhury}, \binits{S.}},
\bauthor{\bsnm{Naik}, \binits{V.}},
\bauthor{\bsnm{Waites}, \binits{L.}},
\bauthor{\bsnm{Weigel}, \binits{P.}}:
\batitle{High intensity cyclotrons for neutrino physics}.
\bjtitle{Nuclear Instruments and Methods in Physics Research Section A: Accelerators, Spectrometers, Detectors and Associated Equipment}
\bvolume{907},
\bfpage{231}--\blpage{243}
(\byear{2018})
\doiurl{10.1016/j.nima.2018.07.036}
\end{barticle}
\endbibitem

\bibitem[\protect\citeauthoryear{Alonso et~al.}{2022}]{alonsoIsoDARYemilabReportTechnology2022}
\begin{barticle}
\bauthor{\bsnm{Alonso}, \binits{J.R.}},
\bauthor{\bsnm{Conrad}, \binits{J.M.}},
\bauthor{\bsnm{Winklehner}, \binits{D.}},
\bauthor{\bsnm{Spitz}, \binits{J.}},
\bauthor{\bsnm{Bartoszek}, \binits{L.}},
\bauthor{\bsnm{Adelmann}, \binits{A.}},
\bauthor{\bsnm{Bang}, \binits{K.M.}},
\bauthor{\bsnm{Barlow}, \binits{R.}},
\bauthor{\bsnm{Bungau}, \binits{A.}},
\bauthor{\bsnm{Calabretta}, \binits{L.}},
\bauthor{\bsnm{Kim}, \binits{Y.D.}},
\bauthor{\bsnm{Mishins}, \binits{D.}},
\bauthor{\bsnm{Park}, \binits{K.S.}},
\bauthor{\bsnm{Seo}, \binits{S.H.}},
\bauthor{\bsnm{Shaevitz}, \binits{M.}},
\bauthor{\bsnm{Voirin}, \binits{E.A.}},
\bauthor{\bsnm{Waites}, \binits{L.H.}}:
\batitle{{{IsoDAR}}@{{Yemilab}}: {{A}} report on the technology, capabilities, and deployment}.
\bjtitle{Journal of Instrumentation}
\bvolume{17}(\bissue{09}),
\bfpage{09042}
(\byear{2022})
\doiurl{10.1088/1748-0221/17/09/P09042}
\end{barticle}
\endbibitem

\bibitem[\protect\citeauthoryear{Kim and Lee}{2024}]{kimYemilabNewUnderground2024}
\begin{barticle}
\bauthor{\bsnm{Kim}, \binits{Y.}},
\bauthor{\bsnm{Lee}, \binits{H.S.}}:
\batitle{Yemilab, a new underground laboratory in {{Korea}}}.
\bjtitle{AAPPS Bulletin}
\bvolume{34}(\bissue{1}),
\bfpage{25}
(\byear{2024})
\doiurl{10.1007/s43673-024-00132-8}
\end{barticle}
\endbibitem

\bibitem[\protect\citeauthoryear{Alonso et~al.}{2022}]{alonso_neutrino_2022}
\begin{barticle}
\bauthor{\bsnm{Alonso}, \binits{J.R.}},
\bauthor{\bsnm{Arg\"uelles}, \binits{C.A.}},
\bauthor{\bsnm{Bungau}, \binits{A.}},
\bauthor{\bsnm{Conrad}, \binits{J.M.}},
\bauthor{\bsnm{Dutta}, \binits{B.}},
\bauthor{\bsnm{Kim}, \binits{Y.D.}},
\bauthor{\bsnm{Marzec}, \binits{E.}},
\bauthor{\bsnm{Mishins}, \binits{D.}},
\bauthor{\bsnm{Seo}, \binits{S.H.}},
\bauthor{\bsnm{Shaevitz}, \binits{M.}},
\bauthor{\bsnm{Spitz}, \binits{J.}},
\bauthor{\bsnm{Thompson}, \binits{A.}},
\bauthor{\bsnm{Waites}, \binits{L.}},
\bauthor{\bsnm{Winklehner}, \binits{D.}}:
\batitle{Neutrino physics opportunities with the {IsoDAR} source at {Yemilab}}.
\bjtitle{Physical Review D}
\bvolume{105}(\bissue{5}),
\bfpage{052009}
(\byear{2022})
\doiurl{10.1103/PhysRevD.105.052009} .
\bcomment{Publisher: American Physical Society}
\end{barticle}
\endbibitem

\bibitem[\protect\citeauthoryear{Bungau et~al.}{2012}]{bungau_proposal_2012}
\begin{barticle}
\bauthor{\bsnm{Bungau}, \binits{A.}},
\bauthor{\bsnm{Adelmann}, \binits{A.}},
\bauthor{\bsnm{Alonso}, \binits{J.R.}},
\bauthor{\bsnm{Barletta}, \binits{W.}},
\bauthor{\bsnm{Barlow}, \binits{R.}},
\bauthor{\bsnm{Bartoszek}, \binits{L.}},
\bauthor{\bsnm{Calabretta}, \binits{L.}},
\bauthor{\bsnm{Calanna}, \binits{A.}},
\bauthor{\bsnm{Campo}, \binits{D.}},
\bauthor{\bsnm{Conrad}, \binits{J.M.}},
\bauthor{\bsnm{Djurcic}, \binits{Z.}},
\bauthor{\bsnm{Kamyshkov}, \binits{Y.}},
\bauthor{\bsnm{Shaevitz}, \binits{M.H.}},
\bauthor{\bsnm{Shimizu}, \binits{I.}},
\bauthor{\bsnm{Smidt}, \binits{T.}},
\bauthor{\bsnm{Spitz}, \binits{J.}},
\bauthor{\bsnm{Wascko}, \binits{M.}},
\bauthor{\bsnm{Winslow}, \binits{L.A.}},
\bauthor{\bsnm{Yang}, \binits{J.J.}}:
\batitle{Proposal for an {Electron} {Antineutrino} {Disappearance} {Search} {Using} {High}-{Rate} $^{\textrm{8}}${Li} {Production} and {Decay}}.
\bjtitle{Physical Review Letters}
\bvolume{109}(\bissue{14}),
\bfpage{141802}
(\byear{2012})
\doiurl{10.1103/PhysRevLett.109.141802} .
\bcomment{Publisher: American Physical Society}
\end{barticle}
\endbibitem

\bibitem[\protect\citeauthoryear{Adelmann et~al.}{2012}]{adelmann_cost-effective_2012}
\begin{botherref}
\oauthor{\bsnm{Adelmann}, \binits{A.}},
\oauthor{\bsnm{Alonso}, \binits{J.R.}},
\oauthor{\bsnm{Barletta}, \binits{W.}},
\oauthor{\bsnm{Barlow}, \binits{R.}},
\oauthor{\bsnm{Bartoszek}, \binits{L.}},
\oauthor{\bsnm{Bungau}, \binits{A.}},
\oauthor{\bsnm{Calabretta}, \binits{L.}},
\oauthor{\bsnm{Calanna}, \binits{A.}},
\oauthor{\bsnm{Campo}, \binits{D.}},
\oauthor{\bsnm{Conrad}, \binits{J.M.}},
\oauthor{\bsnm{Djurcic}, \binits{Z.}},
\oauthor{\bsnm{Kamyshkov}, \binits{Y.}},
\oauthor{\bsnm{Owen}, \binits{H.}},
\oauthor{\bsnm{Shaevitz}, \binits{M.H.}},
\oauthor{\bsnm{Shimizu}, \binits{I.}},
\oauthor{\bsnm{Smidt}, \binits{T.}},
\oauthor{\bsnm{Spitz}, \binits{J.}},
\oauthor{\bsnm{Toups}, \binits{M.}},
\oauthor{\bsnm{Wascko}, \binits{M.}},
\oauthor{\bsnm{Winslow}, \binits{L.A.}},
\oauthor{\bsnm{Yang}, \binits{J.J.}}:
Cost-effective {Design} {Options} for {IsoDAR}.
arXiv: 1210.4454
(2012).
\url{http://arxiv.org/abs/1210.4454}
\end{botherref}
\endbibitem

\bibitem[\protect\citeauthoryear{J.S.~Nico et~al.}{2005}]{Nico:2004ie}
\begin{barticle}
\bauthor{\bsnm{J.S.~Nico}, \binits{M.S.D.}},
\bauthor{\bsnm{Gilliam}, \binits{D.M.}},
\bauthor{\bsnm{Wietfeldt}, \binits{F.E.}},
\bauthor{\bsnm{Fei}, \binits{X.}},
\bauthor{\bsnm{Snow}, \binits{W.M.}},
\bauthor{\bsnm{Greene}, \binits{G.L.}},
\bauthor{\bsnm{Pauwels}, \binits{J.}},
\bauthor{\bsnm{Eykens}, \binits{R.}},
\bauthor{\bsnm{Lamberty}, \binits{A.}},
\bauthor{\bsnm{Gestel}, \binits{J.V.}},
\bauthor{\bsnm{Scott}, \binits{R.D.}}:
\batitle{{Measurement of the neutron lifetime by counting trapped protons in a cold neutron beam}}.
\bjtitle{Phys. Rev. C}
\bvolume{71},
\bfpage{055502}
(\byear{2005})
\doiurl{10.1103/PhysRevC.71.055502}
{\href{https://arxiv.org/abs/nucl-ex/0411041}{{arXiv:nucl-ex/0411041}}}
\end{barticle}
\endbibitem

\bibitem[\protect\citeauthoryear{Yue et~al.}{2013}]{Yue:2013qrc}
\begin{barticle}
\bauthor{\bsnm{Yue}, \binits{A.T.}},
\bauthor{\bsnm{Dewey}, \binits{M.S.}},
\bauthor{\bsnm{Gilliam}, \binits{D.M.}},
\bauthor{\bsnm{Greene}, \binits{G.L.}},
\bauthor{\bsnm{Laptev}, \binits{A.B.}},
\bauthor{\bsnm{Nico}, \binits{J.S.}},
\bauthor{\bsnm{Snow}, \binits{W.M.}},
\bauthor{\bsnm{Wietfeldt}, \binits{F.E.}}:
\batitle{{Improved Determination of the Neutron Lifetime}}.
\bjtitle{Phys. Rev. Lett.}
\bvolume{111}(\bissue{22}),
\bfpage{222501}
(\byear{2013})
\doiurl{10.1103/PhysRevLett.111.222501}
{\href{https://arxiv.org/abs/1309.2623}{{arXiv:1309.2623}}}
{[nucl-ex]}
\end{barticle}
\endbibitem

\bibitem[\protect\citeauthoryear{Serebrov et~al.}{2018}]{Serebrov:2017bzo}
\begin{barticle}
\bauthor{\bsnm{Serebrov}, \binits{A.P.}},
\bauthor{\bsnm{Kolomensky}, \binits{E..A.}},
\bauthor{\bsnm{Fomin}, \binits{A.K.}},
\bauthor{\bsnm{Krasnoschekova}, \binits{I.A.}},
\bauthor{\bsnm{Vassiljev}, \binits{A.V.}},
\bauthor{\bsnm{Prudnikov}, \binits{D.M.}},
\bauthor{\bsnm{Shoka}, \binits{I.V.}},
\bauthor{\bsnm{Chechkin}, \binits{A.V.}},
\bauthor{\bsnm{Chaikovskiy}, \binits{M.E.}},
\bauthor{\bsnm{Varlamov}, \binits{V.E.}},
\bauthor{\bsnm{Ivanov}, \binits{S.N.}},
\bauthor{\bsnm{Pirozhkov}, \binits{A.N.}},
\bauthor{\bsnm{Geltenbort}, \binits{P.}},
\bauthor{\bsnm{Zimmer}, \binits{O.}},
\bauthor{\bsnm{Jenke}, \binits{T.}},
\bauthor{\bsnm{Grinten}, \binits{M.V.}},
\bauthor{\bsnm{Tucker}, \binits{M.}}:
\batitle{{Neutron lifetime measurements with a large gravitational trap for ultracold neutrons}}.
\bjtitle{Phys. Rev. C}
\bvolume{97}(\bissue{5}),
\bfpage{055503}
(\byear{2018})
\doiurl{10.1103/PhysRevC.97.055503}
{\href{https://arxiv.org/abs/1712.05663}{{arXiv:1712.05663}}}
{[nucl-ex]}
\end{barticle}
\endbibitem

\bibitem[\protect\citeauthoryear{Gonzalez et~al.}{2021}]{UCNt:2021pcg}
\begin{barticle}
\bauthor{\bsnm{Gonzalez}, \binits{F.M.}}, \betal:
\batitle{{Improved Neutron Lifetime Measurement with UCN\ensuremath{\tau}}}.
\bjtitle{Phys. Rev. Lett.}
\bvolume{127}(\bissue{16}),
\bfpage{162501}
(\byear{2021})
\doiurl{10.1103/PhysRevLett.127.162501}
{\href{https://arxiv.org/abs/2106.10375}{{arXiv:2106.10375}}}
{[nucl-ex]}
\end{barticle}
\endbibitem

\bibitem[\protect\citeauthoryear{Hostert et~al.}{2022}]{Hostert:2022ntu}
\begin{botherref}
\oauthor{\bsnm{Hostert}, \binits{M.}},
\oauthor{\bsnm{McKeen}, \binits{D.}},
\oauthor{\bsnm{Pospelov}, \binits{M.}},
\oauthor{\bsnm{Raj}, \binits{N.}}:
{Dark sectors in neutron-shining-through-a-wall and nuclear absorption signals}
(2022)
{\href{https://arxiv.org/abs/2201.02603}{{arXiv:2201.02603}}}
{[hep-ph]}
\end{botherref}
\endbibitem

\bibitem[\protect\citeauthoryear{Bungau et~al.}{2024}]{bungauNeutrinoYieldNeutron2024}
\begin{botherref}
\oauthor{\bsnm{Bungau}, \binits{A.}},
\oauthor{\bsnm{Alonso}, \binits{J.}},
\oauthor{\bsnm{Barlow}, \binits{R.}},
\oauthor{\bsnm{Bartozsek}, \binits{L.}},
\oauthor{\bsnm{Conrad}, \binits{J.}},
\oauthor{\bsnm{Shaevitz}, \binits{M.}},
\oauthor{\bsnm{Spitz}, \binits{J.}},
\oauthor{\bsnm{Winklehner}, \binits{D.}}:
Neutrino Yield and Neutron Shielding Calculations for a High-Power Target Installed in an Underground Setting.
arXiv
(2024).
\doiurl{10.48550/arXiv.2409.10211}
\end{botherref}
\endbibitem

\bibitem[\protect\citeauthoryear{Diaz et~al.}{2020}]{Diaz:2019fwt}
\begin{barticle}
\bauthor{\bsnm{Diaz}, \binits{A.}},
\bauthor{\bsnm{Arg\"uelles}, \binits{C.A.}},
\bauthor{\bsnm{Collin}, \binits{G.H.}},
\bauthor{\bsnm{Conrad}, \binits{J.M.}},
\bauthor{\bsnm{Shaevitz}, \binits{M.H.}}:
\batitle{{Where Are We With Light Sterile Neutrinos?}}
\bjtitle{Phys. Rept.}
\bvolume{884},
\bfpage{1}--\blpage{59}
(\byear{2020})
\doiurl{10.1016/j.physrep.2020.08.005}
{\href{https://arxiv.org/abs/1906.00045}{{arXiv:1906.00045}}}
{[hep-ex]}
\end{barticle}
\endbibitem

\bibitem[\protect\citeauthoryear{Arg{\"u}elles et~al.}{2023}]{Arguelles:2022tki}
\begin{barticle}
\bauthor{\bsnm{Arg{\"u}elles}, \binits{C.A.}}, \betal:
\batitle{{Snowmass white paper: beyond the standard model effects on neutrino flavor: Submitted to the proceedings of the US community study on the future of particle physics (Snowmass 2021)}}.
\bjtitle{Eur. Phys. J. C}
\bvolume{83}(\bissue{1}),
\bfpage{15}
(\byear{2023})
\doiurl{10.1140/epjc/s10052-022-11049-7}
{\href{https://arxiv.org/abs/2203.10811}{{arXiv:2203.10811}}}
{[hep-ph]}
\end{barticle}
\endbibitem

\bibitem[\protect\citeauthoryear{Waites et~al.}{2023}]{waites_axionlike_2023}
\begin{barticle}
\bauthor{\bsnm{Waites}, \binits{L.}},
\bauthor{\bsnm{Thompson}, \binits{A.}},
\bauthor{\bsnm{Bungau}, \binits{A.}},
\bauthor{\bsnm{Conrad}, \binits{J.M.}},
\bauthor{\bsnm{Dutta}, \binits{B.}},
\bauthor{\bsnm{Huang}, \binits{W.-C.}},
\bauthor{\bsnm{Kim}, \binits{D.}},
\bauthor{\bsnm{Shaevitz}, \binits{M.}},
\bauthor{\bsnm{Spitz}, \binits{J.}}:
\batitle{Axionlike particle production at beam dump experiments with distinct nuclear excitation lines}.
\bjtitle{Physical Review D}
\bvolume{107}(\bissue{9}),
\bfpage{095010}
(\byear{2023})
\doiurl{10.1103/PhysRevD.107.095010} .
\bcomment{Publisher: American Physical Society}
\end{barticle}
\endbibitem

\bibitem[\protect\citeauthoryear{Berezhiani}{2021}]{Berezhiani:2020vbe}
\begin{barticle}
\bauthor{\bsnm{Berezhiani}, \binits{Z.}}:
\batitle{{A possible shortcut for neutron\textendash{}antineutron oscillation through mirror world}}.
\bjtitle{Eur. Phys. J. C}
\bvolume{81}(\bissue{1}),
\bfpage{33}
(\byear{2021})
\doiurl{10.1140/epjc/s10052-020-08824-9}
{\href{https://arxiv.org/abs/2002.05609}{{arXiv:2002.05609}}}
{[hep-ph]}
\end{barticle}
\endbibitem

\bibitem[\protect\citeauthoryear{Batell et~al.}{2022}]{Batell:2022xau}
\begin{bchapter}
\bauthor{\bsnm{Batell}, \binits{B.}}, \betal:
\bctitle{{Dark Sector Studies with Neutrino Beams}}.
In: \bbtitle{{Snowmass 2021}}
(\byear{2022})
\end{bchapter}
\endbibitem

\bibitem[\protect\citeauthoryear{Peccei and Quinn}{1977}]{Peccei:1977hh}
\begin{barticle}
\bauthor{\bsnm{Peccei}, \binits{R.D.}},
\bauthor{\bsnm{Quinn}, \binits{H.R.}}:
\batitle{{CP Conservation in the Presence of Instantons}}.
\bjtitle{Phys. Rev. Lett.}
\bvolume{38},
\bfpage{1440}--\blpage{1443}
(\byear{1977})
\doiurl{10.1103/PhysRevLett.38.1440}
\end{barticle}
\endbibitem

\bibitem[\protect\citeauthoryear{Backhouse}{2021}]{Backhouse:2021qca}
\begin{botherref}
\oauthor{\bsnm{Backhouse}, \binits{D.}}:
{The Phenomenological Motivation of Axions: A Review}.
PhD thesis,
Oxford U.
(2021)
\end{botherref}
\endbibitem

\bibitem[\protect\citeauthoryear{Zyla et~al.}{2020}]{ParticleDataGroup:2020ssz}
\begin{barticle}
\bauthor{\bsnm{Zyla}, \binits{P.A.}}, \betal:
\batitle{{Review of Particle Physics}}.
\bjtitle{PTEP}
\bvolume{2020}(\bissue{8}),
\bfpage{083}--\blpage{01}
(\byear{2020})
\doiurl{10.1093/ptep/ptaa104}
\end{barticle}
\endbibitem

\bibitem[\protect\citeauthoryear{Zeller et~al.}{2002}]{NuTeV:2001whx}
\begin{barticle}
\bauthor{\bsnm{Zeller}, \binits{G.P.}}, \betal:
\batitle{{A Precise Determination of Electroweak Parameters in Neutrino Nucleon Scattering}}.
\bjtitle{Phys. Rev. Lett.}
\bvolume{88},
\bfpage{091802}
(\byear{2002})
\doiurl{10.1103/PhysRevLett.88.091802}
{\href{https://arxiv.org/abs/hep-ex/0110059}{{arXiv:hep-ex/0110059}}}.
\bcomment{[Erratum: Phys.Rev.Lett. 90, 239902 (2003)]}
\end{barticle}
\endbibitem

\bibitem[\protect\citeauthoryear{Forero and Guzzo}{2011}]{guzzo}
\begin{barticle}
\bauthor{\bsnm{Forero}, \binits{D.V.}},
\bauthor{\bsnm{Guzzo}, \binits{M.M.}}:
\batitle{Constraining nonstandard neutrino interactions with electrons}.
\bjtitle{Phys. Rev. D}
\bvolume{84},
\bfpage{013002}
(\byear{2011})
\doiurl{10.1103/PhysRevD.84.013002}
\end{barticle}
\endbibitem

\bibitem[\protect\citeauthoryear{Stammbach et~al.}{2001}]{Stammbach}
\begin{barticle}
\bauthor{\bsnm{Stammbach}, \binits{T.}},
\bauthor{\bsnm{Adam}, \binits{S.}},
\bauthor{\bsnm{Blumer}, \binits{T.}},
\bauthor{\bsnm{George}, \binits{D.}},
\bauthor{\bsnm{Mezger}, \binits{A.}},
\bauthor{\bsnm{Schmelzbach}, \binits{P.A.}},
\bauthor{\bsnm{Sigg}, \binits{P.}}:
\batitle{{The PSI 2mA beam and future applications}}.
\bjtitle{AIP Conference Proceedings}
\bvolume{600}(\bissue{1}),
\bfpage{423}--\blpage{427}
(\byear{2001})
\doiurl{10.1063/1.1435294}
{\href{https://arxiv.org/abs/https://pubs.aip.org/aip/acp/article-pdf/600/1/423/12144003/423\_1\_online.pdf}{{https://pubs.aip.org/aip/acp/article-pdf/600/1/423/12144003/423\_1\_online.pdf}}}
\end{barticle}
\endbibitem

\bibitem[\protect\citeauthoryear{Axani et~al.}{2015}]{axani_high_2015}
\begin{barticle}
\bauthor{\bsnm{Axani}, \binits{S.}},
\bauthor{\bsnm{Winklehner}, \binits{D.}},
\bauthor{\bsnm{Alonso}, \binits{J.}},
\bauthor{\bsnm{Conrad}, \binits{J.M.}}:
\batitle{A high intensity {H2}+ multicusp ion source for the isotope decay-at-rest experiment, {IsoDAR}}.
\bjtitle{Review of Scientific Instruments}
\bvolume{87}(\bissue{2}),
\bfpage{02}--\blpage{704}
(\byear{2015})
\doiurl{10.1063/1.4932395} .
\bcomment{Publisher: American Institute of Physics}
\end{barticle}
\endbibitem

\bibitem[\protect\citeauthoryear{Winklehner et~al.}{2018}]{winklehner_first_2018}
\begin{bchapter}
\bauthor{\bsnm{Winklehner}, \binits{D.}},
\bauthor{\bsnm{Axani}, \binits{S.}},
\bauthor{\bsnm{Bedard}, \binits{P.}},
\bauthor{\bsnm{Conrad}, \binits{J.}},
\bauthor{\bsnm{Corona}, \binits{J.}},
\bauthor{\bsnm{Hartwell}, \binits{F.}},
\bauthor{\bsnm{Smolsky}, \binits{J.}},
\bauthor{\bsnm{Tripathee}, \binits{A.}},
\bauthor{\bsnm{Waites}, \binits{L.}},
\bauthor{\bsnm{Weigel}, \binits{P.}},
\bauthor{\bsnm{Wester}, \binits{T.}},
\bauthor{\bsnm{Yampolskaya}, \binits{M.}}:
\bctitle{First commissioning results of the multicusp ion source at {MIT} ({MIST}-1) for {H2}+}.
In: \bbtitle{{AIP} {Conference} {Proceedings}},
vol. \bseriesno{2011},
p. \bfpage{030002}.
\bpublisher{American Institute of Physics}, \blocation{???}
(\byear{2018}).
\doiurl{10.1063/1.5053263} .
\burl{https://aip.scitation.org/doi/abs/10.1063/1.5053263}
\end{bchapter}
\endbibitem

\bibitem[\protect\citeauthoryear{Winklehner et~al.}{2021}]{winklehner_high-current_2021-1}
\begin{barticle}
\bauthor{\bsnm{Winklehner}, \binits{D.}},
\bauthor{\bsnm{Conrad}, \binits{J.M.}},
\bauthor{\bsnm{Smolsky}, \binits{J.}},
\bauthor{\bsnm{Waites}, \binits{L.H.}}:
\batitle{High-current {H2}+ beams from a filament-driven multicusp ion source}.
\bjtitle{Review of Scientific Instruments}
\bvolume{92}(\bissue{12}),
\bfpage{123301}
(\byear{2021})
\doiurl{10.1063/5.0063301} .
\bcomment{Publisher: American Institute of Physics}
\end{barticle}
\endbibitem

\bibitem[\protect\citeauthoryear{Winklehner et~al.}{2022}]{winklehner_new_2022}
\begin{barticle}
\bauthor{\bsnm{Winklehner}, \binits{D.}},
\bauthor{\bsnm{Conrad}, \binits{J.}},
\bauthor{\bsnm{Smolsky}, \binits{J.}},
\bauthor{\bsnm{Waites}, \binits{L.}},
\bauthor{\bsnm{Weigel}, \binits{P.}}:
\batitle{New {Commissioning} {Results} of the {MIST}-1 {Multicusp} {Ion} {Source}}.
\bjtitle{Journal of Physics: Conference Series}
\bvolume{2244}(\bissue{1}),
\bfpage{012013}
(\byear{2022})
\doiurl{10.1088/1742-6596/2244/1/012013} .
\bcomment{Publisher: IOP Publishing}
\end{barticle}
\endbibitem

\bibitem[\protect\citeauthoryear{Weigel et~al.}{2023}]{weigel_epics_2023}
\begin{barticle}
\bauthor{\bsnm{Weigel}, \binits{P.}},
\bauthor{\bsnm{Busza}, \binits{M.}},
\bauthor{\bsnm{Namazov}, \binits{A.}},
\bauthor{\bsnm{Park}, \binits{J.}},
\bauthor{\bsnm{Villarreal}, \binits{J.}},
\bauthor{\bsnm{Waites}, \binits{L.H.}},
\bauthor{\bsnm{Winklehner}, \binits{D.}}:
\batitle{The {EPICS} control system for {IsoDAR}}.
\bjtitle{Nuclear Instruments and Methods in Physics Research Section A: Accelerators, Spectrometers, Detectors and Associated Equipment}
\bvolume{1056},
\bfpage{168590}
(\byear{2023})
\doiurl{10.1016/j.nima.2023.168590}
\end{barticle}
\endbibitem

\bibitem[\protect\citeauthoryear{Ehlers and Leung}{1983}]{ehlers:multicusp1}
\begin{barticle}
\bauthor{\bsnm{Ehlers}, \binits{K.}},
\bauthor{\bsnm{Leung}, \binits{K.}}:
\batitle{{High-concentration H2+ or D2+ ion source}}.
\bjtitle{Review of Scientific Instruments}
\bvolume{54}(\bissue{6}),
\bfpage{677}--\blpage{680}
(\byear{1983})
\end{barticle}
\endbibitem

\bibitem[\protect\citeauthoryear{Waites}{2022}]{waites_high_2022}
\begin{botherref}
\oauthor{\bsnm{Waites}, \binits{L.}}:
High {Power} {Cyclotrons}: {The} {Bridge} {Between} {Beyond} the {Standard} {Model} {Physics}, {Computation}, and {Medical} {Applications}.
{PhD},
Massachusetts Institute of Technology,
Cambridge
(2022).
\url{https://arxiv.org/abs/2212.11114}
\end{botherref}
\endbibitem

\bibitem[\protect\citeauthoryear{Kalvas et~al.}{2010}]{kalvasIBSIMUThreedimensionalSimulation2010a}
\begin{barticle}
\bauthor{\bsnm{Kalvas}, \binits{T.}},
\bauthor{\bsnm{Tarvainen}, \binits{O.}},
\bauthor{\bsnm{Ropponen}, \binits{T.}},
\bauthor{\bsnm{Steczkiewicz}, \binits{O.}},
\bauthor{\bsnm{Ärje}, \binits{J.}},
\bauthor{\bsnm{Clark}, \binits{H.}}:
\batitle{{IBSIMU}: {A} three-dimensional simulation software for charged particle optics}.
\bjtitle{Review of Scientific Instruments}
\bvolume{81}(\bissue{2}),
\bfpage{02}--\blpage{703}
(\byear{2010})
\doiurl{10.1063/1.3258608} .
\bcomment{Publisher: American Institute of Physics}
\end{barticle}
\endbibitem

\bibitem[\protect\citeauthoryear{Vay et~al.}{2012}]{vayNovelMethodsParticleInCell2012}
\begin{barticle}
\bauthor{\bsnm{Vay}, \binits{J.-L.}},
\bauthor{\bsnm{Grote}, \binits{D.P.}},
\bauthor{\bsnm{Cohen}, \binits{R.H.}},
\bauthor{\bsnm{Friedman}, \binits{A.}}:
\batitle{Novel methods in the {Particle}-{In}-{Cell} accelerator {Code}-{Framework} {Warp}}.
\bjtitle{Computational Science \& Discovery}
\bvolume{5}(\bissue{1}),
\bfpage{014019}
(\byear{2012})
\doiurl{10.1088/1749-4699/5/1/014019} .
\bcomment{Publisher: IOP Publishing}
\end{barticle}
\endbibitem

\bibitem[\protect\citeauthoryear{Alonso et~al.}{2015}]{alonso_isodar_2015}
\begin{barticle}
\bauthor{\bsnm{Alonso}, \binits{J.}},
\bauthor{\bsnm{Axani}, \binits{S.}},
\bauthor{\bsnm{Calabretta}, \binits{L.}},
\bauthor{\bsnm{Campo}, \binits{D.}},
\bauthor{\bsnm{Celona}, \binits{L.}},
\bauthor{\bsnm{Conrad}, \binits{J.M.}},
\bauthor{\bsnm{Day}, \binits{A.}},
\bauthor{\bsnm{Castro}, \binits{G.}},
\bauthor{\bsnm{Labrecque}, \binits{F.}},
\bauthor{\bsnm{{D. Winklehner}}}:
\batitle{The {IsoDAR} high intensity {H}$_{\textrm{2}}$$^{\textrm{+}}$ transport and injection tests}.
\bjtitle{Journal of Instrumentation}
\bvolume{10}(\bissue{10}),
\bfpage{10003}
(\byear{2015})
\doiurl{10.1088/1748-0221/10/10/T10003}
\end{barticle}
\endbibitem

\bibitem[\protect\citeauthoryear{Waites et~al.}{2021}]{waites_matching_2021}
\begin{bchapter}
\bauthor{\bsnm{Waites}, \binits{L.}},
\bauthor{\bsnm{Conrad}, \binits{J.}},
\bauthor{\bsnm{Smolsky}, \binits{J.}},
\bauthor{\bsnm{Winklehner}, \binits{D.}}:
\bctitle{Matching of an {RFQ} and {Multicusp} {Ion} {Source} with {Compact} {LEBT}}.
In: \bbtitle{Proceedings of the 12th {International} {Particle} {Accelerator} {Conference}},
vol. \bseriesno{IPAC2021}.
\bpublisher{JACoW Publishing, Geneva, Switzerland}, \blocation{???}
(\byear{2021}).
\doiurl{10.18429/JACOW-IPAC2021-MOPAB159} .
\burl{https://jacow.org/ipac2021/doi/JACoW-IPAC2021-MOPAB159.html}
\end{bchapter}
\endbibitem

\bibitem[\protect\citeauthoryear{Waites et~al.}{2022}]{waites_low_2022}
\begin{barticle}
\bauthor{\bsnm{Waites}, \binits{L.}},
\bauthor{\bsnm{Conrad}, \binits{J.}},
\bauthor{\bsnm{Winklehner}, \binits{D.}}:
\batitle{A {Low} {Energy} {Beam} {Transport} to {Match} a {Multicusp} {Ion} {Source} to an {RFQ}}.
\bjtitle{Journal of Physics: Conference Series}
\bvolume{2244}(\bissue{1}),
\bfpage{012086}
(\byear{2022})
\doiurl{10.1088/1742-6596/2244/1/012086} .
\bcomment{Publisher: IOP Publishing}
\end{barticle}
\endbibitem

\bibitem[\protect\citeauthoryear{AlphaMagnetics}{}]{2562NotreDame}
\begin{botherref}
\oauthor{\bsnm{AlphaMagnetics}}:
{N}otre {D}ame {S}teering {C}oil 2.
\url{https://www.alphamag.com/communications-radar-lab-magnets/2562-notre-dame-steering-coil-2/}
Accessed 2025-03-07
\end{botherref}
\endbibitem

\bibitem[\protect\citeauthoryear{Bergoz}{}]{bergozACCT}
\begin{botherref}
\oauthor{\bsnm{Bergoz}}:
{ACCT: Precise Waveform Measurement of Long Pulses}.
{https://www.bergoz.com/products/acct/}.
{Accessed: 2010-09-30}
\end{botherref}
\endbibitem

\bibitem[\protect\citeauthoryear{Winklehner and Leitner}{}]{winklehnerSpaceChargeCompensation2015}
\begin{botherref}
\oauthor{\bsnm{Winklehner}, \binits{D.}},
\oauthor{\bsnm{Leitner}, \binits{D.}}:
A space charge compensation model for positive {{DC}} ion beams
\textbf{10}(10),
10006
\doiurl{10.1088/1748-0221/10/10/T10006}
\end{botherref}
\endbibitem

\bibitem[\protect\citeauthoryear{Ratzinger et~al.}{2020}]{ratzingerTechnicalDesignReport2020}
\begin{botherref}
\oauthor{\bsnm{Ratzinger}, \binits{U.}},
\oauthor{\bsnm{Podlech}, \binits{H.}},
\oauthor{\bsnm{Schütt}, \binits{M.}},
\oauthor{\bsnm{Koubek}, \binits{B.}},
\oauthor{\bsnm{Höltermann}, \binits{H.}},
\oauthor{\bsnm{Koser}, \binits{D.}},
\oauthor{\bsnm{Syha}, \binits{M.}}:
Technical {Design} {Report} {RFQ} for {IsoDAR}.
Technical {Report},
Bevatech, GmbH
(2020)
\end{botherref}
\endbibitem

\bibitem[\protect\citeauthoryear{Winklehner et~al.}{2021}]{winklehner_high-current_2021}
\begin{bchapter}
\bauthor{\bsnm{Winklehner}, \binits{D.}},
\bauthor{\bsnm{Conrad}, \binits{J.}},
\bauthor{\bsnm{Koser}, \binits{D.}},
\bauthor{\bsnm{Smolsky}, \binits{J.}},
\bauthor{\bsnm{Waites}, \binits{L.}}:
\bctitle{High-{Current} {H2}+ {Beams} from a {Compact} {Cyclotron} using {RFQ} {Direct} {Injection}}.
In: \bbtitle{Proceedings of the 12th {International} {Particle} {Accelerator} {Conference}},
vol. \bseriesno{IPAC2021}.
\bpublisher{JACoW Publishing, Geneva, Switzerland}, \blocation{???}
(\byear{2021}).
\doiurl{10.18429/JACOW-IPAC2021-TUXB07} .
\burl{https://jacow.org/ipac2021/doi/JACoW-IPAC2021-TUXB07.html}
\end{bchapter}
\endbibitem

\bibitem[\protect\citeauthoryear{Höltermann et~al.}{2021}]{holtermann_technical_2021}
\begin{bchapter}
\bauthor{\bsnm{Höltermann}, \binits{H.}},
\bauthor{\bsnm{Conrad}, \binits{J.}},
\bauthor{\bsnm{Koser}, \binits{D.}},
\bauthor{\bsnm{Koubek}, \binits{B.}},
\bauthor{\bsnm{Podlech}, \binits{H.}},
\bauthor{\bsnm{Ratzinger}, \binits{U.}},
\bauthor{\bsnm{Schuett}, \binits{M.}},
\bauthor{\bsnm{Smolsky}, \binits{J.}},
\bauthor{\bsnm{Syha}, \binits{M.}},
\bauthor{\bsnm{Waites}, \binits{L.}},
\bauthor{\bsnm{Winklehner}, \binits{D.}}:
\bctitle{Technical {Design} of an {RFQ} {Injector} for the {IsoDAR} {Cyclotron}}.
In: \bbtitle{Proceedings of the 12th {International} {Particle} {Accelerator} {Conference}},
vol. \bseriesno{IPAC2021}.
\bpublisher{JACoW Publishing, Geneva, Switzerland}, \blocation{???}
(\byear{2021}).
\doiurl{10.18429/JACOW-IPAC2021-THPAB167} .
\burl{https://jacow.org/ipac2021/doi/JACoW-IPAC2021-THPAB167.html}
\end{bchapter}
\endbibitem

\bibitem[\protect\citeauthoryear{Sangroula et~al.}{2021}]{sangroula_design_2021}
\begin{bchapter}
\bauthor{\bsnm{Sangroula}, \binits{M.}},
\bauthor{\bsnm{Conrad}, \binits{J.}},
\bauthor{\bsnm{Schuett}, \binits{M.}},
\bauthor{\bsnm{Winklehner}, \binits{D.}}:
\bctitle{Design and {Optimization} of a {Low} {Frequency} {RF}-{Input} {Coupler} for the {IsoDAR} {RFQ}}.
In: \bbtitle{Proceedings of the 12th {International} {Particle} {Accelerator} {Conference}},
vol. \bseriesno{IPAC2021}.
\bpublisher{JACoW Publishing, Geneva, Switzerland}, \blocation{???}
(\byear{2021}).
\doiurl{10.18429/JACOW-IPAC2021-WEPAB195} .
\burl{https://jacow.org/ipac2021/doi/JACoW-IPAC2021-WEPAB195.html}
\end{bchapter}
\endbibitem

\bibitem[\protect\citeauthoryear{Koser et~al.}{2021}]{koser_thermal_2021}
\begin{bchapter}
\bauthor{\bsnm{Koser}, \binits{D.}},
\bauthor{\bsnm{Conrad}, \binits{J.}},
\bauthor{\bsnm{Podlech}, \binits{H.}},
\bauthor{\bsnm{Ratzinger}, \binits{U.}},
\bauthor{\bsnm{Schuett}, \binits{M.}},
\bauthor{\bsnm{Winklehner}, \binits{D.}}:
\bctitle{Thermal {Analysis} of a {Compact} {Split}-{Coaxial} {CW} {RFQ} for the {IsoDAR} {RFQ}-{DIP}}.
In: \bbtitle{Proceedings of the 12th {International} {Particle} {Accelerator} {Conference}},
vol. \bseriesno{IPAC2021}.
\bpublisher{JACoW Publishing, Geneva, Switzerland}, \blocation{???}
(\byear{2021}).
\doiurl{10.18429/JACOW-IPAC2021-WEPAB202} .
\burl{https://jacow.org/ipac2021/doi/JACoW-IPAC2021-WEPAB202.html}
\end{bchapter}
\endbibitem

\bibitem[\protect\citeauthoryear{Koser et~al.}{2022}]{koser_input_2022}
\begin{botherref}
\oauthor{\bsnm{Koser}, \binits{D.}},
\oauthor{\bsnm{Waites}, \binits{L.}},
\oauthor{\bsnm{Winklehner}, \binits{D.}},
\oauthor{\bsnm{Frey}, \binits{M.}},
\oauthor{\bsnm{Adelmann}, \binits{A.}},
\oauthor{\bsnm{Conrad}, \binits{J.}}:
Input {Beam} {Matching} and {Beam} {Dynamics} {Design} {Optimizations} of the {IsoDAR} {RFQ} {Using} {Statistical} and {Machine} {Learning} {Techniques}.
Frontiers in Physics
\textbf{10}
(2022)
\end{botherref}
\endbibitem

\bibitem[\protect\citeauthoryear{Uriot}{2024}]{tracewin}
\begin{botherref}
\oauthor{\bsnm{Uriot}, \binits{D.}}:
TraceWin - Beam Dynamics Simulation Software.
Last accessed: March 7, 2025
(2024).
\url{https://irfu.cea.fr/en/Phocea/Page/index.php?id=780}
\end{botherref}
\endbibitem

\bibitem[\protect\citeauthoryear{Jongen et~al.}{2010}]{jongenCompactSuperconductingCyclotron2010b}
\begin{barticle}
\bauthor{\bsnm{Jongen}, \binits{Y.}},
\bauthor{\bsnm{Abs}, \binits{M.}},
\bauthor{\bsnm{Blondin}, \binits{A.}},
\bauthor{\bsnm{Kleeven}, \binits{W.}},
\bauthor{\bsnm{Zaremba}, \binits{S.}},
\bauthor{\bsnm{Vandeplassche}, \binits{D.}},
\bauthor{\bsnm{Aleksandrov}, \binits{V.}},
\bauthor{\bsnm{Gursky}, \binits{S.}},
\bauthor{\bsnm{Karamyshev}, \binits{O.}},
\bauthor{\bsnm{Karamysheva}, \binits{G.}},
\bauthor{\bsnm{Kazarinov}, \binits{N.}},
\bauthor{\bsnm{Kostromin}, \binits{S.}},
\bauthor{\bsnm{Morozov}, \binits{N.}},
\bauthor{\bsnm{Samsonov}, \binits{E.}},
\bauthor{\bsnm{Shirkov}, \binits{G.}},
\bauthor{\bsnm{Shevtsov}, \binits{V.}},
\bauthor{\bsnm{Syresin}, \binits{E.}},
\bauthor{\bsnm{Tuzikov}, \binits{A.}}:
\batitle{Compact superconducting cyclotron {{C400}} for hadron therapy}.
\bjtitle{Nuclear Instruments and Methods in Physics Research Section A: Accelerators, Spectrometers, Detectors and Associated Equipment}
\bvolume{624}(\bissue{1}),
\bfpage{47}--\blpage{53}
(\byear{2010})
\doiurl{10.1016/j.nima.2010.09.028}
\end{barticle}
\endbibitem

\bibitem[\protect\citeauthoryear{IBA}{2023}]{bib:c400_iron}
\begin{botherref}
\oauthor{\bsnm{IBA}}:
Cast iron specification by Ion Beam Applications S.A., as provided to foundries.
Private Communication
(2023)
\end{botherref}
\endbibitem

\bibitem[\protect\citeauthoryear{Winklehner et~al.}{2017}]{winklehner_realistic_2017}
\begin{barticle}
\bauthor{\bsnm{Winklehner}, \binits{D.}},
\bauthor{\bsnm{Adelmann}, \binits{A.}},
\bauthor{\bsnm{Gsell}, \binits{A.}},
\bauthor{\bsnm{Kaman}, \binits{T.}},
\bauthor{\bsnm{Campo}, \binits{D.}}:
\batitle{Realistic simulations of a cyclotron spiral inflector within a particle-in-cell framework}.
\bjtitle{Physical Review Accelerators and Beams}
\bvolume{20}(\bissue{12}),
\bfpage{124201}
(\byear{2017})
\doiurl{10.1103/PhysRevAccelBeams.20.124201}
\end{barticle}
\endbibitem

\bibitem[\protect\citeauthoryear{Adelmann et~al.}{2019}]{adelmann_opal_2019}
\begin{botherref}
\oauthor{\bsnm{Adelmann}, \binits{A.}},
\oauthor{\bsnm{Calvo}, \binits{P.}},
\oauthor{\bsnm{Frey}, \binits{M.}},
\oauthor{\bsnm{Gsell}, \binits{A.}},
\oauthor{\bsnm{Locans}, \binits{U.}},
\oauthor{\bsnm{Metzger-Kraus}, \binits{C.}},
\oauthor{\bsnm{Neveu}, \binits{N.}},
\oauthor{\bsnm{Rogers}, \binits{C.}},
\oauthor{\bsnm{Russell}, \binits{S.}},
\oauthor{\bsnm{Sheehy}, \binits{S.}},
\oauthor{\bsnm{Snuverink}, \binits{J.}},
\oauthor{\bsnm{Winklehner}, \binits{D.}}:
{OPAL} a {Versatile} {Tool} for {Charged} {Particle} {Accelerator} {Simulations}.
arXiv.
arXiv: 1905.06654
(2019).
\url{http://arxiv.org/abs/1905.06654}
\end{botherref}
\endbibitem

\bibitem[\protect\citeauthoryear{Geuzaine and Remacle}{2009}]{GMSHarticle}
\begin{barticle}
\bauthor{\bsnm{Geuzaine}, \binits{C.}},
\bauthor{\bsnm{Remacle}, \binits{J.-F.}}:
\batitle{Gmsh: A 3-d finite element mesh generator with built-in pre- and post-processing facilities}.
\bjtitle{International Journal for Numerical Methods in Engineering}
\bvolume{79},
\bfpage{1309}--\blpage{1331}
(\byear{2009})
\doiurl{10.1002/nme.2579}
\end{barticle}
\endbibitem

\bibitem[\protect\citeauthoryear{Smigaj et~al.}{2015}]{bempp}
\begin{botherref}
\oauthor{\bsnm{Smigaj}, \binits{W.}},
\oauthor{\bsnm{Betcke}, \binits{T.}},
\oauthor{\bsnm{Arridge}, \binits{S.}},
\oauthor{\bsnm{Phillips}, \binits{J.}},
\oauthor{\bsnm{Schweiger}, \binits{M.}}:
Solving boundary integral problems with bem++.
ACM Trans. Math. Softw.
\textbf{41}
(2015)
\doiurl{10.1145/2590830}
\end{botherref}
\endbibitem

\bibitem[\protect\citeauthoryear{Boris}{1970}]{boris}
\begin{botherref}
\oauthor{\bsnm{Boris}, \binits{J.P.}}:
Relativistic plasma simulation-optimization of a hybrid code.
Proceeding of Fourth Conference on Numerical Simulations of Plasmas
(1970)
\end{botherref}
\endbibitem

\bibitem[\protect\citeauthoryear{Weigel}{2017}]{weigelspiral}
\begin{bchapter}
\bauthor{\bsnm{Weigel}, \binits{P.}}:
\bctitle{{Design and Simulation of the IsoDAR RFQ Direct Injection System and Spiral Inflector}}.
In: \bbtitle{{M}eeting of the {APS} {D}ivision of {P}articles And {F}ields}
(\byear{2017})
\end{bchapter}
\endbibitem

\bibitem[\protect\citeauthoryear{Barnard et~al.}{2021}]{barnard_inflector}
\begin{barticle}
\bauthor{\bsnm{Barnard}, \binits{A.H.}},
\bauthor{\bsnm{Broodryk}, \binits{J.I.}},
\bauthor{\bsnm{Conradie}, \binits{J.L.}},
\bauthor{\bsnm{Villiers}, \binits{J.G.}},
\bauthor{\bsnm{Mira}, \binits{J.P.}},
\bauthor{\bsnm{Nemulodi}, \binits{F.}},
\bauthor{\bsnm{Thomae}, \binits{R.}}:
\batitle{Longitudinal and vertical focusing with a field gradient spiral inflector}.
\bjtitle{Phys. Rev. Accel. Beams}
\bvolume{24},
\bfpage{023501}
(\byear{2021})
\doiurl{10.1103/PhysRevAccelBeams.24.023501}
\end{barticle}
\endbibitem

\bibitem[\protect\citeauthoryear{Edelen et~al.}{2020}]{EdelenMachingLearning}
\begin{barticle}
\bauthor{\bsnm{Edelen}, \binits{A.}},
\bauthor{\bsnm{Neveu}, \binits{N.}},
\bauthor{\bsnm{Frey}, \binits{M.}},
\bauthor{\bsnm{Huber}, \binits{Y.}},
\bauthor{\bsnm{Mayes}, \binits{C.}},
\bauthor{\bsnm{Adelmann}, \binits{A.}}:
\batitle{Machine learning for orders of magnitude speedup in multiobjective optimization of particle accelerator systems}.
\bjtitle{Phys. Rev. Accel. Beams}
\bvolume{23},
\bfpage{044601}
(\byear{2020})
\doiurl{10.1103/PhysRevAccelBeams.23.044601}
\end{barticle}
\endbibitem

\bibitem[\protect\citeauthoryear{Koser et~al.}{2022}]{KoserMachineLearning}
\begin{barticle}
\bauthor{\bsnm{Koser}, \binits{D.}},
\bauthor{\bsnm{Waites}, \binits{L.}},
\bauthor{\bsnm{Winklehner}, \binits{D.}},
\bauthor{\bsnm{Frey}, \binits{M.}},
\bauthor{\bsnm{Adelmann}, \binits{A.}},
\bauthor{\bsnm{Conrad}, \binits{J.}}:
\batitle{{Input Beam Matching and Beam Dynamics Design Optimizations of the IsoDAR RFQ Using Statistical and Machine Learning Techniques}}.
\bjtitle{Front. in Phys.}
\bvolume{10},
\bfpage{875889}
(\byear{2022})
\doiurl{10.3389/fphy.2022.875889}
{\href{https://arxiv.org/abs/2112.02579}{{arXiv:2112.02579}}}
{[physics.acc-ph]}
\end{barticle}
\endbibitem

\bibitem[\protect\citeauthoryear{Villarreal et~al.}{2024}]{villarrealMachineLearning}
\begin{botherref}
\oauthor{\bsnm{Villarreal}, \binits{J.}},
\oauthor{\bsnm{Winklehner}, \binits{D.}},
\oauthor{\bsnm{Koser}, \binits{D.}},
\oauthor{\bsnm{Conrad}, \binits{J.}}:
Neural networks as effective surrogate models of radio-frequency quadrupole particle accelerator simulations.
Machine Learning: Science and Technology
\textbf{5}
(2024)
\doiurl{10.1088/2632-2153/ad3a30}
\end{botherref}
\endbibitem

\bibitem[\protect\citeauthoryear{Conjat and Mandrillon}{2018}]{aima:central}
\begin{botherref}
\oauthor{\bsnm{Conjat}, \binits{M.}},
\oauthor{\bsnm{Mandrillon}, \binits{P.}}:
{Central Region studies for the IsoDAR test cyclotron}.
Technical report,
AIMA Developpement
(2018)
\end{botherref}
\endbibitem

\bibitem[\protect\citeauthoryear{Omniseal}{}]{PrecisionPolymerMetal}
\begin{botherref}
\oauthor{\bsnm{Omniseal}}:
{P}recision {P}olymer \& {M}etal {S}eals | {C}ustom {E}ngineering.
\url{https://www.omniseal-solutions.com/}
Accessed 2025-03-09
\end{botherref}
\endbibitem

\bibitem[\protect\citeauthoryear{CST}{}]{cst:microwave_studio}
\begin{botherref}
\oauthor{\bsnm{CST}}:
{CST} {M}icrowave {S}tudio - 3{D} {EM} {S}imulation {S}oftware.
\url{https://www.cst.com/Products/CSTMWS}
Accessed 2015-09-01
\end{botherref}
\endbibitem

\bibitem[\protect\citeauthoryear{Vacuum}{2024}]{ModulesUHV}
\begin{botherref}
\oauthor{\bsnm{Vacuum}, \binits{S.H.}}:
Modules {UHV}
(2024).
\url{https://www.saesgetters.com/highvacuum/solution/modules-uhv/}
Accessed 2024-07-24
\end{botherref}
\endbibitem

\bibitem[\protect\citeauthoryear{Kim}{}]{MSU_Cyclotron}
\begin{botherref}
\oauthor{\bsnm{Kim}, \binits{J.-W.}}:
Magnetic fields and beam optics studies of a 250 {{MeV}} superconducting proton radiotherapy cyclotron
\textbf{582}(2),
366--373
\end{botherref}
\endbibitem

\bibitem[\protect\citeauthoryear{Mouvers}{2024}]{SPMT}
\begin{botherref}
\oauthor{\bsnm{Mouvers}}:
{SPMT} {V}oyager
(2024).
\url{https://www.mouvers.it/en/self-propelled-modular-transporter/}
Accessed 2024-02-21
\end{botherref}
\endbibitem

\bibitem[\protect\citeauthoryear{Alonso et~al.}{2021}]{alonso_isodaryemilab_2021}
\begin{botherref}
\oauthor{\bsnm{Alonso}, \binits{J.R.}},
\oauthor{\bsnm{Bang}, \binits{K.M.}},
\oauthor{\bsnm{Barlow}, \binits{R.}},
\oauthor{\bsnm{Bartoszek}, \binits{L.}},
\oauthor{\bsnm{Bungau}, \binits{A.}},
\oauthor{\bsnm{Calabretta}, \binits{L.}},
\oauthor{\bsnm{Conrad}, \binits{J.M.}},
\oauthor{\bsnm{Kayser}, \binits{S.}},
\oauthor{\bsnm{Kim}, \binits{Y.D.}},
\oauthor{\bsnm{Park}, \binits{K.S.}},
\oauthor{\bsnm{Seo}, \binits{S.H.}},
\oauthor{\bsnm{Shaevitz}, \binits{M.H.}},
\oauthor{\bsnm{Spitz}, \binits{J.}},
\oauthor{\bsnm{Waites}, \binits{L.H.}},
\oauthor{\bsnm{Winklehner}, \binits{D.}}:
{IsoDAR}@{Yemilab}: {A} {Conceptual} {Design} {Report} for the {Deployment} of the {Isotope} {Decay}-{At}-{Rest} {Experiment} in {Korea}'s {New} {Underground} {Laboratory}, {Yemilab}.
arXiv.
arXiv:2110.10635 [hep-ex, physics:physics]
(2021).
\doiurl{10.48550/arXiv.2110.10635} .
\url{http://arxiv.org/abs/2110.10635}
\end{botherref}
\endbibitem

\bibitem[\protect\citeauthoryear{Richardson}{1972}]{TRIUMF}
\begin{bbook}
\bauthor{\bsnm{Richardson}, \binits{J.R.}}:
\bbtitle{{T}he {P}resent {S}tatus Of {TRIUMF}},
vol. \bseriesno{{Proceedings of 1972 Cyclotron Conference}}
(\byear{1972})
\end{bbook}
\endbibitem

\bibitem[\protect\citeauthoryear{GHCranes}{}]{GHcranes}
\begin{botherref}
\oauthor{\bsnm{GHCranes}}:
GH Cranes.
\url{{https://www.ghcranes.com/en/}}
Accessed 2024-02-22
\end{botherref}
\endbibitem

\end{thebibliography}

\clearpage
\appendix
\section{Appendix A - List of Acronyms}
\begin{itemize}
    \item ACCT: AC Current Transformer  
    \item APS: American Physical Society  
    \item AIMA: AIMA Développement  
    \item BCSI: Best Cyclotron Systems, Inc.  
    \item BEMPP: Boundary Element Method Python Package  
    \item BSM: Beyond Standard Model  
    \item CDR: Conceptual Design Report  
    \item CERN: European Organization for Nuclear Research  
    \item COMSOL: Computational Software for Physics Simulations  
    \item CST: Computer Simulation Technology  
    \item CW: Continuous Wave  
    \item DI: De-Ionized (Water)  
    \item EPICS: Experimental Physics and Industrial Control System  
    \item ESC: Electrostatic Channel  
    \item FEM: Finite Element Method  
    \item FFA: Fixed-Field Alternating Gradient Accelerator  
    \item HCHC: High-Current Hydrogen Cyclotron  
    \item HV: High Voltage  
    \item IBA: Ion Beam Applications  
    \item IBSIMU: Ion Beam Simulation Software  
    \item IBP: International Benchmarking Panel  
    \item IBD: Inverse Beta Decay  
    \item IEEE: Institute of Electrical and Electronics Engineers  
    \item INFN: Istituto Nazionale di Fisica Nucleare  
    \item ISO: International Organization for Standardization  
    \item LEBT: Low Energy Beam Transport  
    \item LDMOS: Laterally Diffused Metal-Oxide Semiconductor  
    \item LLRF: Low-Level Radio Frequency  
    \item LS: Liquid Scintillator  
    \item LSC: Liquid Scintillator Counter (also referred to as vEYE)  
    \item MSU: Michigan State University  
    \item MEBT: Medium Energy Beam Transport  
    \item MIT: Massachusetts Institute of Technology  
    \item NME: Numerical Methods in Engineering  
    \item OPAL: Object-Oriented Parallel Accelerator Library  
    \item PSI: Paul Scherrer Institute  
    \item PID: Proportional Integral Derivative (Control Loop)  
    \item PDR: Preliminary Design Report  
    \item QCD: Quantum Chromodynamics  
    \item RF: Radio Frequency  
    \item RFQ: Radio-Frequency Quadrupole  
    \item SCCM: Standard Cubic Centimeter per Minute  
    \item SI: Spiral Inflector  
    \item SM: Standard Model  
    \item SPMT: Self-Propelled Modular Transporter  
    \item UHV: Ultra-High Vacuum  
    \item VIS: Versatile Ion Source (built at INFN-LNS in Catania) 
\end{itemize}

\end{document}